%% file: main.tex
\documentclass[acmsmall]{acmart}

\usepackage{amsmath,amssymb,amsfonts}
\usepackage{algorithmic}
\usepackage{graphicx}
\usepackage{textcomp}
\usepackage{xcolor}
\usepackage{booktabs}
\usepackage{tikz,pgfplots}
\usepackage{standalone}
\usepackage[most]{tcolorbox}
\usepackage{csquotes}
\usepackage{Figures/pgf-pie}
\usepackage{url}
\usepackage{multirow}
\usetikzlibrary{arrows.meta}
\usepackage{fancyhdr}
\usepackage{threeparttable}
\usepackage{subcaption}
\usepackage{makecell}
\usepackage{hyperref}
\usepackage{dingbat}
\usepackage{newtxmath}
\usepackage{fontawesome5}
\definecolor{beaublue}{rgb}{0.74, 0.83, 0.9}
\definecolor{applegreen}{rgb}{0.55, 0.71, 0.0}
\definecolor{dollarbill}{rgb}{0.52, 0.73, 0.4}
\newtcolorbox{quotebox}{colback=beaublue,boxrule=0.4pt,colframe=black,fonttitle=\bfseries,top=2pt,bottom=2pt}

\newcounter{findingcount}
\newcounter{suggestioncount}
\newcommand{\keybox}[1]{
\begin{tcolorbox}[leftrule=1mm,toprule=0mm,bottomrule=0mm,left=1pt,right=2pt,top=2pt,bottom=2pt, colback=beaublue]
\em #1
\end{tcolorbox}
}

\newcommand{\suggestbox}[1]{
\begin{tcolorbox}[leftrule=1mm,toprule=0mm,bottomrule=0mm,left=1pt,right=2pt,top=2pt,bottom=2pt, colback=dollarbill!50]
\em #1
\end{tcolorbox}
}

\makeatletter
\tcbset{
    myhbox/.style 2 args={%
        enhanced, 
        breakable,
        colback=white,
        colframe=blue!30!black,
        attach boxed title to top left={yshift*=-\tcboxedtitleheight}, 
        title={#2},
        boxed title size=title,
        boxed title style={%
            sharp corners, 
            rounded corners=northwest, 
            colback=tcbcolframe, 
            boxrule=0pt,
        },
        underlay boxed title={%
            \path[fill=tcbcolframe] (title.south west)--(title.south east) 
                to[out=0, in=180] ([xshift=5mm]title.east)--
                (title.center-|frame.east)
                [rounded corners=\kvtcb@arc] |- 
                (frame.north) -| cycle; 
        },
        #1
    }
}   
\makeatother

\newtcolorbox{myhbox}[2][]{%
    myhbox={#1}{#2}
}

\AtBeginDocument{%
  \providecommand\BibTeX{{%
    \normalfont B\kern-0.5em{\scshape i\kern-0.25em b}\kern-0.8em\TeX}}}

\DeclareMathAlphabet\mathbfcal{OMS}{cmsy}{b}{n}
\setcopyright{rightsretained}
\acmJournal{TOSEM}
\acmYear{2024} \acmVolume{1} \acmNumber{1} \acmArticle{1} \acmMonth{1}\acmDOI{10.1145/3702986}

\acmJournal{TOSEM}
\acmVolume{37}
\acmNumber{4}
\acmArticle{111}
\acmMonth{9}

\begin{document}

\title{Deep Configuration Performance Learning: A Systematic Survey and Taxonomy}

\author{Jingzhi Gong}
\authornote{This research was conducted when Jingzhi Going visited the University of Electronic Science and Technology of China.}
\email{j.gong@lboro.ac.uk}
\orcid{0000-0003-4551-0701}
\affiliation{%
\institution{University of Electronic Science and Technology of China}
  \city{Chengdu}
\country{China;}
\institution{Loughborough University}
\city{Loughborough}
\country{UK}
 }

\author{Tao Chen}
\authornote{Corresponding author: Tao Chen, t.chen@bham.ac.uk.}
\email{t.chen@bham.ac.uk}
\orcid{0000-0001-5025-5472}
\affiliation{
  \institution{University of Birmingham}
  \city{Birmingham}
  \country{UK}
}


\input{abstract}



\maketitle

\input{introduction}

\input{background}

\input{methodology}

\input{RQ1}
\input{RQ2}
\input{RQ3}
\input{RQ4}

\input{oppertunities}

\input{threats}

\input{conclusion}

\begin{acks}
This work was supported by an NSFC Grant (62372084) and a UKRI Grant (10054084).
\end{acks}

\bibliographystyle{ACM-Reference-Format}
\bibliography{references}


\end{document}

%% file: abstract.tex
\begin{abstract}
Performance is arguably the most crucial attribute that reflects the quality of a configurable software system. However, given the increasing scale and complexity of modern software, modeling and predicting how various configurations can impact performance becomes one of the major challenges in software maintenance. As such, performance is often modeled without having a thorough knowledge of the software system, but relying mainly on data, which fits precisely with the purpose of deep learning. In this paper, we conduct a comprehensive review exclusively on the topic of deep learning for performance learning of configurable software, covering 1,206 searched papers spanning six indexing services, based on which 99 primary papers were extracted and analyzed. Our results outline key statistics, taxonomy, strengths, weaknesses, and optimal usage scenarios for techniques related to the preparation of configuration data, the construction of deep learning performance models, the evaluation of these models, and their utilization in various software configuration-related tasks. We also identify the good practices and potentially problematic phenomena from the studies surveyed, together with a comprehensive summary of actionable suggestions and insights into future opportunities within the field. To promote open science, all the raw results of this survey can be accessed at our repository: \textcolor{blue}{\url{https://github.com/ideas-labo/DCPL-SLR}}.

\end{abstract}

%% file: introduction.tex
\section{Introduction}
\label{sec:introduciton}

Configuration is pervasive in software systems, ranging from enterprise web applications to robotic software. The intention thereof is simple: by allowing software systems to be flexibly configured in different ways, configuration enables them to have much better applicability over a wide range of domains and a better ability to cope with varying performance requirements, such as latency, throughput, and energy consumption~\cite{DBLP:conf/kbse/LiXCT20,DBLP:conf/icse/JamshidiVKSK17, DBLP:conf/icse/Chen19b, DBLP:journals/tosem/ChengGZ23,DBLP:conf/icse/LiX0WT20}. For example, a popular web server \textsc{Tomcat} can adjust different configuration options (e.g., \texttt{maxThreads}), the values of which are likely to profoundly influence its throughput. Nevertheless, excessive configurability comes with a cost.~\citet{DBLP:conf/esem/HanY16} have discovered that over 59\% of the performance bugs nowadays are due to inappropriate configurations. On the other extreme, software engineers find it generally difficult to adjust the configuration options in order to adapt the performance, therefore most of the options are often ignored, leaving the potential for performance boost untapped~\cite{DBLP:journals/corr/abs-2112-07303,DBLP:conf/sigsoft/0001L21,DBLP:journals/tosem/ChenL23a}. The key cause behind the aforementioned issues is the difficulty of knowing how configurations impact the performance beforehand, since software systems are complex in nature. But what if there is a model that can establish the correlation between configuration and performance?

Configuration performance modeling---a highly active research field over the last decade---has emerged as a crucial topic within the software performance research landscape. The goal therein is exactly to build a model that takes a configuration as input and predicts the likely performance before we deploy that configuration. This holds immense potential for advancements across various domains. For example, in software performance testing, one can easily identify the key configuration options that would likely cause a performance bottleneck by investigating a performance model. Similarly, in configuration tuning, it is straightforward to simply evaluate and compare different configurations on the model, instead of having to deploy and run the configuration on the actual systems, leading to a dramatic reduction in cost.

However, building an accurate performance model for the configuration of software systems is challenging. Classic performance models have been relying on analytical methods~\cite{DBLP:conf/icse/Kumar0BB20,DBLP:journals/tse/GrassiDT92,DBLP:conf/wosp/DidonaQRT15,DBLP:conf/gecco/0001LY18,DBLP:journals/infsof/ChenLY19}, but soon they become ineffective due primarily to the soaring complexity of modern software systems. In particular, there are two key reasons which prevent the success of analytical methods: (1) analytical models often work on a limited set of configurations options~\cite{DBLP:conf/wosp/DidonaQRT15,DBLP:journals/tse/ChenB17}, but the number of configurations options and the complexity continues to increase. For example, \textsc{Hadoop} has only 17 configurations options in 2006, but it was increased by 9$\times$ more to 173 at 2013~\cite{DBLP:conf/sigsoft/XuJFZPT15}; similarly, \textsc{MySQL} has 461 configuration options at 2014, in which around 50\% of them are of complex types. (2) Their effectiveness is highly dependent on assumptions about the internal structure and the environment of the software being modeled. However, many modern scenarios, such as cloud-based systems, and virtualized and multi-tenant software, intentionally hide such information to promote ease of use, which further reduces the reliability of the analytical methods~\cite{DBLP:conf/icse/Chen19b}.

As an alternative, machine learning-based configuration performance models have been explored over the past decade, e.g., liner regression~\cite{DBLP:conf/sigsoft/SiegmundGAK15}, decision tree~\cite{DBLP:journals/ese/GuoYSASVCWY18}, and random forest~\cite{DBLP:conf/msr/GongC22}, which work on arbitrary types of configuration options and do not rely on heavy human intervention~\cite{DBLP:journals/jss/PereiraAMJBV21, DBLP:journals/tse/WangHGGZFSLZN23}. Unlike analytical methods, machine learning is data-driven since it seeks to learn the patterns from the configuration data, hence generalizing the correlation between configuration and performance. This problem, namely configuration performance learning, has been gaining momentum in recent years~\cite{DBLP:conf/icse/HaZ19, DBLP:conf/sigsoft/Gong023, DBLP:conf/esem/ShuS0X20, DBLP:journals/tosem/ChengGZ23,DBLP:journals/dal-ext}. 

Among others, a particular type of data-driven configuration performance learning relies on deep learning, i.e., those that make use of deep neural networks~\cite{DBLP:journals/csur/YangXLG22, DBLP:journals/tosem/WatsonCNMP22}. Indeed, recent studies have demonstrated the benefits of deep learning for modeling configuration performance. For example,~\citet{DBLP:conf/icse/HaZ19} propose \texttt{DeepPerf}, a deep neural network model combined with $L_1$ regularization to address the sparse performance functions, and~\citet{DBLP:journals/tosem/ChengGZ23} invent a hierarchical interaction neural network model called \texttt{HINNPerf} that achieves state-of-the-art accuracy. In Section~\ref{sec:motiv}, we will further elaborate on the importance of deep learning for configuration performance and motivate this survey in detail.

However, despite the importance of such research direction, to the best of our knowledge, there has been little work on a systematic survey that covers the full spectrum of deep configuration performance learning. The current reviews related to this topic mainly focus on either general machine learning models~\cite{DBLP:journals/jss/PereiraAMJBV21}, or deep learning in the general context of software engineering~\cite{DBLP:journals/csur/YangXLG22, DBLP:journals/tosem/WatsonCNMP22}. Undoubtedly, systematically reviewing state-of-the-art studies on this particular research field can provide vast benefits, including summarizing the common categories, revisiting the important concepts, and more importantly, discussing novel perspectives on the good practices and ``bad smells\footnote{In software engineering, the bad smell is a metaphor that denotes the symptoms of code/software that can lead to a deeper problem. In our context, it refers to problematic practices that could lead to serious threats to validity and/or to the sustainability of the research field.}'' of the field, and providing insights for future opportunities. 





\begin{figure}[!t]
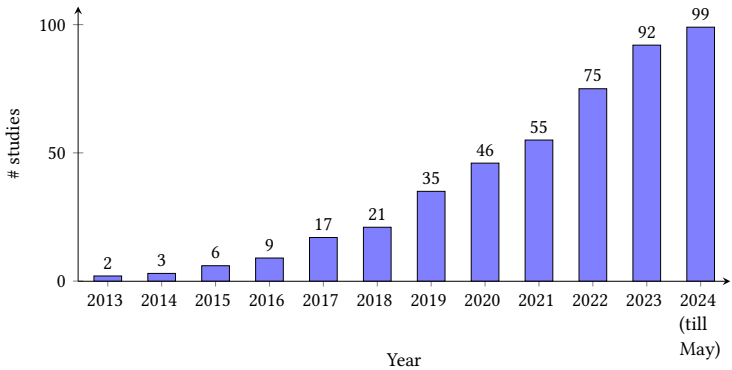

\centering
\includestandalone[width=0.7\columnwidth]{Figures/year_publications_cumulative}
    \caption{Cumulative number of primary studies on deep configuration performance learning models.}
 \label{fig:year_publications}
 \end{figure}


To bridge such a gap, in this paper, we conduct a systematic literature review that covers 1,206 papers from six online repositories and 78 venues, published between 2013 and 2024, based on which we extract 99 prominent studies for data extraction and analysis. The results also confirm that the significance and challenges of deep learning for configuration performance have led to a notable increase in research efforts within this field. Indeed, an overwhelming 79\% of the reviewed publications have emerged since 2019, as shown in Figure~\ref{fig:year_publications}.

In a nutshell, the major contributions of this survey include:

\begin{itemize}
    \item \textcolor{black}{An exhaustive automatic search on six indexing services using a rigorous search string, together with a ``quasi-gold standard'' validation that demonstrates the high sensitivity and sufficient precision of our search strategy.}
    \item \textcolor{black}{An extensive quality assessment of all the primary studies with 18 questions and scoring metrics, which results in a ``golden set'' of studies in the field, serving as a good starting point for new researchers.}

    \item \textcolor{black}{A taxonomy that categorizes the techniques used and key concerns in deep configuration performance learning with up to 13 findings of the trends and 18 actionable suggestions that can be leveraged by future researchers.} 
    
    \item \textcolor{black}{Comprehensive summaries of the key approaches used in the deep learning pipeline for configuration performance, including preparation, modeling, evaluation, and application, together with discussions on their benefits, shortcomings, and best-suited scenarios.}

    

    \item Articulation on the good practices and bad smells observed from the findings.
    
    \item Gaps identified from existing studies, offering insights into the future opportunities for this particular thread of research.
\end{itemize}

\textcolor{black}{In particular, the review results highlight the following key observations that derive several actionable suggestions:}

\begin{itemize}
    \item Random sampling, which is inefficient in finding the most informative configurations, is the most commonly used method, as used in 79 studies. This indicates the need to explore more informative sampling methods to enhance the quality of the collected data.
    
    \item A non-trivial number of studies (34 out of 99) fail to address the issues of sparsity and overfitting in configuration data, while 24 of the rest rely on inefficient manual feature selection methods. This reveals the lack of addressing sparsity issues and encourages researchers to explore alternative approaches.
    
    \item The majority of studies (63 out of 99) apply manual hyperparameter tuning, which relies heavily on human experts. Therefore, an actionable suggestion is that researchers should investigate automatic and heuristic hyperparameter tuning techniques to reduce tuning costs and enhance efficiency.

    \item Nearly half of the primary studies (43 out of 99) do not consider the challenge of dynamic environments. It is crucial for researchers to address this aspect to strengthen the generalizability of configuration performance models across different environments as it has been reported that they can profoundly impact the performance~\cite{DBLP:journals/tse/KrishnaNJM21, DBLP:journals/asc/SinghS20, DBLP:conf/fse/GongC24}.
\end{itemize}

Furthermore, we highlight five specific directions for future research that are promising to produce fruitful outcomes, namely:
\begin{itemize}
    \item Model-based algorithms to enhance configuration sampling.
    \item Explainable deep learning techniques for more reliable configuration performance modeling.
    \item Modeling configuration performance with deep few-shot learning.
    \item Interactive approaches for deep configuration performance prediction.
    \item Configuration performance learning under multiple and dynamic environments.
\end{itemize}

\textcolor{black}{The rest of the paper is organized as follows. Section~\ref{sec:background} presents the background information about deep learning for performance modeling and discusses related work. Section~\ref{sec:methodology} presents our review methodology. Section~\ref{sec:rq1},~\ref{sec:rq2},~\ref{sec:rq3}, and~\ref{sec:rq4} outlines the results of the research questions, the taxonomy and summaries of the strength, weakness, usage scenarios and actionable suggestions, along with the good practices and bad smells discovered. Sections~\ref{sec:gap} and~\ref{sec:threats} discuss future opportunities of the field, and threats to validity, respectively. Finally, Section~\ref{sec:conclusion} concludes the paper.}

%% file: Figures/year_publications_cumulative.tex
\begin{tikzpicture}
\footnotesize
\begin{axis}[
    width=11cm,
    height=5.5cm,
    ybar,
    ymax = 105,
         axis x line  = bottom,
    axis y line  = left  ,
    bar width=0.4cm, 
   ylabel=\# studies,
    enlarge y limits=0.02,
    enlarge x limits=0.05,
    y label style={at={(0.03,0.5)}},
    x label style={at={(0.5,-0.1)}},
    xlabel=Year,
    symbolic x coords={A, B, C, D, E, F, G, H, I, J, K, L}, 
    xtick=data, 
     xticklabels={2013,2014,2015,2016,2017,2018,2019,2020,2021,2022,2023,{2024\\(till May)}},
     xticklabel style={text width=0.6cm},
    nodes near coords, 
]
\addplot[fill=blue!50, draw=black] coordinates {(A, 2) (B, 3) (C, 6) (D, 9) (E, 17) (F, 21) (G, 35) (H, 46) (I, 55) (J, 75) (K, 92) (L, 99)};
\end{axis}
\end{tikzpicture}




%% file: background.tex
\section{Background and Related Work}
\label{sec:background}

In this section, we provide a brief overview of the configurable software systems and the deep learning pipeline. We also differentiate this work and other relevant surveys.

\subsection{Configurations in Software Systems}


To meet the performance requirements, configurable software systems often permit possible configuration options to be adjusted at design time or at runtime. For example, around one-third of the configuration options of \textsc{MySQL}, a popular configurable database system, can be changed at runtime, such as \texttt{max\_connections}; the remaining ones, e.g., \texttt{autocommit}, need to be fixed a priori to the deployment\footnote{\url{https://dev.mysql.com/doc/refman/8.0/en/server-system-variable-reference.html}}. Table~\ref{tb:data_example} presents a dataset for \textsc{VP8}, which includes $m$ configurations, each associated with a measured performance metric (runtime).

It has been widely acknowledged that the configuration options have great impacts on software performance~\cite{DBLP:journals/tosem/ChenLBY18,DBLP:journals/corr/abs-1801-02175,DBLP:conf/esem/HanY16,DBLP:conf/sigsoft/0001L24}. Indeed, inappropriate configurations often cause serious performance bugs, which is the key reason why the users became frustrated enough to (threaten to) switch to another product~\cite{DBLP:conf/msr/ZamanAH12}. At the same time, simply using default configurations does little help, for instance,~\citet{DBLP:conf/cidr/HerodotouLLBDCB11} show that the default settings on \textsc{Hadoop} can actually result in the worst possible performance. In this regard, a fundamental question to ask is: \textit{given a configuration, what is the performance of the software systems?}. 

\input{Tables/data_example}


Indeed, one solution is to directly profile the software system for all possible configurations when needed. This, however, is impractical, because (1) the number of possible configurations may be too high. For example, \textsc{MySQL} has more than a million possible configurations. (2) Even when such a number is small, the profiling of a single set of configurations can be too expensive.~\citet{DBLP:conf/sigsoft/WangHJK13} report that it could take weeks of running time to benchmark and profile even a simple system. All these issues have called for a computational model, which is cheap to evaluate yet accurate, that captures the correlation between configuration options to a performance attribute. 




\subsection{\textcolor{black}{Motivation: Why Deep Learning for Configuration Performance}?}
\label{sec:motiv}

In contrast to traditional analytical models, data-driven configuration performance modeling, such as machine learning, operates in a black-box manner by leveraging observations of a limited set of the software's actual performance behaviors and constructing a statistical model that can predict performance without extensive domain knowledge on configuration options and system characteristics that are prevalent in analytical methods~\cite{DBLP:books/daglib/0020252, DBLP:journals/tse/YuWHH19,DBLP:conf/splc/TempleAPBJR19,DBLP:journals/tosem/ChenLBY18,DBLP:journals/corr/abs-1801-02175}. 

As a result, numerous data-driven machine learning models have been utilized for configuration performance learning, showcasing the efficacy of this approach. For example, linear regression models like \texttt{SPLConqueror} have been utilized in different studies~\cite{DBLP:journals/sqj/SiegmundRKKAS12,DBLP:conf/sigsoft/SiegmundGAK15,DBLP:journals/tc/SunSZZC20,DBLP:journals/is/KangKSGL20}, which combines linear regression with different sampling methods and step-wise feature selection to capture the interactions between configuration options. Tree structure models have also been employed in a number of research works~\cite{DBLP:journals/ese/GuoYSASVCWY18, DBLP:conf/icdcs/HsuNFM18,DBLP:conf/kbse/SarkarGSAC15,DBLP:journals/corr/abs-1801-02175}, e.g., \texttt{DECART}~\cite{DBLP:journals/ese/GuoYSASVCWY18} improves upon \texttt{CART} by incorporating an efficient sampling method.
In the work of Jamshidi et al.~\cite{DBLP:conf/mascots/JamshidiC16}, Gaussian Process (GP) is employed to model configuration performance, which is updated incrementally through Bayesian Optimization.
Moreover, Fourier-learning models have been applied in investigations by two studies~\cite{zhang2015performance,DBLP:conf/icsm/Ha019} to predict configuration performance by learning the Fourier coefficients of the performance function, and transfer learning techniques have been explored in several works~\cite{DBLP:conf/sigsoft/JamshidiVKS18,DBLP:journals/tse/KrishnaNJM21,DBLP:conf/kbse/JamshidiSVKPA17} to reuse configuration data in difference environments on a desired target environment. 

However, as software systems evolve and become more sophisticated, the number of configuration options and their interactions grow exponentially. This poses a significant obstacle for traditional machine learning models, which may struggle to capture the intricate relationships and interactions within the vast configuration space. For example, Siegmund et al.~\cite{DBLP:journals/sqj/SiegmundRKKAS12} illustrate how the complexity of configuration spaces can hinder the performance of linear regression models, necessitating the exploration of alternative approaches.

To address these challenges, researchers have been exploring novel learning paradigms that can handle the complexity and scale of modern software systems. Therein, one promising approach is deep learning, which employs deep neural networks. Deep learning models have shown remarkable success in various domains, including natural language processing, image recognition, and also software performance engineering tasks.

Notably, deep learning models share some commonalities with traditional machine learning models for configuration performance learning, i.e.: 
\begin{itemize}
    \item \textbf{Data-driven nature:} They both rely on historical data to learn patterns and relationships, which enable them to predict new configurations~\cite{DBLP:conf/msr/GongC22}.
    \item \textbf{Common obstacles:} Mitigating overfitting is a common difficulty due to limited training samples and sparse configuration data~\cite{DBLP:conf/sigsoft/Gong023}. Besides, maintaining robustness in unseen environments is also a joint challenge~\cite{DBLP:conf/fse/GongC24}.
    \item \textbf{Similar learning pipeline:} They follow common learning procedures, including sampling, data preprocessing, encoding, hyperparameter tuning, model training, evaluation, and application.
    
\end{itemize}


Despite the above similarities, the specific structure of neural networks in deep learning equips it with some unique characteristics. For example, in hyperparameter tuning, deeper layers in deep learning mean that the dimension of the hyperparameters can be significantly higher than their machine learning counterparts~\cite{DBLP:journals/pr/TranSWQ20}. It is also known that, while the sampling strategy is important for both types, deep learning tends to be less sensitive to the sampled data~\cite{DBLP:conf/icse/HaZ19}. As a result, those differences render deep learning with remarkable advancements, offering numerous advantages over traditional machine learning approaches for configuration performance learning, including:

\begin{itemize}
    \item \textbf{Ease representation handling:} Deep learning is capable of extracting the underlying representations from the configuration options even without careful feature engineering, which allows them to learn directly from raw data, enabling end-to-end learning.
    \item \textbf{Good at handling complex data:} Deep learning models consist of multiple layers of interconnected neurons, enabling them to capture the nonlinear and complex relationships between the configuration options and performance.
    \item \textbf{Better generalizability:} Deep learning can leverage pre-trained knowledge from related tasks and quickly adapt to the target task through fine-tuning, hence they often have better generalizability.
    \item \textbf{More robust to noisy data:} Their ability to learn complex representations helps them generalize well, even in the presence of noisy or incomplete data.
    \item \textbf{More consistent and flexible architectures:} Deep learning offers a wide range of neural network architectures, such as fully connected neural networks, convolutional neural networks (CNNs), and recurrent neural networks (RNNs). Those architectures are from the same origin but differ in their interconnections and can be tailored to specific problem domains and configuration data.
\end{itemize}


Indeed, recent state-of-the-art studies have demonstrated the effectiveness of deep learning models in configuration performance learning, which surpasses the accuracy of traditional machine learning approaches in their empirical experiments~\cite{DBLP:conf/icse/HaZ19,DBLP:journals/tosem/ChengGZ23,DBLP:conf/sigsoft/Gong023,DBLP:conf/fse/GongC24}. For example, in 2019, Ha and Zhang~\cite{DBLP:conf/icse/HaZ19} proposed \texttt{DeepPerf}, a deep neural network model combined with $L_1$ regularization to address the sparse performance functions, and their evaluation results demonstrate that \texttt{DeepPerf} is indeed more accurate than other machine learning models such as \texttt{DECART} and \texttt{SPLConqueror}. Next, in 2023, Cheng et al.~\cite{DBLP:journals/tosem/ChengGZ23} invented a hierarchical interaction neural network model called \texttt{HINNPerf} that achieves state-of-the-art accuracy when compared with \texttt{DeepPerf}, \texttt{DECART}, \texttt{SPLConqueror}, and \texttt{RF}. In the same year, Gong and Chen~\cite{DBLP:conf/sigsoft/Gong023} proposed the idea of divide-and-learn (\texttt{DaL}), which mitigates the sparsity in configuration data and further enhances the accuracy of \texttt{DeepPerf} and other peer machine learning ones. Most recently, Gong and Chen~\cite{DBLP:conf/fse/GongC24} show that training a meta-deep neural network model in an optimal sequence under different environments leads to the best result.

Additionally, there has been an increasing trend from the community in exploiting deep learning for configuration performance learning. Recall from Figure~\ref{fig:year_publications}, we see a significant growth in the number of research papers on deep configuration performance learning during the past decade, further demonstrating the increasing popularity of deep learning in this field. Noteworthily, over the past decade, 78 out of the 99 elected primary studies have emerged since 2019, indicating the rapid emergence of deep learning as a prominent paradigm for configuration performance learning. This trend also emphasizes the widespread adoption and recognition of deep learning's potential in configuration performance modeling.

Yet, despite the importance and potential of such research direction, to the best of our knowledge, there has been little work that focuses explicitly on a systematic survey for deep configuration performance learning, as we will discuss in Section~\ref{sec:related}. This observation serves as the primary motivation behind this survey.

\subsection{Problem Formulation of Deep Configuration Performance Learning}


Without loss of generality, deep configuration performance learning seeks to build a function $f$ such that:
\begin{equation}
    y = f(\mathbf{\overline{x}})
\end{equation}
whereby $y$ is the performance attribute that is of concern; $\mathbf{\overline{x}}$ is a configuration consists of the values for $n$ configuration options, i.e., $\mathbf{\overline{x}}=\{x_1,x_2,...,x_n\}$. Taking the simplest fully connected deep neural network as an example, $f$ is represented as multiple layers of interconnected neurons, where neurons are activated as:
\begin{equation}
    a_j^{l+1}=\sigma(\sum_i a_i^l w_{ij}^{l,l+1} + b_j^{l+1})
\end{equation}
where $\sum$ runs over all the lower layer neurons that
are connected to neuron $j$. $i$ is the activation of a
neuron $i$ in the previous layer, and where $a_i^l w_{ij}^{l,l+1}$ is the contribution of neuron $i$ at layer $l$ to the activation of the neuron $j$ at layer $l + 1$. The function $\sigma$ is a nonlinear monotonously increasing activation function, e.g., a sigmoid function; $w_{ij}^{l,l+1}$ is the weight and $b_j^{l+1}$ is the bias term.

To build function $f$, the training in deep learning aims to find the set of weights for different neurons from all the layers such that a loss function can be minimized. For example, the mean squared error below is one possible loss function since configuration performance learning is essentially a regression problem:
\begin{equation}
    \mathbfcal{L}(\theta) =  {1 \over m} \sum^m_{i=1}{(f(\mathbf{\overline{x}}_i) - y_i')^2}
\end{equation}
$f(\mathbf{\overline{x}}_i)$ and $y_i'$ denote the predicted and actual performance values for the $i$th sample across $m$ data points in the training dataset.

\subsection{\textcolor{black}{Related Surveys}}
\label{sec:related}

The related surveys of this work mainly lie in two big categories, i.e., surveys on analytical/machine learning techniques for software configuration performance learning, or reviews on deep learning approaches for different tasks in software engineering.

\subsubsection{Surveys on Configuration Performance Modeling} There are a few related surveys in the domain of configuration performance modeling. 
Among others,~\citet{DBLP:journals/pe/BalsamoPI03} conduct a review of configuration performance prediction using Queuing Network, and~\citet{DBLP:journals/tse/BalsamoMIS04} review how performance model can be used to help software development.~\citet{DBLP:journals/sigmetrics/NambiarKBSD16} emphasize the significance of configuration performance modeling in SE tasks and calls for further research to enhance the predictive capabilities of performance models, and~\citet{DBLP:conf/wosp/PereiraA0J20} seek to understand to what extent are sampling strategies sensitive to configuration performance prediction of configurable systems. A recent work~\cite{DBLP:journals/spe/HanYY23} classifies software configuration performance learning studies into 6 categories, provides a mapping of them within these categories, and highlights potential weaknesses of the literature. Further, surveys on configuration performance modeling for specific categories of configurable software systems exist, for example, there are reviews on the performance models for distributed systems~\cite{DBLP:conf/cisis/PllanaBB07,DBLP:journals/csur/ChenBY18,DBLP:journals/tpds/Lopez-NovoaMM15} and~\citet{DBLP:journals/tjs/Flores-Contreras21} survey the configuration performance prediction methods for parallel applications published between 2005 and 2020. Similarly,~\citet{DBLP:conf/sc2/FrankHLB17} review 34 studies on configuration performance modeling for multi-core systems, and configuration performance learning and tuning techniques for mobile application systems are reviewed by~\citet{DBLP:journals/tse/HortKSH22}. In recent years,~\citet{DBLP:journals/infsof/StradowskiM23a} emphasize the importance of machine learning techniques for defect prediction in real-life business scenarios by conducting a systematic mapping study on 32 primary studies, and ~\citet{DBLP:journals/infsof/ZainSI23} synthesize existing independent variables, modeling techniques, and performance evaluation criteria used in machine/deep learning software defect prediction research. However, the above work does not focus on how deep learning can be used in the performance model building for configurable software in general. 

\subsubsection{Surveys on Deep Learning for Software Engineering}
On the other hand, several surveys have been conducted on the application of deep learning in software engineering. For example,~\citet{DBLP:journals/csur/YangXLG22} investigates the deep learning-related techniques for software engineering tasks, including the deep learning models, data preprocessing methods, SE tasks, and model optimization methods, while a review on deep learning is also conducted by~\citet{DBLP:journals/tosem/WatsonCNMP22}, which provides a research roadmap and guidelines for future exploration in the context of software engineering.~\citet{DBLP:journals/tosem/LiuGXLGY22} investigates the reproducibility and applicability of deep learning studies in SE, and observes that a significant number of them have overlooked this challenge. A novel work~\cite{DBLP:journals/tse/WangHGGZFSLZN23} explores the use of machine/deep learning techniques in various software engineering tasks and the challenges and differences between machine and deep learning. Yet, these studies do not focus on a specific SE task like configuration performance learning but rather provide a broad review of software engineering with a restricted depth on each topic. 

The most similar prior study to our work is probably a review by~\citet{DBLP:journals/jss/PereiraAMJBV21} that explores the application objectives, sampling, learning and measuring methods, and evaluation related to machine learning models for configuration performance modeling. However, we focus on deep learning models, which is a specific type of machine learning, and cover a wider range of techniques such as preprocessing, hyperparameter tuning, and sparsity handling methods. Besides, they did not discuss and justify the good practices and bad smells in the field, as well as the actionable suggestions and summaries of the strengths, weaknesses, and best-suited usage scenarios.

\subsubsection{Differences of Our Survey}
In summary, our work differs from the above in the following aspects:

\begin{itemize}
    \item We focus explicitly on deep learning-based configuration performance modeling for configurable software in general and capture the latest trends in the past 10 years.
    \item We review aspects that have not been summarized before, e.g., the use of data preprocessing, encoding scheme, sparsity handling, hyper-parameter tuning, statistical test and effect size test, runtime environments, and public artifact.
    \item We present an innovative summary outlining the advantages, disadvantages, and recommended scenarios for each category of deep configuration performance learning techniques.
    \item We disclose and justify the positive and potentially problematic practices, which have not been revealed previously, and offer a list of actionable suggestions and knowledge gaps to shed light on future works in the field.
\end{itemize}

\textcolor{black}{Additionally, we would like to kindly stress that the nature of this study is a systematic literature review specializing in deep learning techniques for software configuration performance learning, instead of a mapping study. While both systematic literature review and mapping studies aim to synthesize the current research literature, they differ in terms of purpose and methodology:}

\begin{itemize}
    \item Firstly, a systematic literature review is conducted by addressing specific research questions through a rigorous and comprehensive process. This involves conducting thorough searches, applying strict inclusion and exclusion criteria, performing quality assessment of primary studies, and conducting detailed data analysis, and the outcome is typically detailed findings and in-depth discussions based on the review results~\cite{keele2007guidelines}. In contrast, a mapping study, or scoping review, typically includes a broader search strategy without strict inclusion and exclusion criteria, producing a mapping of the studies to different concepts, types of studies, and research trends~\cite{DBLP:journals/spe/HanYY23}. In our survey, the concepts themselves have already been unified in the field and we conducted the study on different aspects of the known concepts, following specific criteria and protocols as discussed in Section~\ref{sec:methodology}. 

    \item Secondly, while systematic literature reviews focus on detailed answers and high-quality analysis, mapping studies aim to offer a broader view of the research landscape, without the same depth of analysis or stringent quality assessment. We provided an in-depth analysis of different aspects of deep configuration performance learning, along with discussions of the good practices and bad smells.
\end{itemize}

%% file: Tables/data_example.tex
\begin{table}[t]
\centering
\small
\caption{An example of configurations and performance for \textsc{VP8}. $x_i$ is the $i$th configuration option and $y$ is the performance value (runtime).}
\begin{tabular}{lllllll||c}
\toprule
$x_{1}$ & $x_{2}$ & $x_{3}$ & $\cdots$ & $x_{n-2}$ & $x_{n-1}$ & $x_{n}$ & $y$ \\ \midrule
0 & 0 & 0 & $\cdots$ & 0 & 0 & 1 & 8190.6 seconds \\ 
0 & 1 & 0 & $\cdots$ & 0 & 0 & 2 & 6502.4 seconds \\ 
$\cdots$        & $\cdots$  & $\cdots$       & $\cdots$       & $\cdots$       & $\cdots$ & $\cdots$ & $\cdots$ \\ 
1 & 1 & 1 & $\cdots$ & 1 & 0 & 3 & 29102.2 seconds \\ 
1 & 1 & 1 & $\cdots$ & 1 & 1 & 4 & 25827.4 seconds \\ \bottomrule
\end{tabular}
\label{tb:data_example}
\end{table}

%% file: methodology.tex
\section{Research Methodology}
\label{sec:methodology}

\begin{figure}[t!]
  \centering
  \includegraphics[width=\columnwidth]{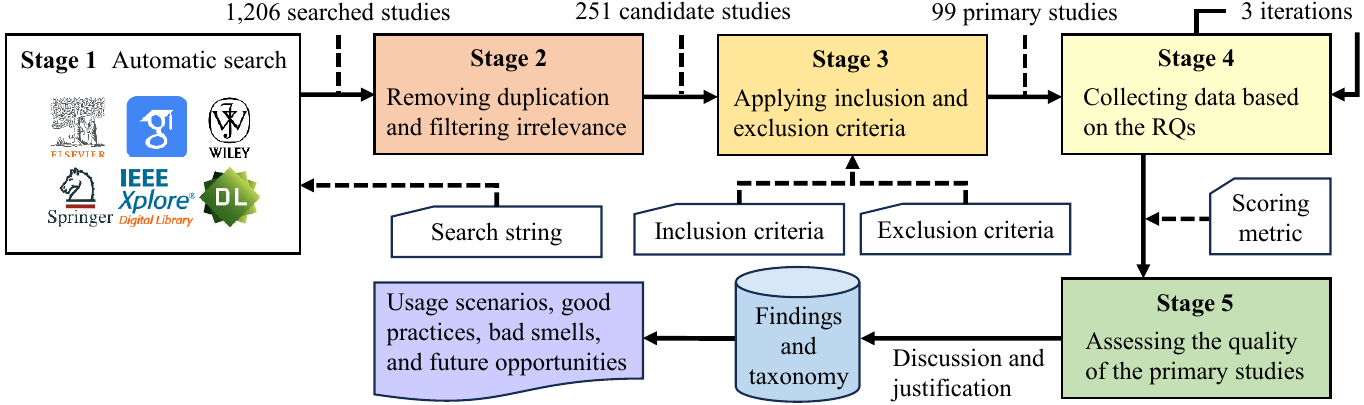}
  ~\vspace{-0.2cm}
  \caption{Overview of the literature review protocol.}
  \label{fig:protocol}
\end{figure}

\textcolor{black}{To bridge the aforementioned gap, we conduct a systematic literature review that covers the relevant papers published between 2013 and 2024.} This period was chosen because we seek to concentrate on the latest trends, mitigating the noises from the old and disappeared practices in the field. In particular, our review methodology follows the best practice of systematic literature review for software engineering~\cite{DBLP:journals/infsof/KitchenhamBBTBL09,keele2007guidelines}, as shown in Figure~\ref{fig:protocol}.

\subsection{Automatic Search}
As can be seen, in Stage 1 we conducted an automatic search over six highly influential indexing services, i.e., ACM Library, IEEE Xplore, Google Scholar, ScienceDirect, SpringerLink, and Wiley Online, as they have been used in a number of systematic literature reviews in the field of software engineering, software configuration, and performance modeling~\cite{DBLP:journals/tse/LiCY22, DBLP:journals/sigmetrics/NambiarKBSD16, DBLP:conf/wosp/PereiraA0J20, DBLP:journals/tosem/ChenL23, DBLP:journals/spe/HanYY23, DBLP:journals/tse/HortKSH22, DBLP:journals/tjs/Flores-Contreras21}. 

\textcolor{black}{By establishing a set of benchmark studies through a manual search over the six indexing services and analyzing their scope and keywords, the aim of the search string was to identify primary studies that:}

\begin{itemize}
    \item are relevant to software systems research;
    \item incorporate deep learning or neural network as part of the model;
    \item utilize configurations as the input for deep configuration performance models;
    \item apply the deep configuration performance models in some practical domains, e.g., predicting the performance of software or systems.
    
\end{itemize}
\textcolor{black}{To that end, different combinations of keywords in these four areas were explored, and the quality of the automated search results was evaluated using the quasi-sensitivity and quasi-precision metrics by~\citet{DBLP:journals/infsof/ZhangBT11}, which will be explained thoroughly in Section~\ref{sec:quasi_gold_standard}.} 
This ensures that the search results exclude redundant and irrelevant works as much as possible, retrieving papers that align with the research focus on the intersection of software configuration performance prediction and deep learning models. 

Specifically, we compile the search string as below:


\begin{tcbitemize}[%
    raster columns=1, 
    raster rows=1
    ]
  \tcbitem[myhbox={}{Search String}]   \textit{(``software" OR ``system") AND ``configuration" AND (``performance prediction" OR ``performance modeling" OR ``performance learning" OR ``configuration tuning") AND (``deep learning" OR ``neural network")}
\end{tcbitemize}


\textcolor{black}{For all indexing services, the search resulted in 1206 studies, accounting for duplicates and excluding non-English documents. Among these, a total of 540 studies were obtained from Google Scholar, 152 studies from SpringerLink, 137 studies from Wiley Online, 190 studies from ACM Library, 132 studies from ScienceDirect, and 55 studies from IEEE Xplore.}

\subsection{Removing Duplication and Filtering Irrelevant Studies}

Next, in Stage 2, we aim to ensure that only unique and highly relevant studies will be considered for further analysis in our review. To achieve this, we first filter out any duplicate studies by carefully examining the titles of the identified papers. This ensures that each study included in our review is distinct and contributes unique insights to the body of knowledge. Subsequently, we conducted a brief evaluation to eliminate any documents that were clearly irrelevant to configuration performance learning. For instance, we excluded studies focused on human and student performance, which is solely relevant to educational outcomes. Similarly, we disregarded papers centered on the performance of, e.g., physical systems, fuel systems, and football systems, as they are not directly aligned with the scope of our investigation.

\textcolor{black}{As a result of these rigorous filtering procedures, we identified 251 highly relevant candidate studies.}

\subsection{Applying Inclusion and Exclusion Criteria}
Through a detailed review of the candidate studies, Stage 3 focuses on applying various criteria to further extract a set of more representative works. 

Firstly, we design a set of inclusion criteria to ensure that studies selected for detailed analysis align closely with the core themes of configuration performance learning. Specifically, in our methodology, a study is temporarily selected as a primary study if it meets all of the following inclusion criteria:

\begin{itemize}
\item The paper presents a configuration performance modeling approach using deep learning algorithm(s).
\item The paper explicitly states how the model built can be used, e.g., for predicting or analyzing the performance of a software system.
\item The paper has at least one section that explicitly specifies the learning algorithm(s) used.
\item The paper contains quantitative experiments in the evaluation with details about how the results were obtained.
\end{itemize}

The above is designed to maintain the relevance of the review and reduce the efforts required in the subsequent analysis. For example, the studies should ensure that they provide explicit details about the learning algorithms employed, as this is important for the understanding, analysis, and application of the model. Besides, it is essential for the inclusion of quantitative experiments to ensure a robust evaluation process, enhancing the depth and reliability of the selected studies.

Next, we applied the following exclusion criteria on the previously included study, which would be removed if it meets any: 
\begin{itemize}
\item The paper is not software or system engineering-related.
\item The paper is not published in a peer-reviewed public venue.
\item The paper is a survey, review, tutorial, case study, or an empirical type of work.
\item The paper is a short and work-in-progress work, i.e., shorter than 8 double-column or 15 single-column pages.
\end{itemize}

By excluding papers unrelated to software or systems, we ensure that the selected studies directly relate to the research focus of this survey. The requirement for peer-reviewed publication enhances the credibility of the included studies, while the exclusion of certain types of works (survey, review, tutorial, case study, or empirical) helps to filter out literature that may not align with our objective of investigating performance prediction using deep learning. Lastly, the limitation on the length of papers ensures that the selected studies possess sufficient depth and detail for a comprehensive analysis.

\textcolor{black}{Finally, we identified 99 primary studies for detailed quality assessment and data extraction}\footnote{Raw data of all primary studies can be found at our repository: \url{https://github.com/ideas-labo/DCPL-SLR}.}.

\subsection{Collecting and Extracting Data}

\begin{figure}[t!]
  \centering
  \includegraphics[width=0.9\columnwidth]{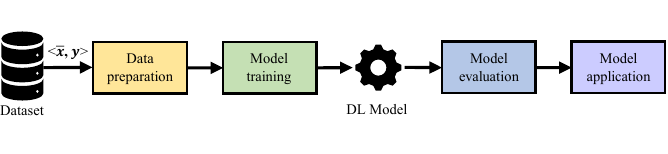}
  ~\vspace{-0.8cm}
  \caption{Pipeline for deep configuration performance learning.}
  \label{fig:process}
\end{figure}

\subsubsection{Research Questions}
\label{subsec:rq}

In the last stage, we derive the research questions (RQs) of this survey following the deep learning pipeline in Figure~\ref{fig:process}. Specifically, we formalize the workflow of deep learning for learning software configurations and performance into four key phases. The first phase is to process the raw configuration data collected and make preparations for their training with the deep learning models. To understand this, we ask the first research question:

\keybox{
\textbf{RQ1:} How do the practitioners prepare raw configuration data for learning?
}


Next, various deep learning algorithms might have been utilized to learn the preprocessed configuration data, handling different challenges in configuration performance prediction, such as sparsity, network structure, and network optimization. To summarize this, we seek to understand: 

\keybox{
\textbf{RQ2:} How do the practitioners build the deep configuration performance model?
}


A crucial part of deep configuration performance learning is how to evaluate the trained deep configuration performance models, with a specific evaluation method, a set of testing samples, and a particular accuracy metric, where this process might be repeated for multiple subject systems. Therefore, the next question is:

\keybox{
\textbf{RQ3:} How are the deep configuration performance models evaluated?
}


Finally, the deep learning models need to be exploited in practical configuration scenarios, adapt to different environmental conditions, and enable researchers to replicate or reproduce the results. To that end, the last research question aims to understand:

\keybox{
\textbf{RQ4:} How to exploit the deep configuration performance model?
}


\subsubsection{Data Items}
\label{subsubsec:data items}

After formulating the research questions, we have developed a comprehensive list of data items that we aim to collect during the review process. The summary of these data items is presented in Table~\ref{tb:data_items}, which includes a total of 22 data items, each serving a specific purpose related to the corresponding research question. In this section, we explain the design rationales behind them and clarify the procedure for extracting and classifying the data from each item.

In the initial stage, we employ data items $I_1$ to $I_4$ to gather meta-information of the reviewed studies, such as the title, author, and publication information. This meta-information enables us to accurately reference and organize the reviewed studies, thereby establishing a robust foundation for further analysis and interpretation.

To address RQ1, we seek to examine the key processes in preparing configuration data, where it is common to pre-process the raw configuration data, encode the data into the most appropriate form, and select a subset of informative samples for model training. To capture this information, we design data items $I_5$ to $I_7$ to extract the corresponding data preparation methods. For instance, by using $I_5$ we extract the pre-processing methods used in each primary study according to the fundamental types of methods, e.g., min-max scaling and z-score, which are all used to normalize the scale of the configuration options, can be classified into the same category. 


For answering RQ2, we first review the methods used to deal with data sparsity in configuration data via $I_8$. Next, $I_9$ to $I_{13}$ are for understanding how the deep learning models are chosen and trained according to their most general type. For example, on $I_{11}$, hyperparameter tuning methods can be set into three categories:

\begin{itemize}
    \item using the default hyperparameter settings without any tuning;
    \item tuning the hyperparameters via manual effort and domain knowledge;
    \item or relying on automated heuristic methods in the tuning.
\end{itemize}


To examine the evaluation-related techniques in RQ3, we do not only examine the evaluation procedure ($I_{14}$) and metrics ($I_{15}$) but also the methods of statistical validation that ensure the statistical difference between the comparisons ($I_{16}$ and $I_{17}$). The number and domain of subject systems used in the evaluation, which may influence the generalizability of the conclusions, are also examined using $I_{18}$ and $I_{19}$, respectively.


Lastly, for RQ4, we identify three data items to examine the exploitation information regarding the deep configuration performance models. In particular, $I_{20}$ is employed to understand the application domains of the performance models, $I_{21}$ summarizes the environmental conditions considered, as they significantly decide the generalizability and robustness of the deep configuration performance models, and $I_{22}$ is applied to explore the data for replication and reproduction of the proposed models. 


\input{Tables/data_items}

\subsubsection{Data Collection}
\label{subsubsec:data collection}

The data collection process aims to collect detailed information from the primary works with the data items presented in Table~\ref{tb:data_items}, and the collection protocol is the same as commonly used by the literature reviews in Software Engineering~\cite{DBLP:journals/tse/LiCY22, zou2018practitioners}. Specifically, the protocol involves all the authors with three iterations. 

\begin{itemize}
    \item \textcolor{black}{In the first iteration, each of the authors conducted initial data collection independently, read carefully throughout the 99 primary studies, extracted the data according to each data item, summarized the data into a table of 22 columns and 99 rows, and conducted preliminary classification.} Notably, there were situations where certain data items could not be found (e.g., the encoding schemes were often ignored); the data items were not clearly stated (e.g., the sampling methods for training/testing data for evaluation were sometimes vague, with only abstract descriptive words like ``case study'' or ``real dataset''); or the data needed to be conjectured from the contexts (e.g., the domains of subject software systems were often not specified and extra searches were needed to be performed for data collection). In these cases, the corresponding data items were marked for further investigation in the next iteration.
    \item In the second iteration, the authors reviewed and cross-checked each other's data summary tables, ultimately integrating them into a unified table. The unclear data items that were marked in the previous iteration were rechecked and we arranged meetings to reach a common agreement for each table cell where the collected data from the authors were different. This is achieved through author discussions; further investigations from the existing literature; or consultation with external researchers/authors of the reviewed studies. For instance, the authors discussed and decided to add a category of ``unknown'' for encoding schemes since a large number of studies omitted the justification of the choice of their encoding method, and investigations on search engines like Google were conducted to collect the domain of subject systems. 
    \item Finally, the goals for the third iteration are to summarize the statistics of the integrated table from the second iteration; count the number of studies for each technique associated with every data item; and aggregate similar techniques into broader categories. For example, in the domain of subject systems, it was found that the number of studies that apply configuration performance models for video encoding software and image processing software are four and three, respectively, and they are combined as the category ``multimedia processing'' as these software systems share much more similar characteristics comparing with others.
 
\end{itemize}

\textcolor{black}{This has finally led to the data for this work, which was discussed among all authors to make quality assessments and comprehensive classifications. }

\subsection{Quality Assessment}

\input{Tables/quality_assessment}

\begin{figure}[!t]
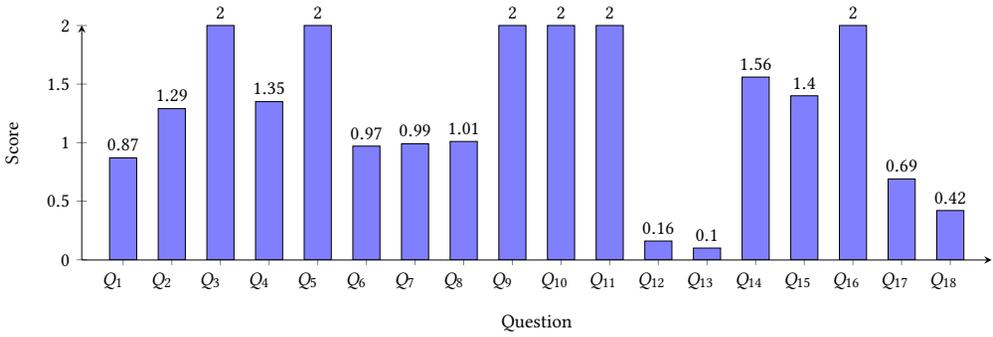

\centering
\includestandalone[width=0.95\columnwidth]{Figures/quality_assessment}
~\vspace{-0.2cm}
    \caption{Average scores for each quality assessment question.}
 \label{fig:avg_score}
 \end{figure}

\textcolor{black}{Based on the extracted information in the previous stage, we performed a quality assessment to evaluate the reliability and validity of the primary studies, which helps ensure that the studies selected for the review are of high methodological quality and provide reliable evidence for further data analysis.}

\textcolor{black}{In particular, following the guideline of SLR by~\citet{keele2007guidelines}, we defined a list of questions corresponding to the data items to quantify the comprehensiveness in each phase of the deep learning pipeline. Then, we scored each study based on the results of the assessment questions. In a nutshell, the questions and scoring metrics for assessing the primary studies are summarized in Table~\ref{tb:quality_assessment}. }

Remarkably, as depicted in Figure~\ref{fig:avg_score}, all 99 studies achieved full scores in six questions, and the average score across all questions was 1.27 out of 2, indicating a high standard of the primary studies. Moreover, Table~\ref{tb:top10_score} provides a breakdown of the scores for the top 10 primary studies, where five of them reach a score of 1.72 or higher, accounting for 86\% of the maximum score. It is worth noting that due to space constraints, the detailed scores of the 99 primary studies are included in the supplementary documents available on our public repository.

\input{Tables/gold_set}

\subsection{Quasi-Gold Standard Validation}
\label{sec:quasi_gold_standard}

\textcolor{black}{Additionally, we conducted a validation of the sensitivity and precision of our automatic search strategy using the ``quasi-golden standard'' (QGS) by Zhang et al.~\cite{DBLP:journals/infsof/ZhangBT11}. Note that since the search covers a wide range of 78 venues, six indexing services, and 11 years of duration, it could be extremely expensive to conduct a grid search over the scope. Therefore, we only focused on primary studies that are published within the most recent five years (from 2019 to May 2024, where 79\% of the primary studies are published according to Figure~\ref{fig:year_publications}) and on five of the most representative and influential venues in the domain of software engineering for assessment~\cite{research_com}, i.e., conferences including ESEC/FSE, ICSE and ASE, and journals including JSS and TOSEM (note that we did not choose TSE because none of the primary studies was published since 2019).}

\textcolor{black}{In particular, by following the QGS procedures, we seek to verify the sensitiveness and exactness of our search string. We first manually defined a set of eight studies that are published in the selected venues and widely recognized as high quality, also known as the quasi-golden set~\cite{DBLP:journals/infsof/ZhangBT11}, as specified in Table~\ref{tb:sensitivity_precision}. Then, by performing an automatic search with our search string exclusively on the five conferences/journals using the ACM Digital Library, we retrieved a collection of 31 searched studies. After applying the inclusion and exclusion criteria, we identified eight relevant studies retrieved from the search. Subsequently, we calculated the sensitivity (\%) $s$ and the precision (\%) $p$ using the following formula as proposed by~\citet{DBLP:journals/infsof/ZhangBT11}:}

\begin{equation}
s = \frac{\text{\# relevant studies retrieved}}{\text{Total \# relevant studies in the QGS}} \times 100
\end{equation}

\begin{equation}
p = \frac{\text{\# relevant studies retrieved}}{\text{Total \# studies retrieved}} \times 100
\end{equation} 

\textcolor{black}{Notably, the sensitivity of our automatic search among the selected five venues is 100\%---significantly better than the suggested 70\% by~\citet{DBLP:journals/infsof/ZhangBT11}, demonstrating the outstanding effectiveness of our search strategy in identifying the most influential primary studies. On the other hand, the precision is 26\%, which means there are some false positives. However,~\citet{DBLP:journals/infsof/ZhangBT11} state that as a verification procedure, sensitivity is of the top importance in the literature review process.}

 \input{Tables/sensitivity_precision}

\begin{figure}[t!]
  \centering
  \includegraphics[width=\columnwidth]{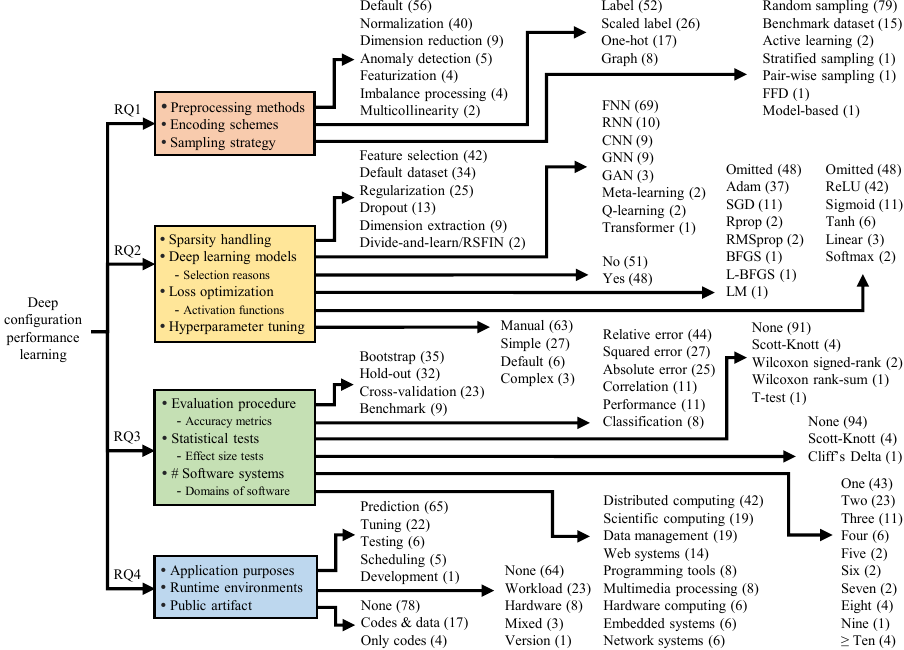}
  \caption{A taxonomy of deep configuration performance learning. The number in the bracket denotes the corresponding number of studies.}
  \label{fig:overview}
\end{figure}

\subsection{The Taxonomy}

With the collected statistics using the RQs, data items, and data collection protocol mentioned in the previous sections, we formalize into a taxonomy of deep configuration performance learning as shown in Figure~\ref{fig:overview}. As can be seen, we found a variety of categories for the RQs and their data items. Some of these are specific algorithms (e.g., for the deep learning models used) while some others are of more general types (e.g., for the preprocessing methods). We will present detailed data statistics and introduce each category of techniques for deep configuration performance modeling in what follows.

%% file: Tables/data_items.tex
 \begin{table}[t!]
  \centering
\small
\caption{The data items considered in this survey.}
  \begin{subtable}[t]{0.45\textwidth}
    \centering
\small
\begin{tabular}{lll}
\toprule
\multicolumn{1}{c}{\textbf{ID}} & \multicolumn{1}{c}{\textbf{Data Item}} & \multicolumn{1}{c}{\textbf{RQ}} \\ \hline
$I_{1}$ & Authors & N/A \\ \hline
$I_{2}$ & Year & N/A \\ \hline
$I_{3}$ & Title & N/A \\ \hline
$I_{4}$ & Venue (conference or journal) & N/A \\ \hline
$I_{5}$ & Data preprocessing methods & RQ1 \\ \hline
$I_{6}$ & Data encoding schemes & RQ1 \\ \hline
$I_{7}$ & Data sampling strategies & RQ1 \\ \hline
$I_{8}$ & Sparsity handling mechanisms & RQ2 \\ \hline
$I_{9}$ & Deep learning models & RQ2 \\ \hline
$I_{10}$ & Reasons for model selection & RQ2 \\ \hline
$I_{11}$ & Optimization algorithms & RQ2 \\ 
\bottomrule
\end{tabular}
  \end{subtable}
  \hspace{0.6cm}
  \begin{subtable}[t]{0.45\textwidth}
    \centering
\small
\begin{tabular}{lll}
\toprule
\multicolumn{1}{c}{\textbf{ID}} & \multicolumn{1}{c}{\textbf{Data Item}} & \multicolumn{1}{c}{\textbf{RQ}} \\ \hline
$I_{12}$ & Activation functions & RQ2 \\ \hline
$I_{13}$ & Hyperparameter tuning methods & RQ2 \\ \hline
$I_{14}$ & Evaluation procedures & RQ3 \\ \hline
$I_{15}$ & Accuracy metrics & RQ3 \\ \hline
$I_{16}$ & Statistical tests & RQ3 \\ \hline
$I_{17}$ & Effect size measurements & RQ3 \\ \hline
$I_{18}$ & Number of subject systems & RQ3 \\ \hline
$I_{19}$ & Domain of subject systems & RQ3 \\ \hline
$I_{20}$ & Application categories & RQ4 \\ \hline
$I_{21}$ & Handling of dynamic environments & RQ4 \\ \hline
$I_{22}$ & Availability of code and dataset & RQ4
\\ \bottomrule
\end{tabular}
  \end{subtable}
  \label{tb:data_items}
\end{table}


%% file: Tables/quality_assessment.tex
\begin{table}[t!]
  \caption{Quality assessment questions and scoring metrics.}
\centering
\footnotesize
\begin{adjustbox}{width=\linewidth,center}
\begin{tabular}{lp{8cm}p{5cm}}
\toprule

\textbf{ID} & \textbf{Question} & \textbf{Scoring Metric} \\ \hline
\emph{Q$_{1}$} & Does the study preprocess the configuration data to enhance learning? & No=0, Yes=2. \\ \hline
\emph{Q$_{2}$} & Is the choice of data encoding methods clearly explained in the study? & Not explained=0, Implicitly explained=1, Clearly explained=2. \\ \hline
\emph{Q$_{3}$} & Is the sampling strategy for training data outlined? & Not outlined=0, Implicitly outlined=1, Clearly outlined=2. \\ \hline
\emph{Q$_{4}$} & Does the study address the sparsity/overfitting problem? & No=0, Yes=2. \\ \hline
\emph{Q$_{5}$} & Does the study employ domain knowledge to improve the model architecture? & No=0, Yes=2. \\ \hline
\emph{Q$_{6}$} & Are the reasons for selecting specific deep learning models provided? & Not provided=0, Implicitly provided=1, Clearly provided=2. \\ \hline
\emph{Q$_{7}$} & Is the optimizer used in the deep learning models clearly stated? & Not stated=0, Implicitly stated=1, Clearly stated=2. \\ \hline
\emph{Q$_{8}$} & Is the activation function employed in the deep models specified? & Not specified=0, Implicitly specified=1, Clearly specified=2. \\ \hline
\emph{Q$_{9}$} & Does the study introduce the hyperparameter tuning method? & Not introduced=0, Implicitly introduced=1, Clearly introduced=2. \\ \hline
\emph{Q$_{10}$} & Does the study outline the evaluation procedures for model performance? & Not outlined=0, Implicitly outlined=1, Clearly outlined=2. \\ \hline
\emph{Q$_{11}$} & Are the accuracy metrics used for evaluation clearly defined? & Not defined=0, Implicitly defined=1, Clearly defined=2. \\ \hline
\emph{Q$_{12}$} & Does the study employ statistical tests for significance analysis? & No=0, Yes=2. \\ \hline
\emph{Q$_{13}$} & Does the study conduct effect size tests conducted to measure the practical significance? & No=0, Yes=2. \\ \hline
\emph{Q$_{14}$} & Does the study consider multiple subject systems for evaluations? & Not mentioned=0, Only one system=1, More than one systems=2. \\ \hline
\emph{Q$_{15}$} & Does the study consider multiple domains of subject systems? & Not mentioned=0, Only one domain=1, More than one domains=2. \\ \hline
\emph{Q$_{16}$} & Does the study clarify the application purpose of & Not clarified=0, Implicitly clarified=1, Clearly clarified=2. \\ \hline
\emph{Q$_{17}$} & Does the study consider the challenge of dynamic environments? & No=0, Yes=2. \\ \hline
\emph{Q$_{18}$} & Are the codes and datasets available publicly for reproducibility? & No=0, Yes=2.
\\ \bottomrule
\end{tabular}
\end{adjustbox}
\label{tb:quality_assessment}
\end{table}

%% file: Figures/quality_assessment.tex
\begin{tikzpicture}
\footnotesize
\begin{axis}[
    width=15cm,
    height=5cm,
    ybar,
    ymax = 2,
    ymin = 0,
         axis x line  = bottom,
    axis y line  = left  ,
    bar width=0.4cm, 
   ylabel=Score,
    enlarge y limits=0.001,
    enlarge x limits=0.05,
    y label style={at={(0.015,0.5)}},
    x label style={at={(0.5,-0.05)}},
    xlabel=Question,
    symbolic x coords={A, B, C, D, E, F, G, H, I, J, K, L, M, N, O, P, Q, R}, 
    xtick=data, 
     xticklabels={\emph{Q$_{1}$},\emph{Q$_{2}$},\emph{Q$_{3}$},\emph{Q$_{4}$},\emph{Q$_{5}$},\emph{Q$_{6}$},\emph{Q$_{7}$},\emph{Q$_{8}$},\emph{Q$_{9}$},\emph{Q$_{10}$},\emph{Q$_{11}$},\emph{Q$_{12}$},\emph{Q$_{13}$},\emph{Q$_{14}$},\emph{Q$_{15}$},\emph{Q$_{16}$},\emph{Q$_{17}$},\emph{Q$_{18}$}},
     xticklabel style={text width=0.6cm},
    nodes near coords, 
]
\addplot[fill=blue!50, draw=black] coordinates {(A, 0.87) (B, 1.29) (C, 2) (D, 1.35) (E, 2) (F, 0.97) (G, 0.99) (H, 1.01) (I, 2) (J, 2) (K, 2) (L, 0.16) (M, 0.10) (N, 1.56) (O, 1.40) (P, 2) (Q, 0.69) (R, 0.42)};
\end{axis}
\end{tikzpicture}

%% file: Tables/gold_set.tex
\begin{table}[t!]
  \caption{Detailed scoring of the top 10 primary studies in the quality assessment. Full scoring data can be found at: \textcolor{blue}{\url{https://github.com/ideas-labo/DCPL-SLR/blob/main/Scoring.xlsb}}}
\centering
\footnotesize
\begin{adjustbox}{width=\linewidth,center}
\begin{tabular}{llllllllllllllllllll}
\toprule

\textbf{Ref.} & \textbf{\emph{Q$_{1}$}} & \textbf{\emph{Q$_{2}$}} & \textbf{\emph{Q$_{3}$}} & \textbf{\emph{Q$_{4}$}} & \textbf{\emph{Q$_{5}$}} & \textbf{\emph{Q$_{6}$}} & \textbf{\emph{Q$_{7}$}} & \textbf{\emph{Q$_{8}$}} & \textbf{\emph{Q$_{9}$}} & \textbf{\emph{Q$_{10}$}} & \textbf{\emph{Q$_{11}$}} & \textbf{\emph{Q$_{12}$}} & \textbf{\emph{Q$_{13}$}} & \textbf{\emph{Q$_{14}$}} & \textbf{\emph{Q$_{15}$}} & \textbf{\emph{Q$_{16}$}} & \textbf{\emph{Q$_{17}$}} & \textbf{\emph{Q$_{18}$}} & \textbf{Avg.} \\ \hline
\cite{DBLP:conf/fse/GongC24} & 2 & 1 & 2 & 2 & 2 & 2 & 1 & 1 & 2 & 2 & 2 & 2 & 2 & 2 & 2 & 2 & 2 & 2 & 1.83 \\ \hline
\cite{DBLP:conf/sigsoft/Gong023} & 2 & 2 & 2 & 2 & 2 & 2 & 1 & 1 & 2 & 2 & 2 & 2 & 2 & 2 & 2 & 2 & 0 & 2 & 1.78 \\ \hline
\cite{DBLP:conf/msr/GongC22} & 0 & 2 & 2 & 2 & 2 & 2 & 2 & 2 & 2 & 2 & 2 & 2 & 2 & 2 & 2 & 2 & 0 & 2 & 1.78 \\ \hline
\cite{DBLP:conf/icse/HaZ19} & 2 & 1 & 2 & 2 & 2 & 2 & 2 & 2 & 2 & 2 & 2 & 2 & 0 & 2 & 2 & 2 & 0 & 2 & 1.72 \\ \hline
\cite{DBLP:journals/tosem/ChengGZ23} & 2 & 1 & 2 & 2 & 2 & 2 & 2 & 2 & 2 & 2 & 2 & 2 & 0 & 2 & 2 & 2 & 0 & 2 & 1.72 \\ \hline
\cite{betting2023oikonomos} & 2 & 2 & 2 & 2 & 2 & 2 & 2 & 2 & 2 & 2 & 2 & 0 & 0 & 2 & 2 & 2 & 2 & 0 & 1.67 \\ \hline
\cite{DBLP:conf/sigmod/ZhangLZLXCXWCLR19} & 2 & 1 & 2 & 2 & 2 & 2 & 2 & 2 & 2 & 2 & 2 & 0 & 0 & 2 & 1 & 2 & 2 & 2 & 1.67 \\ \hline
\cite{DBLP:journals/access/TousiL22} & 2 & 2 & 2 & 2 & 2 & 2 & 2 & 2 & 2 & 2 & 2 & 0 & 0 & 1 & 1 & 2 & 2 & 2 & 1.67 \\ \hline
\cite{DBLP:journals/jss/LiBHWL24} & 2 & 1 & 2 & 2 & 2 & 2 & 2 & 2 & 2 & 2 & 2 & 0 & 0 & 2 & 2 & 2 & 0 & 2 & 1.61 \\ \hline
\cite{DBLP:journals/tse/ChenB17} & 2 & 1 & 2 & 2 & 2 & 2 & 2 & 2 & 2 & 2 & 2 & 2 & 0 & 2 & 2 & 2 & 0 & 0 & 1.61
\\ \bottomrule
\end{tabular}
\end{adjustbox}
\label{tb:top10_score}
\end{table}

%% file: Tables/sensitivity_precision.tex
\begin{table}[t!]
  \caption{The sensitivity and precision test using the quasi-gold standard in the latest five years and five top conferences/journals.}
\centering
\footnotesize
\begin{adjustbox}{width=\linewidth,center}
\begin{tabular}{p{7cm}p{1.5cm}p{1.5cm}p{1.5cm}p{1.5cm}}
\toprule
\textbf{Publication Venue} & \textbf{Total \# Studies Retrieved} & \textbf{\# Relevant Studies Retrieved} & \textbf{\# QGS Studies} & \textbf{QGS Studies} \\ \hline
Joint Meeting on European Software Engineering Conference and Symposium on the Foundations of Software Engineering (ESEC/FSE) & 10 & 2 & 2 & \cite{DBLP:conf/fse/GongC24}, \cite{DBLP:conf/sigsoft/Gong023} \\ \hline
International Conference on Software Engineering (ICSE) & 3 & 2 & 2 & \cite{DBLP:conf/icse/HaZ19}, \cite{DBLP:conf/icse/GaoGZLY23} \\ \hline
International Conference on Automated Software Engineering (ASE) & 3 & 2 & 2 & \cite{DBLP:conf/kbse/BaoLWF19}, \cite{DBLP:conf/kbse/ChenHL022} \\ \hline
Journal of Systems and Software (JSS) & 8 & 1 & 1 & \cite{DBLP:journals/jss/LiBHWL24} \\ \hline
Transactions on Software Engineering and Methodology (TOSEM) & 7 & 1 & 1 & \cite{DBLP:journals/tosem/ChengGZ23} \\ \bottomrule
\end{tabular}
\end{adjustbox}
\label{tb:sensitivity_precision}
\end{table}

%% file: RQ1.tex
\section{RQ1: How to Prepare Raw Configuration Data for Learning? }
\label{sec:rq1}

To improve the accuracy and reliability of the deep configuration performance model, data preparation is necessary~\cite{DBLP:journals/infsof/HuangL015}. In this section, we particularly review the preprocessing, encoding, and sampling when preprocessing configuration data that are commonly used by the approaches of deep configuration learning in the community.

\subsection{Preprocessing Configuration Data}
\label{subsec: preprocess}

\input{Tables/RQ-preprocessing_methods}

Data preprocessing methods play a fundamental role in preparing data for deep learning problems~\cite{DBLP:journals/infsof/HuangL015}, and understanding the preprocessing techniques employed in the context of configuration performance learning is crucial for researchers to gain insights and enhance the quality and reliability of the data. 


In general, data preprocessing is the process of transforming raw data into a computable and noise-free format. This can be crucial for deep configuration performance learning since the configuration options might be of diverse scale, nature, and types. As an example, for the video encoding software \textsc{VP8}, there are binary configuration options like \texttt{allowResize} (0 or 1), categorical options like \texttt{sharpness} (no/low/medium/high sharpening), and \textcolor{black}{numerical options such as \texttt{minGopSize} (from 0 to 1000).}

Table~\ref{tb:preprocessing} summarizes different preprocessing methods identified from our review. As can be seen, while various methods have been employed, a significant number of studies (56 out of 99) opt to use the \textbf{raw configuration performance dataset} without any preprocessing. Although this way may save effort and costs in the short term, it raises extra requirements on the quality of the datasets to ensure reliable results.

In contrast, \textbf{normalization} techniques have been applied in 40 studies, which are fundamental in data preprocessing to ensure that the scales of configuration options are brought to a standardized range, thereby preventing any particular options from dominating the learning due to its larger values~\cite{DBLP:journals/asc/SinghS20}. Several widely used normalization techniques for deep configuration performance learning include: {Min-max scaling}~\cite{DBLP:journals/taco/MoolchandaniKS22, DBLP:conf/sigsoft/Gong023, DBLP:conf/icse/HaZ19, DBLP:conf/msr/GongC22}, which transforms numerical options to a specific range, typically between 0 and 1, by subtracting the minimum value and dividing by the range of the option. {Z-score}~\cite{DBLP:journals/fgcs/LiLTWHQD19, DBLP:journals/mam/JooyaDB19} transforms numerical options by subtracting the mean and dividing by the standard deviation of the option, resulting in a distribution with a mean of 0 and a standard deviation of 1, and allowing for easier comparison and interpretation across different options. Although normalization techniques have gained popularity and offer several advantages, it is crucial to acknowledge their limitations. For example, scaling the data might disrupt the original distribution and decrease the performance of certain deep learning algorithms that depend on the inherent characteristics of the original data. Additionally, normalization techniques like min-max scaling can be sensitive to outliers, as extremely big values of the configuration options can significantly impact the range of the scaled values, potentially compressing the majority of the data into a narrow range. {Indeed, the \texttt{numCore} option of the stencil-grid solver \textsc{HSMGP} can range from 64 to 4096, while the largest value of the remaining configuration options is 6.}


Methods for \textbf{dimensionality reduction} have been used in nine studies, which help to simplify the data representation, remove redundant or irrelevant options, and overcome the curse of dimensionality, thereby improving computational efficiency and visualization of the configuration data. For example, {Principal Component Analysis (PCA)}~\cite{DBLP:conf/middleware/GrohmannNIKL19, DBLP:journals/mam/JooyaDB19, DBLP:journals/jss/ZhuYZZ21, DBLP:journals/concurrency/OuaredCD22, DBLP:conf/icccnt/KumarMBCA20, DBLP:conf/cluster/IsailaBWKLRH15} transforms the original options into a new set of uncorrelated variables called principal components, which are linear combinations of the original options and are ordered by the amount of variance they explain. 
\textcolor{black}{Notably,~\citet{DBLP:journals/ipm/LeeSCP24} leverage an independent method, namely \texttt{knob2vec}, to process the workload-specific configuration option vectors that capture the unique characteristics and relationships of the configuration options in \textsc{RocksDB}, and then the data is fed into a separate deep learning model. We classify it as a preprocessing method because it is used to process the configuration data before the training phase. }
Indeed,~\citet{DBLP:journals/tosem/ChengGZ23} also state that it is crucial to embed the extreme high-dimensional configuration space into low-dimension to better model the interactions between the options and performance.  However, these methods may also require additional resources to find the best hyperparameters for the best reduction outcomes. They also could make the model difficult to interpret since the original option space is destroyed.

Further, five studies have focused on \textbf{anomaly detection} techniques, aiming at identifying data absences, errors, and outliers within a configuration dataset. For example,~\citet{DBLP:conf/wosp/CengizFAM23} filter out samples where the runtime is zero in the SPEC CPU 2017 dataset, and~\citet{DBLP:conf/nsdi/FuGMR21} opt to remove configurations with performance values beyond the $99$th percentile. Examples include: {Smoothing}~\cite{DBLP:journals/fgcs/LiLTWHQD19}, which aims at capturing the underlying trends, patterns, or regularities in the configuration data while suppressing or filtering out the noise, and {Isolation Forest}~\cite{DBLP:journals/tecs/TrajkovicKHZ22} algorithm that constructs random decision trees to measure the average number of steps required to isolate an instance, allowing it to identify anomalies as instances of configuration that require fewer steps for isolation. Their limitations may include the potential loss of important details.

Another four studies have explored \textbf{featurization} approaches, where raw data unsuitable for machine/deep learning are transformed into meaningful options that can be used for modeling. For instance, {discretization}, as highlighted by~\citet{DBLP:journals/fgcs/LiLTWHQD19}, involves converting continuous options into discrete and representative categories, to save efforts on exploring the huge configuration space. Particularly, the authors discretize the reach rate of a Content Delivery Network by calculating its fifth power, such that the reach rate is projected into several discrete intervals, and each interval is assigned a unique value. Yet, these methods may result in information loss, especially in situations where the software system is highly complex, and they often rely heavily on domain experts. 

Besides, \textbf{imbalance processing} approaches have also been applied in four studies, which are crucial in dealing with imbalanced performance datasets with regard to the configurations, where the number of samples for different configuration values is significantly uneven. Indeed, in the configuration dataset of the widely used software \textsc{Apache}~\cite{DBLP:conf/icse/HaZ19, DBLP:conf/sigsoft/Gong023, DBLP:journals/tosem/ChengGZ23}, there are 128 samples with value 1 for the configuration option \texttt{InMemory}, while only 64 samples with value 0, resulting in imbalance with this option. To handle this, oversampling methods could be used, which increase the number of instances from the minority class to balance the dataset. For example,~\citet{DBLP:conf/sigsoft/Gong023} employ {SMOTE (Synthetic Minority Over-sampling Technique)}, which enhances the minority group by interpolating between neighboring instances, to balance the imbalanced divisions of samples within their ``divide-and-learn'' framework.

Lastly, two studies \textbf{remove multicollinearity} by addressing highly correlated options, e.g., {Kendall’s rank correlation}  \cite{DBLP:conf/wosp/CengizFAM23, DBLP:conf/splc/Acher0LBJKBP22}, which quantifies the strength and direction of the association between two ranked options, is used to remove highly correlated options and ensure that the predictors are independent or have a minimal correlation, which eventually reduces the cost of building deep configuration performance model.

\keybox{
\faIcon{search} \textit{\textbf{Finding \thefindingcount:} The default datasets, without any preprocessing, are used by most studies (56 out of
99). Among the rest preprocessing methods, normalization is the most popular in handling configuration data, i.e., in 40 studies.}
\addtocounter{findingcount}{1}
}

\subsubsection{Usage Scenario}
\textcolor{black}{Based on the review results, we summarize the strengths, weaknesses, and best-suited scenarios for each pre-processing method in Table~\ref{tb:scenario_preprocessing}, where researchers and practitioners can prepare their data in the most suitable way under different situations. }

\input{Tables/scenario_preprocessing}

\subsubsection{Good practice} Configuration performance data often exhibits some characteristics that are difficult to learn by deep learning models, and addressing these challenges can lead to improved model performance, which puts high demands on data preprocessing to improve the data quality. Therefore, it is a good practice that a variety of characteristics have been tackled in the primary studies. For example, the configuration space of software is demonstrated to be high-dimensional and sparse, which has been mitigated by regularization and dimension extraction techniques; and the primary studies also reveal that raw performance datasets often have outliers, imbalance, and multicollinearity issues. By addressing these issues, these studies have enhanced the effectiveness and efficiency of performance models and provided insights for future studies.

\subsubsection{Bad Smell} It is observed that a majority of the primary studies (56 out of 99) have paid inadequate attention to the importance of preprocessing methods. Neglecting these crucial steps can severely limit the performance and effectiveness of deep learning models~\cite{DBLP:journals/jss/PereiraAMJBV21}. Therefore, it is imperative for future studies to recognize the significance of preprocessing techniques and incorporate them systematically to ensure optimal model performance.

\suggestbox{
\faIcon{thumbs-up} \textit{\textbf{Actionable Suggestion \thesuggestioncount:} Incorporate different preprocessing techniques in the deep learning pipeline to address different data properties and ensure optimal deep configuration performance prediction performance.}
\addtocounter{suggestioncount}{1}
}

\subsection{Encoding Configuration Options}
\label{sec:rq1.2}

The procedure immediately followed by data preprocessing is encoding, which is concerned with converting the configuration data into a more appropriate format for learning. Indeed, as discovered by~\citet{DBLP:conf/msr/GongC22}, the impacts of encoding schemes are non-trivial, thereby the encoding scheme can considerably influence the learning outcome. 


Figure~\ref{fig:encoding} provides a summary of the encoding schemes surveyed. Among others, \textbf{label encoding}~\cite{DBLP:journals/taco/WangLWB19, chai2023perfsage, DBLP:conf/kbse/BaoLWF19, DBLP:conf/splc/Acher0LBJKBP22, DBLP:journals/tecs/TrajkovicKHZ22, DBLP:conf/pldi/SinghHLS22}, emerges as one of the most popular schemes, employed by 52 primary studies. Label encoding assigns a unique numerical label to each value of the option. As such, label encoding preserves the ordinal relationship between categories but does not introduce additional dimensions like one-hot encoding. Although label encoding can simplify the representations of configuration options, it also introduces a random numerical order for those categorical options, e.g., \textcolor{black}{for the highly configurable database system \textsc{MongoDB}, its configuration option \texttt{data\_strategy} has a set of configurable values $(str\_l1,$ $str\_l2,str\_l3)$, and label encoding will convert them into numeric values of $(0,1,2)$~\cite{DBLP:conf/msr/GongC22}}. This might potentially disrupt the inherent relationships between the configuration options and lead to worse predictions.

\textcolor{black}{Similarly, \textbf{scaled label encoding}~\cite{DBLP:conf/cluster/IsailaBWKLRH15, DBLP:journals/tnse/CaoPC22, DBLP:conf/icpp/DouWZC22, DBLP:conf/icse/HaZ19, DBLP:journals/tosem/ChengGZ23, DBLP:journals/pr/TranSWQ20}, which is a common variation of label encoding, is chosen by 26 primary studies in our survey. Different from classic label encoding, the encoded configuration options of software systems are normalized into a specific range, e.g., between 0 and 1, to overcome the limitations of label encoding. For instance, the \texttt{minGopSize} option of \textsc{VP8} will be label encoded from 0 to 1000, while the binary option can be only 0 and 1. This difference in scale may result in significant sparsity for performance models, which can be effectively handled by scaled label encoding.}

In contrast, \textbf{one-hot encoding}~\cite{DBLP:journals/access/TousiL22, DBLP:journals/fgcs/LiLTWHQD19, DBLP:journals/infsof/WartschinskiNVK22, ding2021portable, DBLP:conf/icse/GaoGZLY23, DBLP:conf/msr/GongC22, DBLP:conf/middleware/GrohmannNIKL19}, employed in 17 studies, transforms discrete options into a vector of binary values, where each element is represented as a binary form. For instance, for the video encoding system \textsc{VP8}, options like \texttt{sharpness}, which has categorical values no/low/medium/high sharpening, will be one-hot encoded into four binary variables. In this way, it enables deep learning models to appropriately interpret and capture relationships between different values. While one-hot encoding has been proven to be beneficial for the accuracy of performance models, it also leads to the highest training overheads~\cite{DBLP:conf/msr/GongC22}.

We find that eight studies employ \textbf{graph encoding} that leverages graph representations, such as the computational graphs of neural networks~\cite{DBLP:conf/pldi/SinghHLS22, DBLP:journals/corr/abs-2304-13032,wang2024slapp}, the directed acyclic graph (DAG) extracted from the workflows~\cite{DBLP:journals/jiii/YuGLZIY22, li2024pattern}, and structural graphs of applications~\cite{DBLP:journals/concurrency/RosarioSZNB23, chokri2022performance, DBLP:journals/pvldb/ZhouSLF20}, to represent configurations. \textcolor{black}{For example,~\citet{DBLP:conf/pldi/SinghHLS22} consider deep neural networks as a configurable software system and use graphs to represent the computational stages and operations of a DNN to better capture the internal interactions while modeling the runtime of the deep learning pipeline.} Graph encoding can lead to high-dimensional representations when dealing with a large set of options, potentially increasing the complexity and computational requirements.

Notably, a substantial portion of the work (i.e., 69 studies) opt to not explicitly specify encoding schemes used. Therefore, we have to inspect their codes or contact the authors to collect the data.


\keybox{
\faIcon{search} \textit{\textbf{Finding \thefindingcount:} Label encoding is applied to convert configuration values for learning in 52 studies, being the most common one, followed by scaled label encoding and one-hot encoding, used in 26 and 17 primary studies, respectively. Noteworthy, the majority of studies, 69 in particular, do not explicitly state their encoding schemes.}
\addtocounter{findingcount}{1}
}

\subsubsection{Usage Scenario}
\textcolor{black}{Based on our review, we have summarized the strengths, weaknesses, and optimal use cases for each encoding method in Table~\ref{tb:scenario_encoding}. It serves as a guide for researchers and practitioners, helping them choose the most appropriate configuration data encoding scheme for various scenarios.}

\input{Tables/scenario_encoding}

\subsubsection{Good Practice} Encouragingly, it has been demonstrated that the choice of encoding scheme is critical since it profoundly impacts the prediction accuracy and time consumption of deep learning models~\cite{DBLP:conf/msr/GongC22}. For instance, one-hot encoding tends to yield the highest accuracy albeit with longer training times compared to other schemes, while label encoding often leads to faster training times. Thereby, they provide practical rules for choosing the best encoding scheme and enable future researchers to make more informative decisions.

\subsubsection{Bad Smell} The influence of encoding schemes is underestimated by a majority of the studies, which solely rely on an ad-hoc way without providing justifications for the encoding schemes. This oversight can be detrimental, as it leaves other researchers unclear of which encoding schemes is suitable, potentially resulting in ambiguous data formats for deep learning tasks~\cite{DBLP:journals/jmlr/AlayaBGG19, DBLP:conf/kbse/BaoLWF19}. 

\textcolor{black}{This bad smell can be attributed to several reasons: (1) Researchers may lack awareness of the importance of encoding schemes in configuration performance learning, leading to a disregard for proper consideration and justification. (2) The absence of established guidelines or best practices regarding encoding schemes in the field can contribute to ad-hoc approaches and a lack of justification. (3) Time and resource constraints, limited emphasis on publications, and limited domain expertise in encoding schemes further compound this issue. These factors collectively leave researchers unsure of suitable encoding schemes, potentially resulting in ambiguous data formats for deep learning tasks, and highlighting the need for increased awareness, guidelines, and clearer justifications in this area.} 

\suggestbox{
\faIcon{thumbs-up} \textit{\textbf{Actionable Suggestion \thesuggestioncount:} Justify the rationals behind the choice of encoding schemes for deep configuration performance prediction to facilitate the understanding of this topic.}
\addtocounter{suggestioncount}{1}

\faIcon{thumbs-up} \textit{\textbf{Actionable Suggestion \thesuggestioncount:} Establish guidelines or best practices for encoding schemes for researchers in the community to avoid ad-hoc approaches and save trial-and-error costs.}
\addtocounter{suggestioncount}{1}
}

\begin{figure}[!t]
\centering
\begin{minipage}{0.45\textwidth}
\includestandalone[width=\columnwidth]{Figures/RQ-encoding}
    \caption{Distribution of the encoding schemes \textcolor{black}{(one study might use items from multiple categories)}.}
 \label{fig:encoding}
  \end{minipage}
~\hspace{-0.3cm}
\begin{minipage}{0.55\textwidth}
\includestandalone[width=\columnwidth]{Figures/RQ-sampling}
~\vspace{0.2cm}
\centering
\captionsetup{justification=centering}
    \caption{Distribution of the sampling strategies \newline
    \textcolor{black}{(one study might use items from multiple categories)}.}
 \label{fig:sampling}
  \end{minipage}
 \end{figure}

\subsection{Sampling Configurations for Learning}
\label{sec:rq1.3}

Sampling, which is typically performed to guide the raw configuration data collection, is also a critical step in data preparation, as the quality of the collected configuration samples determines the prediction accuracy of the deep learning models~\cite{DBLP:conf/wosp/PereiraA0J20}. 

Figure~\ref{fig:sampling} shows the categories of the sampling strategies used in the reviewed studies. We can first notice that an overwhelming majority of the studies surveyed, specifically 79 out of 99, employ \textbf{random sampling} as a prevalent data sampling technique in their research. Random sampling involves selecting data instances from the original configuration dataset in an unbiased and random manner. The widespread adoption of random sampling highlights its efficiency and effectiveness in creating balanced and diverse samples for training~\cite{DBLP:journals/jsa/ZhangLWWZH18, DBLP:conf/icse/GaoGZLY23, DBLP:journals/taco/WangLWB19, DBLP:conf/infocom/LiHZLT20, DBLP:journals/access/ThraneZC20, DBLP:journals/ese/VituiC21a}. However, random sampling may not efficiently capture the less frequent values of options, and these values may be underrepresented in the selected set, leading to bad generalizability of the deep learning models. \textcolor{black}{For instance, for the web server software \textsc{Apache}, the value of the option \texttt{InMemory} is 1 in 128 configurations, while the value 0 only appears 64 times in this option, therefore the training samples collected by random sampling will have a higher possibility to miss the value 0, especially when the training sample size is small, leading to bad accuracy when the model needs to predict new configurations with 0 for \texttt{InMemory}. }

In contrast, 15 of the primary studies follow a common practice of utilizing the default sets of training and testing samples that are pre-defined from the \textbf{benchmark} dataset~\cite{DBLP:journals/taco/MoolchandaniKS22, du2013performance, DBLP:conf/wosp/CengizFAM23, DBLP:journals/cee/KundanSA21, ding2021portable}, 
which allows for a direct comparison of their models' performance against the existing results. \textcolor{black}{For instance,~\citet{DBLP:conf/wosp/CengizFAM23} only consider the training and testing workloads from the SPEC CPU 2017 benchmark.} While they provide a standardized reference for representing specific scenarios or workloads, their applicability to real-world or diverse environments may be limited. 

Two studies apply \textbf{active learning} to obtain the samples~\cite{DBLP:journals/corr/abs-2304-13032, DBLP:journals/taco/LiL22}, which seek to select the most informative samples to be measured for training, thereby accelerating the learning process and reducing the costs of data collection. Yet, active learning methods may put high demand on uncertainty estimates or heuristics to select samples.

For the strategies that are part of the minority, each of the following is adopted by one primary study: \textbf{Stratified sampling}~\cite{DBLP:journals/pr/TranSWQ20} involves dividing the population into homogeneous subgroups based on specific characteristics and then sampling from each subgroup proportionally to its representation in the overall population. It also requires prior knowledge about the population's stratification variables and their distribution. \textbf{Pair-wise sampling}~\cite{DBLP:journals/tsc/KumaraACMHT23} seek to sample data that achieve maximum coverage of interaction effects between pairs of options while minimizing the number of samples required to prevent biases. Indeed, when dealing with systems that have a large number of configurations, pair-wise sampling may become impractical due to the sheer number of possible combinations. \textbf{Fractional factorial design}~\cite{DBLP:conf/cluster/IsailaBWKLRH15}, on the other hand, provides a way to select a subset of factor combinations that represent the most important effects. While it allows for investigating a large number of options with a reduced number of samples, its generalizability is limited by the assumption of linearity and consistency of the configurations. Lastly, \textbf{Model-based sampling}~\cite{wang2021morphling} builds a surrogate model capable of estimating uncertainty to effectively guide the search for selecting optimal training configurations. As a result, the accuracy of the surrogate model is critical.

\keybox{
\faIcon{search} \textit{\textbf{Finding \thefindingcount:} Random sampling is the most widely employed strategy for selecting configuration samples as observed in 79 studies, followed by using the predefined samples from the benchmark as exploited by 15 studies. Other sampling strategies form the minority.}
\addtocounter{findingcount}{1}
}

\subsubsection{Usage Scenario}
\textcolor{black}{Based on the comprehensive review, we have identified the advantages, limitations, and best-suited scenarios for each sampling method, which are summarized in Table~\ref{tb:scenario_sampling}. Thereby, this table serves as a valuable reference for researchers and practitioners to choose the most configuration data sampling approach based on specific situations and requirements.}

\input{Tables/scenario_sampling}

\subsubsection{Good Practice} The configuration space can be enormous, while the resources available for collecting training samples are often limited. Therefore, it is crucial to carefully select the most appropriate configurations to train performance models. However, one of the challenges lies in determining the criteria for sample selection. In the current literature, a good practice is that a few studies have addressed this issue by employing heuristic sampling methods tailored to specific requirements. Especially,~\citet{wang2021morphling} leverage SMBO to guide the sampling process with a surrogate Gaussian Process model to minimize the uncertainty of the samples in each iteration. By combining with various surrogate models, SMBO can potentially yield diverse sampling results, providing potential for future studies to investigate.

\subsubsection{Bad Smell} Despite the availability of various sampling strategies, a significant portion of the studies tends to focus on the most straightforward approach, namely random sampling. While random sampling is effective in representing the distribution of the dataset, it can lead to overfitting in highly sparse systems, as many of the features may prove ineffective for software performance analysis~\cite{DBLP:conf/wosp/PereiraA0J20}. This can be even more problematic considering the fact that there exist invalid configurations~\cite{DBLP:conf/kbse/JamshidiSVKPA17}. Therefore, researchers need to explore alternative sampling methods in such scenarios to enhance the accuracy of the performance models.

\suggestbox{
\faIcon{thumbs-up} \textit{\textbf{Actionable Suggestion \thesuggestioncount:} Avoid over-reliance on random sampling and explore alternative sampling methods to improve the quality of training data for better configuration performance prediction accuracy.}
\addtocounter{suggestioncount}{1}
}




%% file: Tables/RQ-preprocessing_methods.tex
\begin{table}[t!]
\caption{Distribution of the preprocessing methods \textcolor{black}{(one study might use items from multiple categories)}.}
\centering
\footnotesize
\begin{adjustbox}{width=\linewidth,center}

\begin{tabular}{p{2.9cm}lp{3cm}lp{5cm}}
\toprule

\textbf{Category} & \textbf{Total \#} & \textbf{Example} & \textbf{\# Studies} & \textbf{References} \\ \hline
Default & 56 & N/A & 56 & \cite{DBLP:conf/pldi/SinghHLS22},   \cite{DBLP:conf/mascots/KarniavouraM17},   \cite{DBLP:journals/corr/abs-2304-13032}, \cite{said2021accurate},   \cite{DBLP:conf/kbse/BaoLWF19},   \cite{DBLP:journals/tompecs/MakraniSNDSMRH21},   \cite{DBLP:journals/concurrency/YinH00C23},   \cite{DBLP:journals/access/LiLSJ20}, \cite{DBLP:conf/splc/GhamiziCPT19},   \cite{DBLP:conf/infocom/LiHZLT20}, \cite{DBLP:conf/sc/MalakarBVMK18},   \cite{DBLP:journals/comcom/AteeqAAK22}, \cite{DBLP:conf/spects/KimK17},   \cite{DBLP:journals/concurrency/OuaredCD22}, \cite{DBLP:conf/kbse/ChenHL022},   \cite{DBLP:conf/cf/LiuMCV20}, \cite{DBLP:conf/msr/GongC22},   \cite{DBLP:journals/fgcs/KousiourisMKGV14},   \cite{DBLP:conf/wosp/DidonaQRT15}, \cite{DBLP:conf/sc/MalikFP09},   \cite{DBLP:conf/noms/SanzEJ22}, \cite{DBLP:journals/concurrency/RosarioSZNB23},   \cite{blott2018finn}, \cite{DBLP:journals/tsc/KumaraACMHT23},   \cite{DBLP:journals/jsa/TangLLLZ22},   \cite{DBLP:journals/concurrency/CiciogluC22}, \cite{DBLP:conf/ipps/FalchE15},   \cite{DBLP:conf/IEEEcloud/MarosMSALGHA19},   \cite{DBLP:journals/concurrency/FalchE17}, \cite{myung2021machine},   \cite{wang2021morphling}, \cite{DBLP:conf/nsdi/LiangFXYLZYZ23},   \cite{DBLP:journals/cee/KundanSA21}, \cite{DBLP:journals/access/WangXTW20},   \cite{DBLP:conf/iccad/KimMMSR17}, \cite{DBLP:conf/esem/ShuS0X20},   \cite{du2013performance}, \cite{DBLP:conf/im/JohnssonMS19},   \cite{chokri2022performance}, \cite{chai2023perfsage},   \cite{DBLP:journals/taco/LiL22}, \cite{DBLP:journals/pvldb/MarcusP19},   \cite{ding2021portable}, \cite{DBLP:journals/jsa/ChengCWX17},   \cite{DBLP:conf/ic2e/RahmanL19}, \cite{DBLP:journals/taco/WangLWB19},   \cite{DBLP:journals/tcc/PhamDF20}, \cite{DBLP:conf/springsim/LuxWCBLXBBCH18},   \cite{DBLP:conf/dsrt/DuongZCLZ16},   \cite{DBLP:conf/middleware/MahgoubWGMGHMGB17},   \cite{DBLP:journals/access/NadeemAMFA19},   \cite{DBLP:journals/infsof/WartschinskiNVK22}, \cite{li2024pattern},   \cite{DBLP:conf/qrs/ZhangWZ23}, \cite{DBLP:journals/pomacs/LiPFLWH22},   \cite{DBLP:conf/ics/TrumperBSCH23} \\ \hline
\multirow{7}{*}{Normalization} & \multirow{7}{*}{40} & Min-max scaling & 34 & \cite{DBLP:conf/sigsoft/Gong023},   \cite{DBLP:conf/nsdi/FuGMR21}, \cite{DBLP:journals/pvldb/ZhouSLF20},   \cite{DBLP:journals/taco/MoolchandaniKS22},   \cite{DBLP:journals/jiii/YuGLZIY22}, \cite{DBLP:journals/concurrency/JiZL22},   \cite{DBLP:conf/icse/GaoGZLY23}, \cite{DBLP:journals/jsa/ZhangLWWZH18},   \cite{zain2022software}, \cite{DBLP:journals/tse/ChenB17},   \cite{sabbeh2016performance}, \cite{DBLP:journals/ese/VituiC21a},   \cite{DBLP:journals/pr/TranSWQ20}, \cite{DBLP:journals/tosem/ChengGZ23},   \cite{DBLP:conf/icpp/DouWZC22}, \cite{DBLP:journals/tnse/CaoPC22},   \cite{DBLP:conf/icccnt/KumarMBCA20}, \cite{DBLP:conf/icst/PorresARLT20},   \cite{DBLP:conf/icpp/MadireddyBCLLRS19},   \cite{DBLP:conf/sbac-pad/NemirovskyAMNUC17},   \cite{DBLP:conf/middleware/GrohmannNIKL19},   \cite{DBLP:journals/mam/JooyaDB19}, \cite{DBLP:journals/access/TousiL22},   \cite{DBLP:conf/cluster/IsailaBWKLRH15}, \cite{DBLP:conf/icse/HaZ19},   \cite{DBLP:conf/sigmod/ZhangLZLXCXWCLR19},   \cite{DBLP:journals/access/ThraneZC20}, \cite{betting2023oikonomos},   \cite{DBLP:conf/cluster/AssogbaLRK23}, \cite{DBLP:conf/fse/GongC24},   \cite{DBLP:conf/hpdc/YokelsonCL23}, \cite{DBLP:journals/jss/LiBHWL24},   \cite{wang2024slapp}, \cite{DBLP:conf/hipc/BettingZS23} \\ \cline{3-5} 
 &  & Standardization & 6 & \cite{DBLP:journals/jss/ZhuYZZ21},   \cite{DBLP:conf/sbac-pad/NemirovskyAMNUC17},   \cite{DBLP:conf/sc/MaratheAJBTKYRG17}, \cite{betting2023oikonomos},   \cite{jiang2024ml}, \cite{wyzykowski2024optimizing} \\ \cline{3-5} 
 &  & Z-score & 2 & \cite{DBLP:journals/mam/JooyaDB19},   \cite{DBLP:journals/fgcs/LiLTWHQD19} \\ \cline{3-5} 
 &  & Shift-log transformation & 2 & \cite{DBLP:conf/fse/GongC24},   \cite{wang2024slapp} \\ \cline{3-5} 
 &  & Box-Cox transformation & 1 & \cite{DBLP:conf/cluster/IsailaBWKLRH15} \\ \cline{3-5} 
 &  & Centering & 1 & \cite{DBLP:conf/cluster/IsailaBWKLRH15} \\ \cline{3-5} 
 &  & Make unit consistent & 1 & \cite{DBLP:conf/wosp/CengizFAM23} \\ \hline
\multirow{3}{*}{Dimension reduction} & \multirow{3}{*}{9} & PCA & 8 & \cite{DBLP:conf/cluster/IsailaBWKLRH15},   \cite{DBLP:conf/middleware/GrohmannNIKL19},   \cite{DBLP:journals/mam/JooyaDB19},   \cite{DBLP:journals/concurrency/OuaredCD22},   \cite{DBLP:conf/icccnt/KumarMBCA20}, \cite{DBLP:journals/jss/ZhuYZZ21},   \cite{DBLP:conf/sc/MalikFP09}, \cite{jiang2024ml} \\ \cline{3-5} 
 &  & CCA & 1 & \cite{DBLP:journals/concurrency/OuaredCD22} \\ \cline{3-5} 
 &  & Knob2vec & 1 & \cite{DBLP:journals/ipm/LeeSCP24} \\ \hline
\multirow{3}{*}{Anomaly   detection} & \multirow{3}{*}{5} & Outlier detection & 3 & \cite{DBLP:journals/mam/JooyaDB19},   \cite{DBLP:conf/nsdi/FuGMR21}, \cite{DBLP:conf/wosp/CengizFAM23} \\ \cline{3-5} 
 &  & Isolation   forest algorithm & 1 & \cite{DBLP:journals/tecs/TrajkovicKHZ22} \\ \cline{3-5} 
 &  & Smoothing & 1 & \cite{DBLP:journals/fgcs/LiLTWHQD19} \\ \hline
\multirow{4}{*}{Featurization} & \multirow{4}{*}{4} & Feature construction & 2 & \cite{DBLP:conf/splc/Acher0LBJKBP22},   \cite{DBLP:conf/nsdi/FuGMR21} \\ \cline{3-5} 
 &  & Discretization & 1 & \cite{DBLP:journals/fgcs/LiLTWHQD19} \\ \cline{3-5} 
 &  & Alphanumeric cleaning & 1 & \cite{DBLP:conf/wosp/CengizFAM23} \\ \cline{3-5} 
 &  & Make columns categorical & 1 & \cite{DBLP:conf/wosp/CengizFAM23} \\ \hline
\multirow{2}{*}{Imbalance processing} & \multirow{2}{*}{4} & Simple over-sampling & 2 & \cite{DBLP:journals/fgcs/LiLTWHQD19},   \cite{jiang2024ml} \\ \cline{3-5} 
 &  & SMOTE & 2 & \cite{DBLP:conf/sigsoft/Gong023},   \cite{DBLP:journals/jss/ZhuYZZ21} \\ \hline
Remove   multicollinearity & 2 & Kendall's rank correlation & 2 & \cite{DBLP:conf/wosp/CengizFAM23},   \cite{DBLP:conf/splc/Acher0LBJKBP22}
\\ \bottomrule
\end{tabular}

\end{adjustbox}
\label{tb:preprocessing}
\end{table}

%% file: Tables/scenario_preprocessing.tex
\begin{table}[t!]
  \caption{Summaries of the strengths, weaknesses, and best-suited usage scenarios for the data preprocessing methods.}
\centering
\footnotesize
\begin{adjustbox}{width=\linewidth,center}
\begin{tabular}{p{2cm}p{4cm}p{3.5cm}p{3.5cm}}
\toprule

\textbf{Category} & \textbf{Strength} & \textbf{Weakness} & \textbf{Best-Suited Scenario} \\ \hline
Normalization & (1) Ensures options contribute equally. (2) Speeds up convergence in model training. & Sensitive to outliers. & Datasets with varying option scales. \\ \hline
Dimension reduction & (1) Removes redundant or irrelevant options. (2) Reduces computational cost. & Makes the model difficult to interpret since the original option space is destroyed. & High-dimensional and sparse datasets. \\ \hline
Anomaly detection & Identifies data absences, errors, and outliers. & The potential loss of important details. & Datasets with extreme noises. \\ \hline
Featurization & Captures meaningful features in raw data. & Requires domain knowledge. & When human experts are available. \\ \hline
Imbalance processing & Enhances accuracy when training samples are unevenly distributed. & Increases computational complexity. & Datasets with biased values. \\ \hline
Remove multicollinearity & Reduces variance in regression coefficients. & May exclude useful information. & When some options are highly correlated.
\\ \bottomrule
\end{tabular}
\end{adjustbox}
\label{tb:scenario_preprocessing}
\end{table}

%% file: Tables/scenario_encoding.tex
\begin{table}[t!]
  \caption{Summaries of the strengths, weaknesses, and best-suited usage scenarios for the data encoding methods.}
\centering
\footnotesize
\begin{adjustbox}{width=\linewidth,center}
\begin{tabular}{p{2cm}p{4cm}p{3cm}p{4cm}}
\toprule

\textbf{Category} & \textbf{Strength} & \textbf{Weakness} & \textbf{Best-Suited Scenario} \\ \hline
Label encoding & (1) Preserves ordinal relationships in categorical data. (2) Training efficiency and prediction accuracy were well-balanced. & Introduces a random numerical order for those categorical options. & When categorical options have a meaningful order. \\ \hline
Scaled label encoding & (1) Same strengths as label encoding. (2) Handles options in different scales to prevent dominating. & Sensitive to outliers. & For ordered categorical features with varying magnitudes. \\ \hline
One-hot encoding & (1) Captures the influence of each option value. (2) Beneficial for the accuracy of performance models. & Introduces extra training overheads. & When prediction accuracy is more important than training overhead. \\ \hline
Graph encoding & Captures complex relationships and structures between options. & Increases the complexity and computational requirements. & When data naturally forms a graph, such as computation graphs of deep learning models.
\\ \bottomrule
\end{tabular}
\end{adjustbox}
\label{tb:scenario_encoding}
\end{table}

%% file: Figures/RQ-encoding.tex
\begin{tikzpicture}[scale=0.8]
\tikzstyle{every node}=[font=\large]
    \pie[
        /tikz/every pin/.style={align=center},
        text=pin,
        rotate=65,
        explode=0, 
        ]
        {
        50/\texttt{Label} (52),
        25/\texttt{Scaled label} (26),
        17/\texttt{One-hot} (17),
        8/\texttt{Graph} (8)
        }

    \end{tikzpicture}

%% file: Figures/RQ-sampling.tex
\begin{tikzpicture}
\begin{axis}[
    xbar=0.05cm,
    width=12cm,
    height=6cm,
    xmax=100,
    bar width=0.3cm,
 enlarge y limits=0.09,
 enlarge x limits=0.01,
    legend style={at={(1.05,0.7)},anchor=north,legend columns=1,font=\tiny},
    legend cell align={left},
   xlabel={\# studies},
     x label style={at={(0.5,0.02)},font=\Large},
     nodes near coords,
          point meta=explicit symbolic,
          every node near coord/.append style={anchor=west,font=\large},
        symbolic y coords={
        SMBO,
        FFD,
        Coverage-based,
        Stratified,
        Active learning,
        Benchmark,
        Random
        },
         y tick label style={font=\large},
          x tick label style={font=\large},
          yticklabels={,,},
         ytick=data
    ]

\addplot[fill=blue!50, draw=black] coordinates {(79,Random)[79 (Random)] (15,Benchmark)[15 (Benchmark dataset)] (2,Active learning)[2 (Active learning)]  (1,Stratified)[1 (Stratified sampling)] (1,Coverage-based)[1 (Pair-wise sampling)] (1,FFD)[1 (Fractional factorial design)] (1,SMBO)[1 (Model-based sampling)]};




\end{axis}
\end{tikzpicture}

%% file: Tables/scenario_sampling.tex
\begin{table}[t!]
  \caption{Summaries of the strengths, weaknesses, and best-suited usage scenarios for the data sampling methods.}
\centering
\footnotesize
\begin{adjustbox}{width=\linewidth,center}
\begin{tabular}{p{1.8cm}p{4cm}p{4cm}p{4cm}}
\toprule

\textbf{Category} & \textbf{Strength} & \textbf{Weakness} & \textbf{Best-Suited Scenario} \\ \hline
Random & (1) Simple and easy to implement. (2) Efficient and effective in creating balanced and diverse samples. & Can miss rare but important options. & For exploratory analysis with no prior knowledge. \\ \hline
Benchmark & Enables direct comparison against existing models. & May not be applicable to real-world scenarios. & In well-defined environments with established benchmarks. \\ \hline
Active learning & Can select the most informative samples. & Puts high demand on heuristics to select samples. & When collecting data is expensive. \\ \hline
Stratified & Ensures representation of all subgroups. & Requires knowledge of subgroup proportions. & When the dataset has distinct subgroups. \\ \hline
Pair-wise & Maximum coverage of interaction effects between pairs of options. & Impractical due to the sheer number of possible combinations. & When the options have strong correlations with each other. \\ \hline
Fractional factorial design & Reduces the number of experiments needed. & Generalizability is limited by the assumption of linearity and consistency of the options. & When the configuration space is large, linear, and has consistent options. \\ \hline
Model-based & Efficiently explores the search space. & The choice of the surrogate model is critical. & When there is an accurate surrogate model. 
\\ \bottomrule
\end{tabular}
\end{adjustbox}
\label{tb:scenario_sampling}
\end{table}

%% file: RQ2.tex
\section{RQ2: How can the Deep Configuration Performance Model be Built? }
\label{sec:rq2}

A deep learning model can be trained once the data preparation has been completed. To understand the state-of-the-art at this stage for deep configuration performance learning, this section summarizes details on how deep learning can be effectively built to learn configuration performance, offering insights into best practices and potential problems in this field.

\subsection{Handling Configuration Sparsity and Preventing Overfitting}
\label{sec:rq2.1}

\input{Tables/RQ-overfitting}

As large-scale software systems generate vast amounts of data, sparsity arises when valuable information is scarce in the options or unevenly distributed across the configuration space. As a result, overfitting occurs when deep learning models excessively learn the configuration data used for training, leading to poor generalization on unseen samples~\cite{DBLP:conf/sigsoft/Gong023, DBLP:conf/icse/HaZ19, DBLP:conf/sc/MalikFP09, DBLP:conf/middleware/GrohmannNIKL19, DBLP:journals/corr/abs-2304-13032, DBLP:journals/pvldb/MarcusP19, DBLP:journals/tecs/TrajkovicKHZ22, DBLP:conf/splc/Acher0LBJKBP22, DBLP:conf/esem/ShuS0X20}. \textcolor{black}{For instance, the option \texttt{use\_gpu} which enables GPU processing for the video codec \textsc{x264} has a large influence on the runtime, and deep learning models tend to assign a large weight to this option while ignoring the influence of others, leading to overfitted predictions.}


Table~\ref{tb:overfitting} presents the survey results of mechanisms used to handle sparsity and overfitting. Notably, 34 out of 99 studies have \textbf{not adopted any specific mechanism} to handle sparsity, which could harm the accuracy of deep configuration performance models.

\textcolor{black}{Among the remaining studies, 42 primary research papers have leveraged \textbf{feature selection} techniques as a means to deselect the undesired configuration options, hence mitigating sparsity. In particular, feature sparsity is observed when there exists a set of non-influential configuration options, which adds redundant parameters to the configuration performance model. Therefore, by eliminating these options with feature selection, the performance model will be resilient from sparsity. However, it is worth noting that 24 out of the 42 studies simply rely on {human effort} to select the most important options, which is inefficient and lacks generalizability. Among others, the following statistics are found:}
\begin{itemize}
    \item Five studies have applied {correlation-based feature selection} techniques, such as correlation analysis~\cite{DBLP:journals/ese/VituiC21a, DBLP:conf/sc/MalikFP09, DBLP:journals/tecs/TrajkovicKHZ22, DBLP:journals/jss/ZhuYZZ21, DBLP:conf/hpdc/YokelsonCL23}, that leverages statistical measures to determine the strength of relationships between options. 
    \item {Neural Network-based feature selection} applies neural network layers to identify relevant features, which have the inherent ability to learn and extract informative options from high-level representations, as demonstrated by four studies. For example, CNNs~\cite{DBLP:journals/access/ThraneZC20} (Convolutional Neural Network) are commonly used in image-related tasks to extract local patterns and spatial hierarchies, and GNNs~\cite{chai2023perfsage} (Graph Neural Networks) are particularly effective for structured data represented as graphs, which analyze relationships and dependencies among nodes in a graph to capture important patterns and interactions. 
    \item Four other studies have used {tree-based feature selection}, which inherently enables analyzing the structure and splits of the trees to determine the most informative options for prediction. For instance, extra trees regression~\cite{DBLP:conf/icpp/MadireddyBCLLRS19} involves constructing multiple decision trees with randomized splits to estimate feature importance, and~\citet{DBLP:conf/middleware/GrohmannNIKL19} utilize the ensemble nature of random forest to rank options based on their contribution to the model's predictive power.
    \item  Moreover, each of the following feature selection methods is used by two primary studies: (1) {Filter-based feature selection} techniques evaluate each feature independently of the target option and rank them based on predefined criteria. An example is the minimum Redundancy Maximum Relevance Feature Selection (mRMR)~\cite{DBLP:journals/jsa/ZhangLWWZH18}, which seeks a balance between selecting options that have high relevance to the performance while minimizing redundancy among selected options; (2) {Wrapper-based feature selection}~\cite{DBLP:journals/access/TousiL22} uses Recursive Feature Elimination (RFE) to iteratively eliminate options with the least importance, while evaluating the model's performance using cross-validation. \textcolor{black}{In particular, Tousi et al.~\cite{DBLP:journals/access/TousiL22} select 10 most important options like \texttt{max\_mhz}, \texttt{log\_cpus} and \texttt{sockets} in the SPEC CPU 2017 dataset using RFE with four different estimators.} Hybrid model also exists as proposed by Chen and Bahsoon~\cite{DBLP:journals/tse/ChenB17}. The core is to combine the outputs of multiple learning algorithms to guide the feature selection process. 
    \item  Finally, {metaheuristic feature selection} \cite{DBLP:journals/jss/ZhuYZZ21}, which combines a whale optimization algorithm and simulated annealing to construct a search-based option selection algorithm named EMWS, is explored in one study. 
\end{itemize}

Alternatively, 25 studies have employed \textbf{regularization} techniques, which introduce additional constraints or penalties to the model's objective function, aiming to control sparsity, reduce overfitting, and enhance the robustness of the performance models. 
Several commonly used examples are: 

\begin{itemize}
    \item {$L_1$ regularization}~\cite{DBLP:conf/sigsoft/Gong023, DBLP:journals/access/ThraneZC20, DBLP:journals/taco/WangLWB19, DBLP:conf/iccad/KimMMSR17, DBLP:conf/esem/ShuS0X20, DBLP:conf/icse/HaZ19}, also known as Lasso regularization, adds a penalty term proportional to the absolute value of the model's coefficients, encouraging sparsity in the model by driving some coefficients to zero and reducing the impact of irrelevant features. \textcolor{black}{For instance, after performing $L_1$ regularization on the configuration options of the video encoding software \textsc{VP8}, the coefficients of options like \texttt{autoAltRef} and \texttt{allowResize} are reduced to zero because they have trivial influence on the performance (runtime), while the coefficient of options like \texttt{bestQuality} is changed from $10^{16}$ to $1.9 \times 10^{4}$, significantly reducing the sparsity of configurations.} 
    \item {$L_2$  regularization}~\cite{DBLP:journals/access/ThraneZC20, DBLP:journals/jsa/TangLLLZ22, DBLP:conf/im/JohnssonMS19, DBLP:conf/esem/ShuS0X20, DBLP:journals/access/TousiL22}, also called Ridge regularization, adds a penalty term proportional to the squared magnitude of the model's coefficients. It encourages smaller coefficients and can help mitigate multicollinearity issues by reducing the impact of highly correlated options. However, regularization methods come with the trade-off of introducing additional parameters that require tuning to achieve optimal results, which may incur higher costs in terms of time, computational resources, and expertise needed to find the best parameter values~\cite{DBLP:conf/icse/HaZ19, DBLP:conf/sigsoft/Gong023}. 
\end{itemize}

\textbf{Dropout} techniques are widely used to eliminate insignificant model parameters during the neural network training process in 13 prominent studies. Particularly, dropout~\cite{DBLP:journals/access/ThraneZC20, DBLP:conf/cf/LiuMCV20, DBLP:journals/infsof/WartschinskiNVK22, said2021accurate, zain2022software, DBLP:conf/esem/ShuS0X20, DBLP:journals/fgcs/LiLTWHQD19} works by randomly deactivating a fraction of neurons during the training phase of a neural network. By doing so, dropout prevents the network from relying too heavily on specific neurons, forcing it to learn more robust and generalized features. In addition,~\citet{DBLP:conf/middleware/MahgoubWGMGHMGB17} applied ensemble pruning of networks to prune the top 30\% of the ensemble networks to prevent overfitting. Compared to feature selection and regularization, the advantages of dropout are its simplicity and computational efficiency, which is particularly useful for training large neural networks. Yet, this also causes a lack of interpretability, as dropout does not provide explicit information about which specific options or neurons are important or unimportant.

Meanwhile, data preprocessing techniques like \textbf{dimension extraction} have also been leveraged to eliminate unnecessary representations and information from the original dataset in nine studies of deep configuration performance learning. Unlike feature selection techniques, which only focus on mitigating specific options, dimension extraction approaches can achieve simultaneous handling of features, which is crucial when there are interactions or dependencies among features. This is because they work on the overall weight magnitudes, allowing the deep learning models to strike a balance between the negative impacts of the features and their potential interactions with others.  \textcolor{black}{For example,~\citet{DBLP:journals/ipm/LeeSCP24} embed knobs of configurable software systems into a latent space representation to extract the most informative features and use together for configuration performance modeling, thereby reducing the sparse representations of configuration data.}


Furthermore, although a number of studies have focused on addressing feature sparsity, where a small subset of features holds significant influence, there are very few studies that deal with sample sparsity, in which samples exhibit substantial variations in performance, resulting in a non-smooth distribution. \textcolor{black}{For example,~\citet{DBLP:conf/sigsoft/Gong023} have shown that by dividing the samples of the video codec \textsc{VP8} according to the option \texttt{rtQuality}, there are two clusters of samples with significantly diverse distributions, i.e., the division with \texttt{rtQuality} = 0 has runtime between 0 and 10000 ms, while the other ranges from 10000 to 60000.} In light of this gap, they proposed a framework based on the idea of \textbf{``divide-and-learn''}, i.e., \texttt{DaL}, which divides the original dataset into distinct divisions based on hidden characteristics and learns a local deep learning model for each division to mitigate sample sparsity~\cite{DBLP:conf/sigsoft/Gong023}. \textcolor{black}{In another study by~\citet{DBLP:journals/jss/LiBHWL24}, \textbf{\texttt{RSFIN} (Rule Search-based Fuzzy Inference Network)} is designed to capture configurations with low information entropy, as they are more likely to represent common patterns or structures in the data, thereby, \texttt{RSFIN} can effectively capture the hidden structures and distributions in the configuration space, leading to a better understanding of the system behavior. }

\keybox{
\faIcon{search} \textit{\textbf{Finding \thefindingcount:} A considerable number of studies do not explicitly handle the problem of sparsity and overfitting
(34 out of 99). Among the studies that deal with this challenge, 
manual feature selection is the most common method (24 studies).}
\addtocounter{findingcount}{1}
}

\subsubsection{Usage Scenario}
\textcolor{black}{Based on our comprehensive review, we have identified the distinct advantages, limitations, and optimal usage scenarios for each sparsity handling method. These findings are concisely summarized in Table~\ref{tb:scenario_sparsity}, providing researchers and practitioners with a valuable reference to select the most appropriate approach for addressing the sparsity problem in their specific scenarios.}

\input{Tables/scenario_sparsity}

\subsubsection{Good Practice} Encouragingly, a majority of studies recognize the significance of addressing sparsity and overfitting, which are critical for causing the degradation of predictive models, urging the application of a diverse range of handling techniques~\cite{DBLP:conf/sigsoft/Gong023, DBLP:conf/icse/HaZ19, DBLP:conf/sc/MalikFP09, DBLP:conf/middleware/GrohmannNIKL19, DBLP:journals/corr/abs-2304-13032, DBLP:journals/pvldb/MarcusP19, DBLP:journals/tecs/TrajkovicKHZ22, DBLP:conf/splc/Acher0LBJKBP22, DBLP:conf/esem/ShuS0X20}. However, it is worth noting that most of these techniques are option selection methods as the configuration options often have redundant information and most of them have little influence on the software performance. Yet,~\citet{DBLP:conf/sigsoft/Gong023} discover that except for feature/option sparsity, which has been realized by most studies, the distribution of samples in the configuration landscape is also sparse---a significant property in the configuration performance learning problem that is worth further investigating.

\subsubsection{Bad Smell} Given that sparsity has been addressed by 65 out of 99 papers, 24 of them rely on human efforts to select the correct feature sets to reduce sparsity, which is of low efficiency and low generalizability. In addition, there are still 34 works that did not acknowledge the sparsity issue. \textcolor{black}{Moreover, only one study focused on solving the sparsity problem from the perspective of sample distributions by dividing the sparse clusters into more focused divisions each learned by a local deep model~\cite{DBLP:conf/fse/GongC24}, which could be a key aspect of future research..} 

\suggestbox{
\faIcon{thumbs-up} \textit{\textbf{Actionable Suggestion \thesuggestioncount:} Future researchers should avoid relying heavily on human efforts for feature selection, but explore automatic approaches to improve efficiency and generalizability.}
\addtocounter{suggestioncount}{1}

\faIcon{thumbs-up} \textit{\textbf{Actionable Suggestion \thesuggestioncount:} It would be highly beneficial to address sample sparsity by handling the sparse distribution of samples in the configuration landscape, instead of solely focusing on feature sparsity.}
\addtocounter{suggestioncount}{1}
}

\subsection{Choosing Deep Learning Model}
\label{sec:rq2.2}

\input{Tables/RQ-learning_model}

\subsubsection{\textcolor{black}{Deep Learning Models}}

It is not hard to anticipate that there will be a variety of models applied to deep configuration performance learning, as shown in Table~\ref{tb:learning}. Specifically, the majority of primary studies (69 out of 99) employ a \textbf{Feedforward Neural Network (FNN)}. Among these, the most prevalent approach is Multilayer Perceptron (MLP) in 60 studies. MLP is a fundamental neural network consisting of interconnected neurons organized in multiple layers. These neurons apply activation functions to the weighted sum of their inputs, introducing non-linear transformations to the data.~\citet{DBLP:conf/icse/HaZ19} propose the utilization of regularization techniques in regularized Deep Neural Networks (rDNN), such that additional penalties are introduced during training to remove insignificant features, control feature sparsity of the model and thereby prevent overfitting.~\citet{DBLP:conf/sigsoft/Gong023, DBLP:conf/fse/GongC24} combines rDNN with the divide-and-learn framework to further address the sample sparsity issues. In addition, Kernel Extreme Learning Machines (KELM) are also used in two studies, aiming at simplifying the training process of neural networks, which randomly initializes the weights between the input and hidden layers, then applies a kernel function to obtain the transformed feature representation, and analytically determines the weights between the hidden and output layers based on the transformed features. A most recent work~\cite{DBLP:journals/tosem/ChengGZ23} employs a Hierarchical Interaction Neural Network (HINN) that leverages a hierarchical structure for performance learning, where lower-level layers typically capture low-level options, and higher-level layers learn to combine these low-level options into more complex representations. Further, dynamic neural networks and radial basis function neural networks are each utilized in one study.

However, FNNs cannot handle sequential or temporal data effectively in many real-world tasks, such as natural language processing and time series prediction. To mitigate this, \textbf{Recurrent Neural Networks (RNN)} are specifically designed to maintain an internal memory or hidden state that persists across time steps, allowing them to capture information from past inputs. Among our review scope, RNNs have been employed in 10 works. For example, Long Short-Term Memory (LSTM) is a dedicated version of the recurrent neural network with internal memory cells, which allows them to selectively retain or forget information based on the input and previous context by using specialized units called gates, and therefore overcome the limitations of traditional RNNs in capturing and remembering long-term dependencies in sequential data. \textcolor{black}{For example,~\citet{DBLP:journals/jsa/TangLLLZ22} employ LSTM to tune the performance of file systems like \textsc{Ext4}, \textsc{F2FS}, and \textsc{PMFS}, where the operations and workloads are dynamically changing over time and it is crucial to apply time-related deep learning models to model the temporal data.}

On the other hand, \textbf{Convolutional Neural Networks (CNN)} have been studied in nine research studies to process grid-like structured data, such as images and videos, which can not be handled by FNNs. The hierarchical feature learning capability of the convolutional layers allows CNNs to automatically learn and extract meaningful representations from the input data. Notably, a particular variant of CNNs, known as Residual Networks (ResNets), has been exploited by~\citet{DBLP:conf/wosp/CengizFAM23}, which incorporate novel architectural designs with residual connections to address the issue of vanishing gradients during training. \textcolor{black}{As an example,~\citet{DBLP:conf/cf/LiuMCV20} leverage the ability of CNN that can disclose the hidden and complex correlations among different options to predict the performance of high-performance computing systems like \textsc{RISC-V}.}

In addition, \textbf{Graph Neural Networks (GNN)} and its variant DAG-transformer have also been used in nine studies, which utilize graph convolutions and message-passing techniques to capture complex relationships and dependencies within graph structures composed of nodes and edges. \textcolor{black}{For example,~\citet{chai2023perfsage} utilize GNN to predict the inference latency, energy, and memory footprint of DNN inference, because it can directly process arbitrary DNN TFlite graphs and has good generalizability.}

Further, Generative Adversarial Networks (GAN) is an \textbf{adversarial learning} approach that is primarily designed for generative modeling tasks, aiming at learning the underlying distribution of the training data and generating new samples that resemble the training data. It consists of two neural networks: a generator network and a discriminator network, which are trained in an adversarial manner, with the generator trying to produce more realistic samples to deceive the discriminator, and the discriminator trying to become better at distinguishing real from generated samples. \textcolor{black}{For instance, GAN is used by~\citet{DBLP:conf/kbse/BaoLWF19} to automatically generate configuration samples for different software systems like \textsc{Kafka}, \textsc{Spark}, and \textsc{MySQL} to save performance measurement costs.}

\textcolor{black}{\textbf{Deep meta-learning}, where meta-learning paradigms are used jointly with deep learners, has been explored in two studies. Specifically, Model-Agnostic Meta-Learning (\texttt{MAML})~\cite{wang2021morphling} pre-trains a neural network's parameters using the known software environments to facilitate fast adaptation to new environments, allowing for efficient generalization and rapid adaptation. Most recently, a state-of-the-art sequential meta-learning model (\texttt{SeMPL}) is proposed by~\citet{DBLP:conf/fse/GongC24}, which builds a regularized deep neural network to learn a set of meta-environments in a specific sequence that can discriminate the influence of different environments and maximize the usage of meta-data.}

On the other hand, \textbf{deep reinforcement learning} approaches like Q-learning networks are used in two studies to approximate the action-value function (Q-function), which represents the expected cumulative reward for taking a specific action from a given state and following a particular policy. \textcolor{black}{As an example, \citet{DBLP:journals/concurrency/YinH00C23} utilizes a Q-network to deal with the dynamic changes of workloads while predicting the configuration performance for multitier web systems like \textsc{RUBiS}.}

\textcolor{black}{Lastly,~\citet{wyzykowski2024optimizing} employ \textbf{transformer} models, which are based on a self-attention mechanism that allows the model to weigh the importance of different input elements when making predictions, capturing the long-range dependencies in the sequential data of high-performance computing workloads more effectively compared to RNNs or CNNs.}


\begin{figure}[!t]
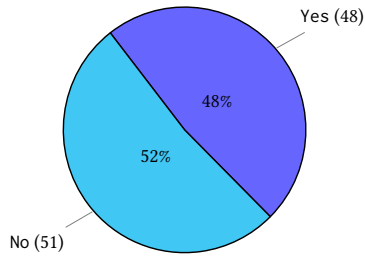

  \centering
\includestandalone[width=0.35\columnwidth]{Figures/reasons}
\caption{Distribution of the reasons for choosing deep learning model.}
 \label{fig:reasons}
\end{figure}

\subsubsection{\textcolor{black}{Rational Behind the Choices}}
In addition to the statistics of deep learning models, our survey also investigates whether the studies have provided an explanation for their choice of specific learning models, which is captured in Figure~\ref{fig:reasons}. Among the 99 studies analyzed, it was found that only 48 of them have provided \textbf{explicit reasons} for the utilization of a specific learning model, accounting for approximately 48\% of the total studies examined. On the other hand, a significant portion of the studies, 51 in total, do \textbf{not offer any justification} or rationale behind their choice of learning model.



\keybox{
\faIcon{search} \textit{\textbf{Finding \thefindingcount:} Various deep learning models have been applied for deep configuration performance learning, within which
the most common one is the multilayer perception (a kind of FNN) from 60 out of 99 studies.
However, it is worth noting that nearly 52\% of the primary studies do not provide
explicit justifications for their choice of deep learning model.}
\addtocounter{findingcount}{1}
}

\subsubsection{Usage Scenario}
\textcolor{black}{Based on the above review results, we summarize the strengths, weakness, and best-suited scenario for each deep learning model in Table~\ref{tb:scenario_learning}, where researchers and practitioners can discover the most suitable deep learning model under different situations. }

\input{Tables/scenario_learning_model}

\subsubsection{Good Practice} Section~\ref{sec:rq2.2} shows that the very basic version of DNN, i.e., MLP is the most commonly used learning model being employed by a total of 60 studies. This is evidence that a simple deep learning model is useful for learning software configurations and performance. This derives the exploitation of more complex DNN, for example, rDNN~\cite{DBLP:conf/icse/HaZ19, DBLP:conf/sigsoft/Gong023} is a DNN armed with regularization to address the problem of feature sparsity, and ResNet~\cite{DBLP:conf/wosp/CengizFAM23}---a type of neural network that incorporates residual connections to overcome the challenges of vanishing gradient for configurations embed into high dimensional representation. This demonstrates the potential of the deep learning models by addressing specific challenges and sheds light on future studies on improving the existing models.

\subsubsection{Bad Smell} Recall from Section~\ref{sec:rq2.2}, there are 52\% studies (51 out of 99) that provide no justification for their choices of deep learning models, which is a concern. It is essential for researchers to clarify their reasons for selecting specific models, as this enhances the understanding of the research motivations and contributions~\cite{DBLP:journals/tosem/LiuGXLGY22}. In the meantime, the lack of explicit explanations of model selection could be harmful to the community as researchers may miss out on important considerations or potential limitations.

\textcolor{black}{The absence of justifications can be attributed to researchers assuming that the chosen models are widely accepted or commonly used, hence not realizing the need to provide explicit justifications. To address this issue, it is crucial for researchers to recognize the importance of providing clear explanations for their model selection, enhancing transparency, and promoting a deeper understanding within the research community.}

\suggestbox{
\faIcon{thumbs-up} \textit{\textbf{Actionable Suggestion \thesuggestioncount:}
Leverage domain knowledge in configuration performance learning to design dedicated DNNs like \texttt{HINNPerf} for handling hierarchical configuration data and \texttt{DaL} for addressing sparsity in configuration options and sample distribution.}
\addtocounter{suggestioncount}{1}

\faIcon{thumbs-up} \textit{\textbf{Actionable Suggestion \thesuggestioncount:} Avoid omitting justifications for deep learning model choices and provide clear explanations to enhance research transparency and understanding.}
\addtocounter{suggestioncount}{1}
}

\subsection{Optimizing Learning Loss}
\label{sec:rq2.4}


The effective training of deep learning models heavily relies on the choice of optimization methods that reduce the loss of training the activation functions, as the loss determines how the model's parameters are updated during training, while activation functions introduce non-linearity and enable complex representations within the neural network. 

\begin{figure}[!t]
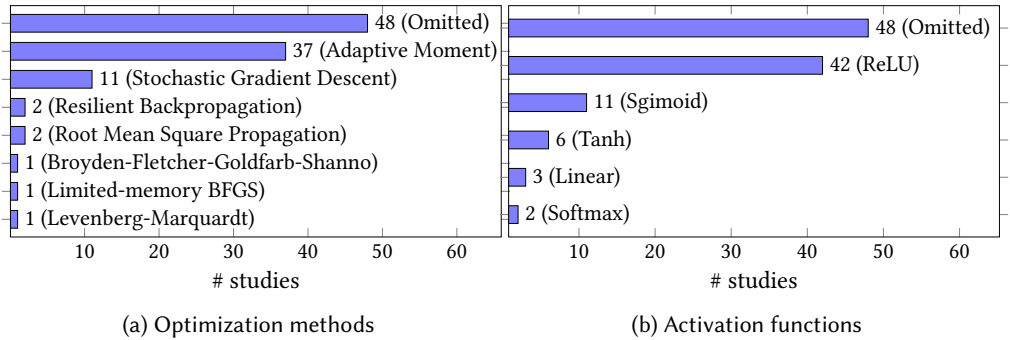

\centering
\begin{subfigure}[t]{0.5\columnwidth}
\includestandalone[width=\columnwidth]{Figures/RQ-optimization}
  \caption{Optimization methods}
 \label{subfig:optimization}
  \end{subfigure}
~\hspace{-0.5cm}
\begin{subfigure}[t]{0.5\columnwidth}
\includestandalone[width=\columnwidth]{Figures/RQ-activation}
  \caption{Activation functions}
 \label{subfig:activation}
  \end{subfigure}
\caption{Distribution of the loss function optimization techniques \textcolor{black}{(one study might use items from multiple categories)}.}
\label{fig:optimization_activation}
 \end{figure}

Figure~\ref{subfig:optimization} provides data on the optimization methods employed in the examined studies. A notable observation is that: a significant number of studies, 48 in total, \textbf{omitted} the specification of the loss optimization method employed~\cite{DBLP:conf/ic2e/RahmanL19, DBLP:journals/tsc/KumaraACMHT23, DBLP:conf/nsdi/LiangFXYLZYZ23, DBLP:conf/middleware/GrohmannNIKL19, DBLP:conf/splc/Acher0LBJKBP22, DBLP:journals/mam/JooyaDB19, DBLP:journals/concurrency/RosarioSZNB23}. 
Despite this omission, \textbf{Adaptive Moment Estimation (Adam)} optimizer emerges as the most prevalent optimization method, being utilized in 37 studies~\cite{zain2022software, DBLP:journals/access/TousiL22, DBLP:conf/icse/HaZ19, DBLP:journals/tosem/ChengGZ23, DBLP:conf/sigsoft/Gong023, DBLP:conf/msr/GongC22}, 
which is an adaptive optimization algorithm that maintains adaptive learning rates based on the first and second moments of the gradients. \textbf{Stochastic gradient descent (SGD)}, which is an iterative optimization algorithm that updates the model parameters by computing gradients on small randomly sampled subsets of the training data, stands as the second most frequently used method, appearing in 11 studies~\cite{DBLP:conf/sbac-pad/NemirovskyAMNUC17, DBLP:journals/comcom/AteeqAAK22, DBLP:conf/IEEEcloud/MarosMSALGHA19, myung2021machine, DBLP:journals/concurrency/JiZL22, said2021accurate}. 
\textbf{Resilient backpropagation (Rprop)}, which individually determines the appropriate step size for each parameter based on the sign of the gradient during training~\cite{DBLP:journals/tse/ChenB17, du2013performance}, and \textbf{root mean square propagation (RMSprop)}, an extension of SGD that adapts the learning rate for each parameter based on the root mean square of recent gradients~\cite{DBLP:conf/icst/PorresARLT20, DBLP:journals/taco/WangLWB19}, are both applied in two studies. Besides, all the rest methods are used by one study, i.e., Broyden-Fletcher-Goldfarb-Shanno (BFGS)~\cite{DBLP:conf/springsim/LuxWCBLXBBCH18}, Limited-memory BFGS (L-BFGS)~\cite{myung2021machine}, and Levenverg-Marquardt~\cite{DBLP:conf/spects/KimK17}.  \textcolor{black}{Among others, when dealing with the high-dimensional, sparse, and non-linear configuration performance data of highly configurable software, Adam is a promising choice among others as it combines adaptive learning rates and momentum, efficiently handling sparse gradients and noisy data~\cite{DBLP:journals/corr/KingmaB14}.}

In Figure~\ref{subfig:activation}, an overview of the activation functions employed in the surveyed studies is provided. Similar to the data of optimization methods, a significant number of studies, 48 out of 99, did not explicitly report the activation function employed. Despite this omission, \textbf{Rectified Linear Unit (ReLU)} emerges as the most prevalent, utilized in 42 studies~\cite{DBLP:conf/springsim/LuxWCBLXBBCH18, DBLP:journals/comcom/AteeqAAK22, DBLP:conf/sbac-pad/NemirovskyAMNUC17, DBLP:journals/jss/ZhuYZZ21, DBLP:journals/concurrency/RosarioSZNB23}, 
which is known for its simplicity and effectiveness in handling non-linearities. \textbf{Sigmoid} activation function follows as the second most commonly used function appearing in 11 studies~\cite{zain2022software, DBLP:journals/concurrency/FalchE17, DBLP:conf/dsrt/DuongZCLZ16, DBLP:journals/tse/ChenB17, du2013performance}, 
and it can map the output of a deep learning model to a probability-like value between 0 and 1. In addition, the \textbf{hyperbolic tangent (tanh)} activation function, which is a smooth, symmetric, and nonlinear function that maps the input to a continuous range between -1 and 1~\cite{DBLP:journals/access/TousiL22, DBLP:conf/nsdi/FuGMR21, DBLP:conf/sc/MalikFP09, DBLP:journals/taco/WangLWB19}, and \textbf{linear} activation functions are employed in six and three studies, respectively~\cite{DBLP:journals/concurrency/JiZL22, DBLP:journals/tsc/KumaraACMHT23, DBLP:conf/middleware/GrohmannNIKL19}, while \textbf{softmax} is used in two studies~\cite{DBLP:conf/splc/GhamiziCPT19, DBLP:conf/middleware/GrohmannNIKL19}, which ensures the output values lie in the range of 0 to 1 and sum up to 1, representing valid probabilities.  \textcolor{black}{Notably, in the context of configuration performance learning, where it is often the case that only a subset of configuration options holds significant influence over performance, ReLU offers a natural solution by effectively zeroing out negative inputs, effectively focusing on the important configuration options~\cite{DBLP:conf/icml/NairH10}.}

\keybox{
\faIcon{search} \textit{\textbf{Finding \thefindingcount:} Despite that 48 of the primary studies omit the justifications of optimization and activation
functions, we observe several different techniques, where the Adam optimizer and
ReLU activation function stands out as the most common one, used
in 37 and 42 studies, respectively.}
\addtocounter{findingcount}{1}
}

\subsubsection{Usage Scenario}
\textcolor{black}{Based on our comprehensive review, we have identified the distinct advantages, limitations, and optimal usage scenarios for each model optimization method and activation function. These findings are concisely summarized in Table~\ref{tb:scenario_optimizer} and~\ref{tb:scenario_activation}, providing researchers and practitioners with a valuable reference to select the most appropriate approach for training the deep learning performance model in their specific scenarios.}

\input{Tables/scenario_optimizer}

\input{Tables/scenario_activation}

\subsubsection{Good Practice} 
In configuration performance learning tasks, the best optimization method for reducing the loss in training differs depending on the systems and performance attribute. For example, tasks with sparse features may benefit from adaptive optimization methods like Adam, while SGD is often used in situations where computational resources are limited. Thus, given that the relationship between configurations and performance of highly configurable software is often sparse, it is encouraging to find that 24 studies address this problem by leveraging Adam optimizer.

Similarly, choosing the appropriate activation function can impact the DL model's performance, e.g., ReLU is effective in capturing complex, non-linear relationships. Hence, it is a promising behavior to see that ReLU is used in 42 primary studies to handle the non-linearity specializing in the problem of configuration performance prediction.

\subsubsection{Bad Smell} Surprisingly, nearly half of the studies (48 out of 99) omitted information on their choices of optimization and activation approaches. This omission poses a significant challenge to understanding and reproducing the deep learning models' results and findings~\cite{DBLP:conf/msr/GongC22, DBLP:conf/icse/HaZ19, DBLP:conf/sigsoft/Gong023}. Therefore, justification for these decisions should be provided in future works, as it allows practitioners to comprehend the rationale behind the decision-making.

\suggestbox{
\faIcon{thumbs-up} \textit{\textbf{Actionable Suggestion \thesuggestioncount:} Do not omit details on the choices of model optimization methods and activation functions in studies, instead, provide justifications to aid understanding and reproducibility.}
\addtocounter{suggestioncount}{1}
}


\subsection{Tuning Hyperparameters}
\label{sec:rq2.3}


Hyperparameter tuning plays a critical role in the accuracy and generalization ability of deep learning models, \textcolor{black}{especially for configuration performance learning, where the interactions between the configurations and performance are complex and can change largely and nonlinearly across software~\cite{DBLP:conf/icse/JamshidiVKSK17, DBLP:conf/kbse/JamshidiSVKPA17}}. However, determining the optimal values for hyperparameters is a challenging task as the hyperparameter space is often huge and the tuning is time-consuming. 

\begin{figure}[!t]
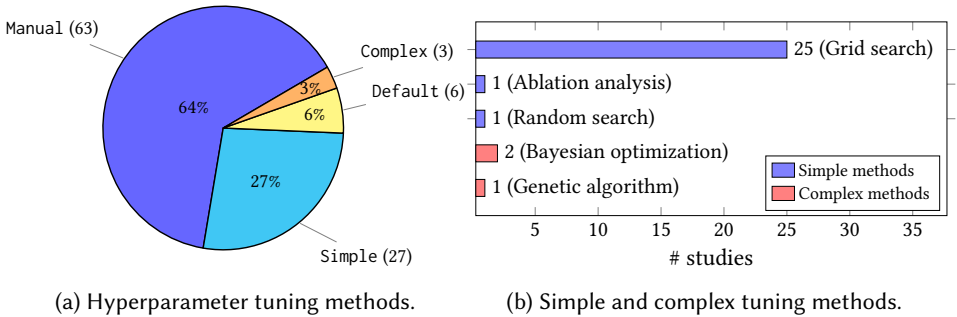

\centering
\begin{subfigure}[t]{0.45\columnwidth}
\includestandalone[width=\columnwidth]{Figures/RQ-hyperparameter}
  \caption{Hyperparameter tuning methods.}
 \label{subfig:tuning}
  \end{subfigure}
~\hspace{-0.4cm}
\begin{subfigure}[t]{0.48\columnwidth}
\includestandalone[width=\columnwidth]{Figures/tuning_simple_complex}
  \caption{Simple and complex tuning methods.}
 \label{subfig:simple_complex}
  \end{subfigure}
\caption{Distribution of the hyperparameter tuning methods.}
\label{fig:hyperparameter}
 \end{figure}

Figure~\ref{fig:hyperparameter} presents an analysis of the hyperparameter tuning methods utilized in the examined studies. It is most worth noting that, a great portion of 63 studies, rely on \textbf{human experts} and domain knowledge to tune the hyperparameters. Among others, six primary studies simply rely on the \textbf{default} hyperparameter settings~\cite{DBLP:conf/icpp/MadireddyBCLLRS19, DBLP:journals/access/LiLSJ20, DBLP:journals/tecs/TrajkovicKHZ22, DBLP:conf/cluster/IsailaBWKLRH15, DBLP:conf/wosp/DidonaQRT15, said2021accurate}, where models are trained using pre-defined configurations without any explicit tuning. 

In addition, three \textbf{simple search methods} have been applied to tune the hyperparameters. For instance, grid search, which exhaustively tests over a predefined grid of values to find the best configuration, is the most widely used technique and is applied in 27 studies~\cite{DBLP:conf/icse/HaZ19, DBLP:journals/corr/abs-2304-13032, DBLP:journals/tosem/ChengGZ23, DBLP:conf/pldi/SinghHLS22, DBLP:conf/mascots/KarniavouraM17, DBLP:conf/sigsoft/Gong023}; 
ablation analysis~\cite{DBLP:journals/jss/ZhuYZZ21}, a way that gradually increases the number of layers and selects the hyperparameter setting with the highest AUC value, and random search~\cite{DBLP:journals/access/ThraneZC20}, where the search path is randomly sampled from the hyperparameter space, are each employed by one study.

Then, more \textbf{complex hyperparameter tuning} methods, which leverage more complicated heuristics such as probabilistic models, evolutionary algorithms, or learning-based approaches to intelligently explore the hyperparameter space, are applied in three studies. For example, Bayesian optimization, a method that uses probability models to efficiently search for optimal hyperparameters by considering the performance of previously evaluated configurations, is employed in two studies~\cite{DBLP:journals/pr/TranSWQ20, zain2022software}. Additionally, genetic algorithm is explored by one study~\cite{DBLP:conf/splc/GhamiziCPT19}, which leverages the principles of natural selection to iteratively search for optimal hyperparameter configurations. 

\keybox{
\faIcon{search} \textit{\textbf{Finding \thefindingcount:} Manual hyperparameter tuning is preferred by the majority of studies (63 out of 99). For the automated tuning method,
grid search is the predominately popular one (25 studies).}
\addtocounter{findingcount}{1}
}

\subsubsection{Usage Scenario}
\textcolor{black}{After conducting a thorough analysis, we have identified the unique benefits, constraints, and optimal application scenarios associated with each hyperparameter tuning technique. The results are presented in Table~\ref{tb:scenario_hyperparameter}, serving as a practical resource for researchers to determine the most suitable approach for optimizing the hyperparameters of deep performance models.}

\input{Tables/scenario_hyperparameter}

\subsubsection{Good Practice} With the complicated architectures of deep learning models, one challenge for hyperparameter tuning is the explosion of search space. As such, we are pleased to reveal that 33 primary studies have addressed this challenge by exploring different heuristic methods to guide the tuning of hyperparameters while training deep learning performance models. For instance,~\citet{DBLP:conf/splc/GhamiziCPT19} employ the idea of natural selection and evolution to find the best population of hyperparameter configurations, which is robust to noise and outliers as the overall fitness is based on the collective behavior of the population rather than individual solutions. This diverse range of heuristic methods reflects the researchers' efforts to explore different approaches for optimizing hyperparameters and enhancing model performance, which should be kept by future researchers.

\subsubsection{Bad Smell} A great portion (63 out of 99) of papers choose to tune the hyperparameters manually, which is time-consuming and not generalizable. Besides, a subset of six studies relies solely on default hyperparameters that may not be optimal for all scenarios. \textcolor{black}{This bad smell can be attributed to several reasons. Firstly, researchers may opt for manual tuning due to familiarity or convenience, as it allows them to have direct control over the hyperparameter settings. Secondly, the lack of awareness or availability of automated tuning methods may contribute to the prevalence of manual tuning. Additionally, some studies may rely on default hyperparameters as a time-saving measure or due to limited resources. } This bad smell may restrict the exploration of the optimal hyperparameter settings and hinder the potential of deep learning models~\cite{DBLP:journals/csur/YangXLG22, DBLP:conf/sc2/FrankHLB17}. As such, it is important for researchers to consider automatic tuning methods in order to maximize the performance and robustness of deep learning models. 

\suggestbox{
\faIcon{thumbs-up} \textit{\textbf{Actionable Suggestion \thesuggestioncount:} Researchers should employ automatic and heuristic approaches for hyperparameter tuning to navigate the vast search space effectively.}
\addtocounter{suggestioncount}{1}
}



%% file: Tables/RQ-overfitting.tex
\begin{table}[t!]
\caption{Distribution of the sparsity handling mechanisms \textcolor{black}{(one study might use items from multiple categories)}.}
\centering
\footnotesize
\begin{adjustbox}{width=\linewidth,center}

\begin{tabular}{p{2.3cm}lp{3.9cm}lp{5cm}}
\toprule

\textbf{Category} & \textbf{Total \#} & \textbf{Example} & \textbf{\# Studies} & \textbf{References} \\ \hline
\multirow{7}{*}{Feature selection} & \multirow{7}{*}{42} & Manual feature selection & 24 & \cite{DBLP:conf/sc/MalakarBVMK18},   \cite{DBLP:journals/access/LiLSJ20},   \cite{DBLP:journals/tompecs/MakraniSNDSMRH21},   \cite{DBLP:journals/jiii/YuGLZIY22}, \cite{blott2018finn},   \cite{DBLP:conf/cf/LiuMCV20}, \cite{DBLP:conf/icpp/DouWZC22}, \cite{DBLP:conf/IEEEcloud/MarosMSALGHA19},   \cite{DBLP:conf/ipps/FalchE15}, \cite{DBLP:conf/nsdi/FuGMR21},   \cite{DBLP:conf/sc/MaratheAJBTKYRG17}, \cite{DBLP:conf/spects/KimK17},   \cite{DBLP:journals/access/WangXTW20},   \cite{DBLP:journals/comcom/AteeqAAK22}, \cite{DBLP:journals/concurrency/FalchE17},   \cite{DBLP:journals/jsa/TangLLLZ22}, \cite{DBLP:journals/pr/TranSWQ20},   \cite{DBLP:journals/tcc/PhamDF20}, \cite{DBLP:journals/tnse/CaoPC22},   \cite{myung2021machine}, \cite{sabbeh2016performance},   \cite{DBLP:journals/fgcs/LiLTWHQD19}, \cite{DBLP:conf/splc/GhamiziCPT19},   \cite{DBLP:journals/tsc/KumaraACMHT23} \\ \cline{3-5} 
 &  & Correlation-based   feature selection & 5 & \cite{DBLP:journals/ese/VituiC21a},   \cite{DBLP:conf/sc/MalikFP09}, \cite{DBLP:journals/tecs/TrajkovicKHZ22},   \cite{DBLP:journals/jss/ZhuYZZ21}, \cite{DBLP:conf/hpdc/YokelsonCL23} \\ \cline{3-5} 
 &  & Tree-based   feature selection & 4 & \cite{DBLP:conf/icpp/MadireddyBCLLRS19},   \cite{DBLP:conf/middleware/GrohmannNIKL19}, \cite{DBLP:conf/im/JohnssonMS19},   \cite{DBLP:conf/splc/Acher0LBJKBP22} \\ \cline{3-5} 
  &  & NN-based   feature selection & 4 & \cite{DBLP:journals/access/ThraneZC20},   \cite{chai2023perfsage}, \cite{ DBLP:conf/icse/GaoGZLY23},   \cite{DBLP:conf/infocom/LiHZLT20} \\ \cline{3-5} 
 &  & Filter-based   feature selection & 2 & \cite{DBLP:journals/jsa/ZhangLWWZH18},   \cite{DBLP:journals/concurrency/OuaredCD22} \\ \cline{3-5} 
 &  & Wrapper-based   feature selection & 2 & \cite{DBLP:journals/access/TousiL22},   \cite{DBLP:journals/tse/ChenB17} \\ \cline{3-5} 
 &  & Metaheuristic   feature selection & 1 & \cite{DBLP:journals/jss/ZhuYZZ21} \\ \hline
Default   dataset & 34 & N/A & 34 & \cite{chokri2022performance},   \cite{DBLP:conf/dsrt/DuongZCLZ16}, \cite{DBLP:conf/noms/SanzEJ22},   \cite{DBLP:conf/springsim/LuxWCBLXBBCH18}, \cite{DBLP:conf/wosp/CengizFAM23},   \cite{DBLP:journals/access/NadeemAMFA19},   \cite{DBLP:journals/cee/KundanSA21}, \cite{DBLP:journals/concurrency/CiciogluC22},   \cite{DBLP:journals/concurrency/JiZL22},   \cite{DBLP:journals/concurrency/RosarioSZNB23},   \cite{DBLP:journals/fgcs/KousiourisMKGV14},   \cite{DBLP:journals/jsa/ChengCWX17}, \cite{DBLP:journals/pvldb/MarcusP19}, \cite{DBLP:journals/taco/LiL22},   \cite{ding2021portable}, \cite{DBLP:conf/icst/PorresARLT20},   \cite{DBLP:conf/kbse/BaoLWF19}, \cite{DBLP:conf/kbse/ChenHL022},   \cite{DBLP:conf/mascots/KarniavouraM17},   \cite{DBLP:conf/nsdi/LiangFXYLZYZ23}, \cite{DBLP:conf/pldi/SinghHLS22},   \cite{DBLP:conf/sbac-pad/NemirovskyAMNUC17},   \cite{DBLP:conf/wosp/DidonaQRT15}, \cite{DBLP:journals/corr/abs-2304-13032},   \cite{wang2021morphling}, \cite{DBLP:journals/taco/MoolchandaniKS22},   \cite{du2013performance}, \cite{DBLP:journals/concurrency/YinH00C23},   \cite{li2024pattern}, \cite{DBLP:conf/cluster/AssogbaLRK23},   \cite{DBLP:journals/ipm/LeeSCP24}, \cite{wyzykowski2024optimizing},   \cite{DBLP:journals/jss/LiBHWL24}, \cite{DBLP:journals/pomacs/LiPFLWH22} \\ \hline
\multirow{6}{*}{Regularization} & \multirow{6}{*}{25} & $L_1$ regularization & 16 & \cite{DBLP:conf/sigsoft/Gong023},   \cite{DBLP:journals/access/ThraneZC20}, \cite{DBLP:journals/taco/WangLWB19},   \cite{DBLP:conf/iccad/KimMMSR17}, \cite{DBLP:conf/esem/ShuS0X20},   \cite{DBLP:conf/icse/HaZ19},  \cite{DBLP:journals/tosem/ChengGZ23},   \cite{DBLP:journals/access/TousiL22}, \cite{myung2021machine},   \cite{DBLP:journals/ese/VituiC21a},    \cite{DBLP:conf/ic2e/RahmanL19}, \cite{DBLP:conf/icccnt/KumarMBCA20},   \cite{DBLP:conf/msr/GongC22}, \cite{DBLP:conf/fse/GongC24},   \cite{DBLP:conf/qrs/ZhangWZ23}, \cite{DBLP:conf/hipc/BettingZS23} \\ \cline{3-5} 
 &  & $L_2$ regularization & 10 & \cite{DBLP:journals/access/ThraneZC20},   \cite{DBLP:journals/jsa/TangLLLZ22}, \cite{DBLP:conf/im/JohnssonMS19},   \cite{DBLP:conf/esem/ShuS0X20}, \cite{DBLP:journals/access/TousiL22},   \cite{myung2021machine}, \cite{DBLP:journals/ese/VituiC21a}, \cite{DBLP:conf/msr/GongC22},   \cite{jiang2024ml}, \cite{DBLP:conf/qrs/ZhangWZ23} \\ \cline{3-5} 
 &  & Early   stopping & 4 & \cite{betting2023oikonomos},   \cite{wang2024slapp}, \cite{DBLP:conf/ics/TrumperBSCH23},   \cite{DBLP:conf/hipc/BettingZS23} \\ \cline{3-5} 
 &  & Noise regularization & 1 & \cite{DBLP:journals/concurrency/OuaredCD22} \\ \cline{3-5} 
 &  & Bayesian   regularization & 1 & \cite{DBLP:conf/middleware/MahgoubWGMGHMGB17} \\ \cline{3-5} 
 &  & Laplacian regularization & 1 & \cite{DBLP:journals/pvldb/ZhouSLF20} \\ \hline
\multirow{2}{*}{Dropout} & \multirow{2}{*}{13} & Dropout & 12 & \cite{DBLP:journals/access/ThraneZC20},   \cite{DBLP:conf/cf/LiuMCV20}, \cite{DBLP:journals/infsof/WartschinskiNVK22},   \cite{said2021accurate}, \cite{zain2022software},   \cite{DBLP:conf/esem/ShuS0X20}, \cite{DBLP:journals/fgcs/LiLTWHQD19},   \cite{DBLP:conf/sigmod/ZhangLZLXCXWCLR19},   \cite{DBLP:conf/hpdc/YokelsonCL23}, \cite{DBLP:conf/qrs/ZhangWZ23},   \cite{wang2024slapp}, \cite{DBLP:conf/hipc/BettingZS23} \\ \cline{3-5} 
 &  & Ensemble   pruning of network & 1 & \cite{DBLP:conf/middleware/MahgoubWGMGHMGB17} \\ \hline
\multirow{3}{*}{Dimension reduction} & \multirow{3}{*}{9} & PCA & 8 & \cite{DBLP:conf/cluster/IsailaBWKLRH15},   \cite{DBLP:conf/middleware/GrohmannNIKL19},   \cite{DBLP:journals/mam/JooyaDB19},   \cite{DBLP:journals/concurrency/OuaredCD22},   \cite{DBLP:conf/icccnt/KumarMBCA20}, \cite{DBLP:journals/jss/ZhuYZZ21},   \cite{DBLP:conf/sc/MalikFP09}, \cite{jiang2024ml} \\ \cline{3-5} 
 &  & CCA & 1 & \cite{DBLP:journals/concurrency/OuaredCD22} \\ \cline{3-5} 
 &  & Knob2vec & 1 & \cite{DBLP:journals/ipm/LeeSCP24} \\ \hline
Divide-and-learn & 1 & N/A & 1 & \cite{DBLP:conf/sigsoft/Gong023} \\ \hline
RSFIN & 1 & N/A & 1 & \cite{DBLP:journals/jss/LiBHWL24}
\\ \bottomrule
\end{tabular}

\end{adjustbox}
\label{tb:overfitting}
\end{table}

%% file: Tables/scenario_sparsity.tex
\begin{table}[t!]
  \caption{Summaries of the strengths, weaknesses, and best-suited usage scenarios for the sparsity handling techniques.}
\centering
\footnotesize
\begin{adjustbox}{width=\linewidth,center}
\begin{tabular}{p{2cm}p{4cm}p{4cm}p{4cm}}
\toprule

\textbf{Category} & \textbf{Strength} & \textbf{Weakness} & \textbf{Best-Suited Scenario} \\ \hline
Feature selection & Removes non-influential options. & Requires domain knowledge to select options. & When dealing with high-dimensional data. \\ \hline
Regularization & Mitigates overfitting by penalizing complex models. & Requires careful tuning of regularization parameters. & When facing overfitting with complex models. \\ \hline
Dropout & Prevents the network from relying too heavily on specific neurons. & Causes a lack of interpretability. & For models with large parameters. \\ \hline
Dimension reduction & Eliminates unnecessary representations and information. & May lose important information if over-reduced. & When handling complex and high-dimensional datasets. \\ \hline
Divide-and-learn & Addresses sample sparsity by dividing the samples into more focused divisions. & Requires tuning the number of divisions to divide. & When the samples are distributed into sparse clusters. \\ \hline
RSFIN & Captures the hidden structures and distributions in the configuration space. & Struggles to achieve ideal accuracy in highly complex and rugged configuration landscapes. & When the configuration space exhibits common patterns or structures.
\\ \bottomrule
\end{tabular}
\end{adjustbox}
\label{tb:scenario_sparsity}
\end{table}

%% file: Tables/RQ-learning_model.tex
\begin{table}[t!]
  \caption{Distribution of the deep learning models \textcolor{black}{(one study might use items from multiple categories)}.}
\centering
\footnotesize
\begin{adjustbox}{width=\linewidth,center}
\begin{tabular}{p{3.4cm}lp{4cm}lp{5cm}}
\toprule

\textbf{Category} & \textbf{Total \#} & \textbf{Example} & \textbf{\# Studies} & \textbf{References} \\ \hline
\multirow{7}{*}{Feedforward neural   network} & \multirow{7}{*}{69} & Multilayer perceptron & 60 & \cite{DBLP:conf/sbac-pad/NemirovskyAMNUC17},   \cite{DBLP:journals/tse/ChenB17}, \cite{du2013performance},   \cite{DBLP:conf/dsrt/DuongZCLZ16},   \cite{DBLP:journals/fgcs/KousiourisMKGV14}, \cite{DBLP:conf/ipps/FalchE15},   \cite{sabbeh2016performance}, \cite{DBLP:journals/concurrency/OuaredCD22},   \cite{DBLP:journals/concurrency/FalchE17},   \cite{DBLP:journals/concurrency/JiZL22},   \cite{DBLP:conf/middleware/MahgoubWGMGHMGB17},   \cite{DBLP:journals/access/ThraneZC20}, \cite{DBLP:conf/ic2e/RahmanL19},   \cite{DBLP:journals/taco/WangLWB19}, \cite{DBLP:journals/comcom/AteeqAAK22},   \cite{chokri2022performance}, \cite{DBLP:conf/wosp/DidonaQRT15},   \cite{DBLP:journals/tecs/TrajkovicKHZ22},   \cite{DBLP:journals/taco/MoolchandaniKS22},   \cite{DBLP:journals/cee/KundanSA21}, \cite{ding2021portable},   \cite{DBLP:journals/tsc/KumaraACMHT23},   \cite{DBLP:conf/springsim/LuxWCBLXBBCH18},   \cite{DBLP:journals/pvldb/ZhouSLF20}, \cite{DBLP:journals/tcc/PhamDF20},   \cite{DBLP:conf/wosp/CengizFAM23}, \cite{DBLP:journals/tnse/CaoPC22}, \cite{DBLP:journals/tompecs/MakraniSNDSMRH21},   \cite{DBLP:conf/sc/MalikFP09}, \cite{myung2021machine},   \cite{DBLP:conf/nsdi/LiangFXYLZYZ23}, \cite{DBLP:journals/ese/VituiC21a},   \cite{DBLP:conf/icst/PorresARLT20}, \cite{DBLP:journals/pr/TranSWQ20},   \cite{DBLP:journals/access/TousiL22}, \cite{DBLP:conf/im/JohnssonMS19},   \cite{DBLP:conf/noms/SanzEJ22}, \cite{DBLP:conf/icpp/DouWZC22},   \cite{DBLP:conf/nsdi/FuGMR21}, \cite{DBLP:conf/sc/MaratheAJBTKYRG17},   \cite{DBLP:conf/spects/KimK17}, \cite{DBLP:conf/cluster/IsailaBWKLRH15},   \cite{DBLP:conf/sc/MalakarBVMK18}, \cite{DBLP:conf/msr/GongC22},   \cite{DBLP:conf/icccnt/KumarMBCA20}, \cite{DBLP:conf/splc/Acher0LBJKBP22},   \cite{DBLP:conf/middleware/GrohmannNIKL19}, \cite{DBLP:conf/iccad/KimMMSR17},   \cite{DBLP:journals/taco/LiL22}, \cite{DBLP:journals/fgcs/LiLTWHQD19},  \cite{DBLP:journals/jsa/ChengCWX17},   \cite{DBLP:journals/pvldb/MarcusP19}, \cite{DBLP:journals/mam/JooyaDB19},   \cite{DBLP:conf/IEEEcloud/MarosMSALGHA19}, \cite{betting2023oikonomos},   \cite{DBLP:conf/hpdc/YokelsonCL23}, \cite{DBLP:conf/qrs/ZhangWZ23},   \cite{DBLP:conf/ics/TrumperBSCH23}, \cite{DBLP:conf/hipc/BettingZS23},   \cite{jiang2024ml} \\ \cline{3-5} 
 &  & Regularized DNN & 3 & \cite{DBLP:conf/icse/HaZ19},   \cite{DBLP:conf/sigsoft/Gong023}, \cite{DBLP:conf/fse/GongC24} \\ \cline{3-5} 
 &  & Kernel extreme learning   machines & 2 & \cite{DBLP:journals/jss/ZhuYZZ21},   \cite{DBLP:journals/access/WangXTW20} \\ \cline{3-5} 
 &  & Fuzzy neural network & 1 & \cite{DBLP:journals/jss/LiBHWL24} \\ \cline{3-5} 
 &  & Hierarchical   interaction neural network & 1 & \cite{DBLP:journals/tosem/ChengGZ23} \\ \cline{3-5} 
 &  & Dynamic neural networks & 1 & \cite{said2021accurate} \\ \cline{3-5} 
 &  & Radial   basis function neural network & 1 & \cite{DBLP:journals/access/NadeemAMFA19} \\ \hline
\multirow{2}{*}{Recurrent neural   network} & \multirow{2}{*}{10} & Long short-term memory & 9 & \cite{DBLP:journals/jsa/TangLLLZ22},   \cite{DBLP:journals/infsof/WartschinskiNVK22},   \cite{DBLP:conf/mascots/KarniavouraM17},   \cite{DBLP:journals/concurrency/CiciogluC22},   \cite{DBLP:conf/icpp/MadireddyBCLLRS19}, \cite{DBLP:conf/infocom/LiHZLT20},   \cite{DBLP:journals/fgcs/LiLTWHQD19}, \cite{DBLP:journals/ese/VituiC21a},   \cite{DBLP:journals/jsa/ZhangLWWZH18} \\ \cline{3-5} 
 &  & Gated   recurrent unit & 1 & \cite{DBLP:journals/ipm/LeeSCP24} \\ \hline
\multirow{2}{*}{Convolutional   neural network} & \multirow{2}{*}{9} & Standard CNN & 8 & \cite{zain2022software},   \cite{DBLP:conf/cf/LiuMCV20}, \cite{blott2018finn},   \cite{DBLP:conf/kbse/ChenHL022}, \cite{DBLP:conf/splc/GhamiziCPT19},   \cite{DBLP:journals/jss/ZhuYZZ21},    \cite{DBLP:journals/ese/VituiC21a},   \cite{DBLP:journals/pomacs/LiPFLWH22} \\ \cline{3-5} 
 &  & Residual neural network & 1 & \cite{DBLP:conf/wosp/CengizFAM23} \\ \hline
\multirow{3}{*}{Graph neural   networks} & \multirow{3}{*}{9} & Standard GNN & 7 & \cite{DBLP:journals/concurrency/RosarioSZNB23},   \cite{DBLP:conf/icse/GaoGZLY23}, \cite{chai2023perfsage},   \cite{DBLP:conf/pldi/SinghHLS22}, \cite{DBLP:journals/corr/abs-2304-13032},   \cite{li2024pattern}, \cite{wang2024slapp} \\ \cline{3-5} 
 &  & Graph hyper network & 1 & \cite{DBLP:conf/cluster/AssogbaLRK23} \\ \cline{3-5} 
 &  & DAG-transformer & 1 & \cite{DBLP:journals/jiii/YuGLZIY22} \\ \hline
Adversarial learning & 3 & Generative adversarial network & 3 & \cite{DBLP:conf/esem/ShuS0X20},   \cite{DBLP:journals/access/LiLSJ20}, \cite{DBLP:conf/kbse/BaoLWF19} \\ \hline
\multirow{2}{*}{Deep meta-learning} & \multirow{2}{*}{2} & Model-agnostic meta-learning & 1 & \cite{wang2021morphling} \\ \cline{3-5} 
 &  & Sequential meta-learning & 1 & \cite{DBLP:conf/fse/GongC24} \\ \hline
Deep reinforcement learning & 2 & Q-learning network & 2 & \cite{DBLP:journals/concurrency/YinH00C23},   \cite{DBLP:conf/sigmod/ZhangLZLXCXWCLR19} \\ \hline
Transformer & 1 & Transformer & 1 & \cite{wyzykowski2024optimizing} 
\\ \bottomrule
\end{tabular}

\end{adjustbox}
\label{tb:learning}
\end{table}

%% file: Figures/reasons.tex
\begin{tikzpicture}[scale=0.8]
\tikzstyle{every node}=[font=\large]
    \pie[
        /tikz/every pin/.style={align=center},
        text=pin,
        rotate=-45,
        explode=0, 
        ]
        {
        48/\texttt{Yes} (48),
        52/\texttt{No} (51)
        }

    \end{tikzpicture}

%% file: Tables/scenario_learning_model.tex
\begin{table}[t!]
  \caption{Summaries of the strengths, weaknesses, and best-suited usage scenarios for the deep learning models.}
\centering
\footnotesize
\begin{adjustbox}{width=\linewidth,center}
\begin{tabular}{p{1.5cm}p{4cm}p{4cm}p{4cm}}
\toprule

\textbf{Category} & \textbf{Strength} & \textbf{Weakness} & \textbf{Best-Suited Scenario} \\ \hline
FNN & (1) Simple architecture. (2) Scalable with more neurons and layers. (3) Flexible to be paired with various techniques~\cite{DBLP:journals/jss/LiBHWL24}. & Poor handling of sequential or temporal data~\cite{DBLP:journals/fgcs/LiLTWHQD19}. & Performance prediction with non-sequential configuration data. \\ \hline
RNN & (1) Effective with sequential data. (2) Captures temporal dependencies~\cite{DBLP:conf/mascots/KarniavouraM17}. & Computationally intensive due to continuous learning. & Continuous performance prediction with a time series of dynamic operations/workloads. \\ \hline
CNN & (1) Effective in capturing structured and hierarchical configuration data. (2) Robust to translations and distortions in input data~\cite{DBLP:conf/cf/LiuMCV20}. & Limited ability to model long-range dependencies. & Performance prediction for configuration data along with other data modalities, such as sequence data of unit codes and defect matrices of the configuration. \\ \hline
GNN & (1) Capable of handling graph-structured data. (2) Ability to capture relational dependencies between nodes~\cite{wang2024slapp}. & High computational cost and complex implementation. & Software dependency analysis of configuration and performance; network topology-based predictions; relationship-driven performance factors. \\ \hline
Adversarial learning & (1) Ability to learn the underlying distribution of the training data. (2) Capable of generating new samples that resemble the training data~\cite{DBLP:conf/kbse/BaoLWF19}. & Computationally expensive due to the need for adversarial training. & Configuration data synthesis when available data is limited. \\ \hline
Deep reinforcement learning & (1) Ability to learn from interactions with an environment. (2) Suitable for sequential decision-making problems~\cite{DBLP:conf/sigmod/ZhangLZLXCXWCLR19}. & Requires a large number of interactions with the environment to learn optimal policies. & Making sequential decisions for tasks like resource scheduling to maximize rewards in an environment. \\ \hline
Deep meta-learning & (1) Ability to learn from multiple related tasks. (2) Fast adaptation to new tasks with limited data~\cite{DBLP:conf/fse/GongC24}. & Requires meta-data from a diverse set of tasks/environments. & Leveraging data on known environments for rapid modeling for unseen environments. \\ \hline
Transformer & (1) Effective in capturing contextual information. (2) Provides insights into the model's decision-making process~\cite{wyzykowski2024optimizing}. & High memory consumption due to the self-attention mechanism's quadratic complexity with respect to input length. & Performance modeling for context-rich configuration data. 
\\ \bottomrule
\end{tabular}
\end{adjustbox}
\label{tb:scenario_learning}
\end{table}

%% file: Figures/RQ-optimization.tex
\begin{tikzpicture}
\begin{axis}[
    xbar=0.05cm,
    width=10cm,
    height=5.5cm,
    xmax=65,
    bar width=0.3cm,
 enlarge y limits=0.09,
 enlarge x limits=0.015,
    legend style={at={(1.05,0.7)},anchor=north,legend columns=1,font=\tiny},
    legend cell align={left},
   xlabel={\# studies},
     x label style={at={(0.5,0.02)},font=\Large},
     nodes near coords,
          point meta=explicit symbolic,
          every node near coord/.append style={anchor=west,font=\large},
     symbolic y coords={ LM, LBFGS, BFGS, RMSprop, RPROP, SGD, adam, not},
         y tick label style={font=\large},
          x tick label style={font=\large},
          yticklabels={,,},
         ytick=data
    ]
    

\addplot[fill=blue!50, draw=black] coordinates { (1,LM)[1 (Levenberg-Marquardt)] (1,LBFGS)[1 (Limited-memory BFGS)] (1,BFGS)[1 (Broyden-Fletcher-Goldfarb-Shanno)] (2,RMSprop)[2 (Root Mean Square Propagation)]  (2,RPROP)[2 (Resilient Backpropagation)] (11,SGD)[11 (Stochastic Gradient Descent)] (37,adam)[37 (Adaptive Moment)] (48,not)[48 (Omitted)]
};



\end{axis}
\end{tikzpicture}

%
%
%

%% file: Figures/RQ-activation.tex
\begin{tikzpicture}
\begin{axis}[
    xbar=0.05cm,
    width=10cm,
    height=5.5cm,
    xmax=64,
    bar width=0.3cm,
 enlarge y limits=0.12,
 enlarge x limits=0.02,
    legend style={at={(1.05,0.7)},anchor=north,legend columns=1,font=\tiny},
    legend cell align={left},
   xlabel={\# studies},
     x label style={at={(0.5,0.02)},font=\Large},
     nodes near coords,
          point meta=explicit symbolic,
          every node near coord/.append style={anchor=west,font=\large},
     symbolic y coords={
     F,
     E,
     D,
     C,
     B,
     A},
         y tick label style={font=\large},
          x tick label style={font=\large},
          yticklabels={,,},
         ytick=data
    ]
    

\addplot[fill=blue!50, draw=black] coordinates {(2,F)[2 (Softmax)] (3,E)[3 (Linear)]
(6,D)[6 (Tanh)] (11,C)[11 (Sgimoid)] (42,B)[42 (ReLU)]  (48,A)[48 (Omitted)] 
};



\end{axis}
\end{tikzpicture}

%
%
%

%% file: Tables/scenario_optimizer.tex
\begin{table}[t!]
  \caption{Summaries of the strengths, weaknesses, and best-suited usage scenarios for the deep learning model optimizer.}
\centering
\footnotesize
\begin{adjustbox}{width=\linewidth,center}
\begin{tabular}{p{1.5cm}p{4cm}p{4cm}p{4cm}}
\toprule

\textbf{Category} & \textbf{Strength} & \textbf{Weakness} & \textbf{Best-Suited Scenario} \\ \hline
Adam & Automatically adjusts learning rates. & Requires more memory due to storing past gradients. & In settings where quick convergence is desired. \\ \hline
SGD & Simple to implement and requires less memory. & Requires careful tuning of the learning rate. & In situations where simplicity and computational efficiency are key. \\ \hline
Rprop & Fast convergence by adapting step sizes. & Requires more memory for storing step sizes. & For batch training scenarios. \\ \hline
RMSprop & Adjusts learning rate based on recent gradient magnitudes. & Requires tuning of the decay hyperparameter. & In scenarios with noisy gradients. \\ \hline
Lavenberg-Marquardt & Combines gradient descent and Gauss-Newton methods. & Memory-intensive and computationally expensive. & Effective for small and medium-sized datasets. \\ \hline
BFGS & Uses second-order information for faster convergence. & Memory-intensive due to second-order derivatives. & For learning tasks with smooth gradients. \\ \hline
L-BFGS & Reduces memory usage by storing limited history. & Requires careful tuning of hyperparameters. & In scenarios where memory efficiency is important.
\\ \bottomrule
\end{tabular}
\end{adjustbox}
\label{tb:scenario_optimizer}
\end{table}

%% file: Tables/scenario_activation.tex
\begin{table}[t!]
  \caption{Summaries of the strengths, weaknesses, and best-suited usage scenarios for the activation functions.}
\centering
\footnotesize
\begin{adjustbox}{width=\linewidth,center}
\begin{tabular}{p{1.5cm}p{4cm}p{4cm}p{4cm}}
\toprule

\textbf{Category} & \textbf{Strength} & \textbf{Weakness} & \textbf{Best-Suited Scenario} \\ \hline
ReLU & (1) Simple and computationally efficient. (2) Zeroing out negative inputs, effectively focusing on the important configuration options. & Can suffer from the ``dying ReLU" problem where neurons become inactive if they fall into the negative region and stop updating. & When some options are redundant and non-influential. \\ \hline
Sigmoid & Produces a smooth and continuous output, which can be interpreted as a probability. & Prone to the vanishing gradient problem. & When interpretability as probability is essential, such as interpreting the predicted reach rates as probabilities of events related to CDN performance. \\ \hline
Tanh & Stronger gradients than sigmoid, reducing vanishing gradient issue. & Computationally more expensive than ReLU. & For hidden layers in recurrent neural networks. \\ \hline
Linear & Does not suffer from vanishing gradient problem. & Limited representation power; can't capture non-linear relationships. & For simple networks where non-linearity is unnecessary. \\ \hline
Softmax & Ensures the output values lie in the range of 0 to 1 and sum up to 1, representing valid probabilities. & Sensitive to outliers and large input values. & Useful to generate probabilities for different performance states.
\\ \bottomrule
\end{tabular}
\end{adjustbox}
\label{tb:scenario_activation}
\end{table}

%% file: Figures/RQ-hyperparameter.tex
\begin{tikzpicture}[scale=0.8]
\tikzstyle{every node}=[font=\large]
    \pie[
        /tikz/every pin/.style={align=center},
        text=pin,
        rotate=30,
        explode=0, 
        ]
        {
        64/\texttt{Manual} (63),
        27/\texttt{Simple} (27),
        6/\texttt{Default} (6),
        3/\texttt{Complex} (3)
        }

    \end{tikzpicture}

%% file: Figures/tuning_simple_complex.tex
\begin{tikzpicture}
\begin{axis}[
    xbar=0.05cm,
    width=10cm,
    height=5cm,
    xmax=37,
    bar width=0.3cm,
    bar shift=0cm, 
 enlarge y limits=0.2,
 enlarge x limits=0.02,
    legend style={at={(0.8,0.31)},anchor=north,legend columns=1,font=\small},
    legend cell align={left},
    legend image code/.code={
        \draw [#1] (0cm,-0.1cm) rectangle (0.4cm,0.15cm); },
   xlabel={\# studies},
     x label style={at={(0.5,0.02)},font=\Large},
     nodes near coords,
          point meta=explicit symbolic,
          every node near coord/.append style={anchor=west,font=\large},
        symbolic y coords={F,G,H,I,J,K},
         y tick label style={font=\large},
          x tick label style={font=\large},
          yticklabels={,,},
         ytick=data
    ]

\addplot[fill=blue!50, draw=black] coordinates {(25,J)[25 (Grid search)] (1,I)[1 (Ablation analysis)] (1,H)[1 (Random search)]};

\addplot[fill=red!50, draw=black] coordinates {(2,G)[2 (Bayesian optimization)] (1,F)[1 (Genetic algorithm)]};

\addlegendentry{Simple methods}
\addlegendentry{Complex methods}




\end{axis}
\end{tikzpicture}

%% file: Tables/scenario_hyperparameter.tex
\begin{table}[t!]
  \caption{Summaries of the strengths, weaknesses, and best-suited usage scenarios for the hyperparameter tuning methods.}
\centering
\footnotesize
\begin{adjustbox}{width=\linewidth,center}
\begin{tabular}{p{2cm}p{4cm}p{4cm}p{4cm}}
\toprule

\textbf{Category} & \textbf{Strength} & \textbf{Weakness} & \textbf{Best-Suited Scenario} \\ \hline
Grid search & Exhaustive search covers all combinations. & Computationally expensive and time-consuming. & Suitable for limited hyperparameter sets. \\ \hline
Ablation analysis & Identifies the impact of each hyperparameter. & Not exhaustive; may miss optimal combinations. & For diagnosing model performance issues. \\ \hline
Random search & Reduces computational cost by sampling randomly. & May miss optimal settings due to randomness. & When computational efficiency is required. \\ \hline
Bayesian optimization & Utilizes past evaluations to guide search. & Sensitive to the choice of acquisition function. & In optimization tasks requiring a balance between exploration and exploitation. \\ \hline
Genetic algorithm & Can escape local optima through mutation and crossover. & Requires tuning of genetic algorithm parameters. & For complex hyperparameter spaces. 
\\ \bottomrule
\end{tabular}
\end{adjustbox}
\label{tb:scenario_hyperparameter}
\end{table}

%% file: RQ3.tex
\section{RQ3: How are the Deep Configuration Performance Models Evaluated? }
\label{sec:rq3}

To understand how to evaluate the deep learning models for configuration performance learning, in this section, we summarize the procedure, metrics, statistical validation, and subject software systems that are commonly used in comparing the models.


\subsection{Following Evaluation Procedures and Metrics}
\label{sec:rq3.1}


 \begin{figure}[!t]
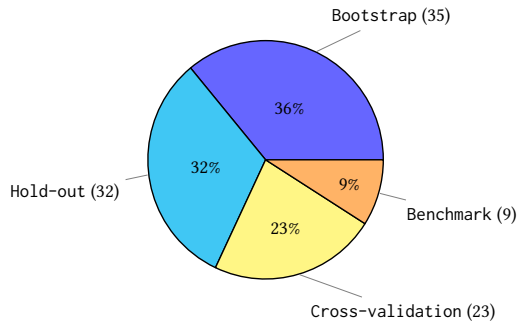

\centering
\includestandalone[width=0.5\columnwidth]{Figures/RQ-evaluation}
\caption{Distribution of the model evaluation procedures \textcolor{black}{(one study might use items from multiple categories)}.}
\label{fig:evaluation}
 \end{figure}

\input{Tables/metric}


In Figure~\ref{fig:evaluation}, we provide a summary of the evaluation procedure and metrics used. Among others, the most prevalent evaluation procedures is \textbf{bootstrap}, utilized by 35 out of the 99 studies~\cite{DBLP:conf/nsdi/LiangFXYLZYZ23, DBLP:journals/jsa/ZhangLWWZH18, DBLP:conf/esem/ShuS0X20, DBLP:conf/spects/KimK17, said2021accurate, DBLP:conf/infocom/LiHZLT20}, 
which is a resampling technique that involves creating multiple datasets by randomly sampling from the original data, allowing for robust estimation of model performance. \textbf{Hold-out} evaluation, another widely used method, is employed in 32 studies~\cite{DBLP:journals/jiii/YuGLZIY22, DBLP:conf/wosp/DidonaQRT15, DBLP:journals/ese/VituiC21a, DBLP:journals/tsc/KumaraACMHT23, DBLP:journals/pvldb/MarcusP19, chai2023perfsage}. 
This approach involves splitting the dataset into training and testing sets based on a specific percentage. \textbf{Cross-validation}, which involves iteratively partitioning the dataset into multiple subsets for training and testing, is used by 23 studies~\cite{DBLP:journals/tompecs/MakraniSNDSMRH21, DBLP:conf/ic2e/RahmanL19, DBLP:journals/tecs/TrajkovicKHZ22, DBLP:journals/cee/KundanSA21, DBLP:journals/concurrency/OuaredCD22, DBLP:journals/concurrency/JiZL22}.
Additionally, we find that nine studies opt to use the default evaluation pipeline provided by \textbf{benchmark} datasets, simplifying the evaluation process~\cite{DBLP:conf/middleware/MahgoubWGMGHMGB17, DBLP:journals/concurrency/YinH00C23, ding2021portable, DBLP:conf/im/JohnssonMS19, DBLP:conf/iccad/KimMMSR17}.
\textcolor{black}{For instance,~\citet{DBLP:conf/iccad/KimMMSR17} run a subset of \textsc{Linpack} benchmarks on different hardware platforms to collect the power consumption data, representing a variety of industry-standard workloads, to build the models in the offline learning stage, while using the rest benchmarks for evaluating the online identification stage.} 

\textcolor{black}{Table~\ref{tb:metric} presents the statistics and taxonomy of the accuracy metrics to evaluate the performance models. Among the identified metrics, \textbf{relative error metrics}, emerges as the most commonly used in 44 studies~\cite{DBLP:journals/concurrency/FalchE17, DBLP:journals/concurrency/OuaredCD22, DBLP:journals/taco/LiL22, DBLP:journals/tosem/ChengGZ23, DBLP:conf/icse/HaZ19, DBLP:conf/icse/GaoGZLY23}. Therein, {Mean Absolute Percentage Error (MAPE)} or Mean Relative Error (MRE), which measures the percentage difference between the predicted and actual values, is seen in 34 primary studies, being the most widely used metric. \textbf{Squared error metrics} follow as the second most frequently employed metric, appearing in 27 studies, e.g., {Root Mean Square Error (RMSE)}~\cite{ DBLP:conf/wosp/DidonaQRT15, DBLP:conf/cluster/IsailaBWKLRH15, DBLP:journals/concurrency/CiciogluC22} 
and {Mean Squared Error (MSE)}~\cite{DBLP:journals/jsa/TangLLLZ22, sabbeh2016performance, DBLP:journals/access/WangXTW20} are utilized in 16 and 14 studies, respectively, providing insights into the model's ability to capture both small and large errors. Besides, \textbf{absolute error metrics} like Mean Absolute Error (MAE)~\cite{DBLP:journals/fgcs/LiLTWHQD19, said2021accurate, DBLP:journals/ese/VituiC21a, DBLP:conf/wosp/CengizFAM23, DBLP:journals/concurrency/CiciogluC22} calculate the average absolute difference between the predicted and actual values, offering a straightforward measure of model accuracy, utilized in 25 reviewed studies. 
\textbf{Correlation metrics} like {coefficient of determination (R$^{2}$)}---a metric indicating the proportion of variance in the dependent variable explained by the model---is used in 11 studies~\cite{DBLP:conf/wosp/CengizFAM23, DBLP:conf/icccnt/KumarMBCA20, DBLP:journals/tecs/TrajkovicKHZ22}. 
Additionally, \textbf{performance metrics} like performance improvement~\cite{DBLP:journals/concurrency/YinH00C23, DBLP:conf/kbse/ChenHL022, DBLP:conf/infocom/LiHZLT20, DBLP:conf/kbse/BaoLWF19, DBLP:conf/icpp/DouWZC22} and optimal performance~\cite{wang2021morphling, ding2021portable, DBLP:journals/jiii/YuGLZIY22, DBLP:conf/sigmod/ZhangLZLXCXWCLR19}, which are computed using the performance resulting from a selected configuration, are used by six and five works, respectively. {Notably, while these two metrics do not quantify the accuracy of deep configuration models directly, they are practical to evaluate how the deep configuration performance models can impact performance optimization, performance testing, and configuration tuning tasks. For instance,~\citet{DBLP:conf/kbse/ChenHL022} measure the increment in the latency and energy consumption to examine the effectiveness of their model.} Lastly, \textbf{classification metrics} such as F-measure are employed in eight studies, which is commonly used to examine the accuracy in classification tasks like predicting the software defect classes~\cite{DBLP:journals/infsof/WartschinskiNVK22, zain2022software, DBLP:journals/jss/ZhuYZZ21} or performance levels~\cite{DBLP:conf/middleware/GrohmannNIKL19, DBLP:journals/pr/TranSWQ20}. }

\keybox{
\faIcon{search} \textit{\textbf{Finding \thefindingcount:} Bootstrap is the most prevalent evaluation procedure, which is utilized in 35 out of 99
studies. On the other hand, the most popular accuracy metric is MAPE/MRE in 33 studies.}
\addtocounter{findingcount}{1}
}

\subsubsection{Usage Scenario}
\textcolor{black}{Based on our thorough review, we have identified the unique advantages, limitations, and optimal usage scenarios for each model evaluation method, as summarized in Table~\ref{tb:scenario_evaluation}. By considering the strengths and weaknesses of each method, researchers and practitioners can make informed decisions to ensure accurate and reliable evaluations of their models, leading to improved understanding and advancement in the field of deep configuration performance learning.}

\input{Tables/scenario_evaluation}

\subsubsection{Good Practice} A variety of metrics for accuracy has been applied is a rather positive sign for the community, as~\citet{mathews1994towards} point out, no single measure gives an unambiguous indication of the modeling performance. More importantly, we have discovered that all the primary studies have leveraged some form of repeated evaluation procedure. This is crucial to ensure that the results were not produced due to chance or biased samples.

\subsubsection{Bad Smell} We found that MAPE is the most widely used metric, due primarily to the fact that it is unit-free and easy to calculate. However, it is known that MAPE suffers the problem of being \emph{asymmetric} on the error above and below the actual value~\cite{goodwin1999asymmetry}, where the former receives a much greater penalty. As a result, such a strong bias tends to be problematic. \textcolor{black}{To address this issue, researchers should be encouraged to explore alternative metrics that provide a more balanced and comprehensive evaluation of error, promoting a more accurate assessment of model performance.}

\suggestbox{
\faIcon{thumbs-up} \textit{\textbf{Actionable Suggestion \thesuggestioncount:} Implement repeated evaluation procedures in the experiments to ensure results and findings are reliable and not due to biased evaluation setup.}
\addtocounter{suggestioncount}{1}

\faIcon{thumbs-up} \textit{\textbf{Actionable Suggestion \thesuggestioncount:} Researchers should consider using multiple metrics that can offer a more balanced evaluation of errors.}
\addtocounter{suggestioncount}{1}
}

\subsection{Leveraging Statistical Validation}
\label{sec:rq3.2}

\begin{figure}[!t]
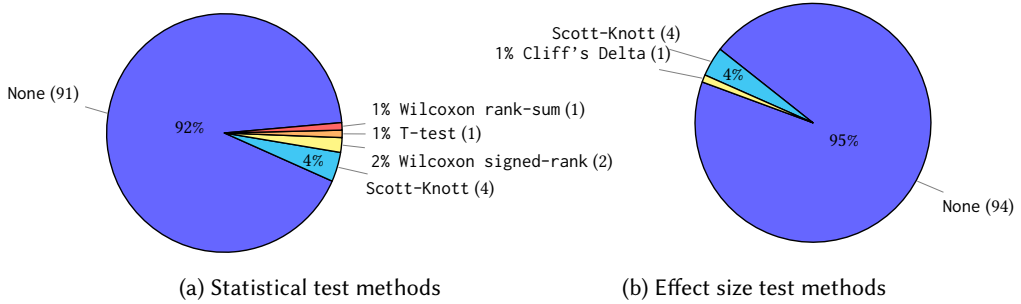

\centering
\begin{subfigure}[t]{0.59\columnwidth}
\includestandalone[width=\columnwidth]{Figures/RQ-statistical}
  \caption{Statistical test methods}
 \label{subfig:statistical}
  \end{subfigure}
~\hspace{-1.9cm}
\begin{subfigure}[t]{0.508\columnwidth}
\includestandalone[width=\columnwidth]{Figures/effect_size}
  \caption{Effect size test methods}
 \label{subfig:effect_size}
  \end{subfigure}
\caption{Distribution of the statistical validation methods.}
\label{fig:statistical_effect_size}
 \end{figure}


Since the deep learning models are stochastic in nature, evaluation in deep configuration performance learning is often conducted with repeated runs. Therefore, understanding their statistical significance is important. Indeed, when evaluating a deep configuration performance model, the results of two runs with exactly the same model settings and training/testing sample sizes could still be different, due to the stochastic loss optimization and the randomness in selecting the training and testing configurations, e.g., when evaluating with the compression tool \textsc{Lrzip}, two runs under the same conditions might resulted a MRE of 17.49\% or 70.44\% --- a nearly $4\times$ difference~\cite{DBLP:conf/sigsoft/Gong023}.

In Figure~\ref{subfig:statistical}, we survey the methods used for statistical validation. Surprisingly, the majority of the studies, 91 out of 99, do not utilize any specific statistical test methods to measure the significance of the results compared. Among the studies that do apply statistical tests, the \textbf{Scott-Knott Effect Size Difference (ESD)} test emerged as the most commonly used, appearing in three studies~\cite{DBLP:conf/sigsoft/Gong023, DBLP:conf/msr/GongC22, DBLP:journals/jss/ZhuYZZ21}, which is a non-parametric test that groups data into distinct subset with significant difference. Additionally, the \textbf{Wilcoxon signed-rank} test was employed in two studies~\cite{DBLP:journals/jss/ZhuYZZ21, DBLP:journals/tse/ChenB17}, providing a non-parametric test for comparing two paired sets of data. Furthermore, the \textbf{t-test}~\cite{DBLP:conf/icse/HaZ19} and the \textbf{Wilcoxon rank-sum}~\cite{DBLP:journals/tosem/ChengGZ23} test were each used in one study, offering comparisons for two independent data groups.

Similarly, Figure~\ref{subfig:effect_size} reveals a phenomenon in the utilization of effect size tests for evaluating deep configuration performance learning, where the majority of the studies, 94 out of 99, do not employ any specific effect size tests. Among the rest studies, the \textbf{Scott-Knott ESD} test is used in three studies~\cite{DBLP:conf/sigsoft/Gong023, DBLP:conf/msr/GongC22, DBLP:journals/jss/ZhuYZZ21}, which can identify significant differences on the effect size between groups in addition to its ability of assessing the statistical significance. Further,~\cite{DBLP:journals/jss/ZhuYZZ21} utilized \textbf{Cliff's Delta}, a non-parametric effect size measure that quantifies the ordinal association between two variables.

 \keybox{
\faIcon{search} \textit{\textbf{Finding \thefindingcount:} A significant portion of 92\% and 95\% studies omit the usage of statistical and effect
size tests, respectively. Scott-Knott ESD test is used in 4 studies, standing out as the most common method for both cases.}
\addtocounter{findingcount}{1}
}

\subsubsection{Good Practice} Almost all studies have conducted comparisons between their models with existing works, and a few of them have utilized statistical tests and effect size tests to ensure the statistical significance and effect size difference of the comparison. This practice is vital in deep learning research as it enhances the reliability and robustness of the findings.

\subsubsection{Bad Smell} A notable concern arises from the fact that a significant majority of studies, around 92\% (91 out of 99), did not conduct any statistical tests, and 95\% of studies did not measure the effect size. This represents a significant problem as it introduces a potential source of randomness and uncertainty in the comparison results, limiting the ability to draw reliable conclusions and increasing the risk of causing severe external threats to validity. This highlights the need for greater awareness and adoption of statistical analysis in deep learning research. 

\suggestbox{
\faIcon{thumbs-up} \textit{\textbf{Actionable Suggestion \thesuggestioncount:} Researchers should pay more attention to the importance of statistical and effect size tests to avoid external validity threats and to strengthen the reliability of their experiments.}
\addtocounter{suggestioncount}{1}
}

\subsection{Evaluating Configurable Software Systems}
\label{sec:rq3.3}

\begin{figure}[!t]
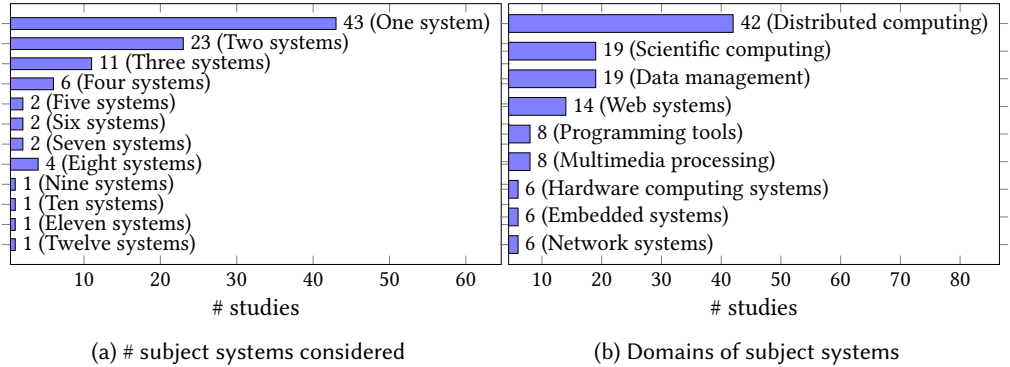

\centering
\begin{subfigure}[t]{0.5\columnwidth}
\includestandalone[width=\columnwidth]{Figures/RQ-subject_system}
  \caption{\# subject systems considered}
 \label{subfig:num_subject}
  \end{subfigure}
~\hspace{-0.5cm}
\begin{subfigure}[t]{0.5\columnwidth}
\includestandalone[width=\columnwidth]{Figures/domain_subject_system}
  \caption{Domains of subject systems}
 \label{subfig:domain_subject}
  \end{subfigure}
\caption{Distribution of the subject systems \textcolor{black}{(one study might use items from multiple categories)}.}
\label{fig:subject_system}
 \end{figure}

A crucial factor in evaluating deep configuration performance models is to ensure the evaluation covers a good range of subject systems. Figure~\ref{subfig:num_subject} provides the distribution of the number of systems considered in the primary studies. Among others, 43 studies evaluate a \textbf{single software}, 23 studies evaluate \textbf{two software}, 11 studies assess \textbf{three systems}, six studies evaluate \textbf{four subject systems}, and the remaining 16 studies evaluate five or more systems.


The taxonomy of the domains of the subject software systems in the primary studies is shown in Figure~\ref{subfig:domain_subject}. Among these domains, 42 prominent studies build deep configuration performance models for \textbf{distributed computing} systems, which typically involve distributing computational tasks across multiple interconnected devices or nodes, enabling collaboration and parallel processing for improved computing performance, including domains like cloud computing~\cite{DBLP:journals/tsc/KumaraACMHT23, myung2021machine, DBLP:conf/im/JohnssonMS19, DBLP:conf/IEEEcloud/MarosMSALGHA19, DBLP:journals/taco/MoolchandaniKS22, DBLP:conf/dsrt/DuongZCLZ16}, big data processing~\cite{DBLP:conf/sigmod/ZhangLZLXCXWCLR19, DBLP:conf/mascots/KarniavouraM17, DBLP:journals/pvldb/MarcusP19, DBLP:journals/pvldb/ZhouSLF20, DBLP:journals/concurrency/OuaredCD22, DBLP:conf/sigsoft/Gong023}, and high-performance computing~\cite{DBLP:journals/taco/LiL22, DBLP:journals/corr/abs-2304-13032, DBLP:journals/concurrency/JiZL22, DBLP:conf/iccad/KimMMSR17, DBLP:journals/access/NadeemAMFA19}. 
\textcolor{black}{In learning configuration performance of distributed computing systems, the key challenges include capturing the dependencies and interactions between multiple interconnected components, the dynamic environments with nodes joining or leaving the system, network conditions changing, and workloads varying over time. To address these, researchers leverage domain knowledge in extracting diverse environmental features, constructing domain-specific data structures to capture the interconnection between the distributed components, and designing specialized deep learning model architectures. For example,~\citet{DBLP:conf/hipc/BettingZS23} propose a contextual multi-armed bandit approach to improve resource recommendations by adapting with application requirements, hardware capabilities, and cost considerations, and~\citet{li2024pattern} utilize Graph Neural Network to identify and learn patterns of the interconnected nodes, enabling efficient performance modeling based on recognized patterns. }

\textbf{Scientific computing} systems, used in 19 primary studies, play a crucial role in solving complex scientific and research problems by utilizing advanced algorithms and numerical methods, including domains like matrix multiplication~\cite{DBLP:conf/ipps/FalchE15}, multi-grid solvers~\cite{DBLP:journals/tosem/ChengGZ23, DBLP:conf/esem/ShuS0X20}, stencil-grid solvers~\cite{DBLP:conf/esem/ShuS0X20}, and the training of deep learning models~\cite{DBLP:conf/icse/GaoGZLY23, DBLP:conf/infocom/LiHZLT20, DBLP:conf/pldi/SinghHLS22, chai2023perfsage, DBLP:journals/access/NadeemAMFA19, DBLP:conf/nsdi/FuGMR21}, enabling researchers and scientists to analyze and model intricate data sets efficiently. \textcolor{black}{Modeling scientific computing software systems faces several key challenges: (1) scientific computing applications frequently leverage parallelism and hardware accelerators to accelerate computations, which requires considering related factors such as load balancing, communication speed, and hardware environments, and (2) scientific computing systems are often optimized for specialized mathematical operations or problem domains, putting requirements on domain-specific knowledge and tailored modeling strategies. The primary studies address these challenges by building tailored configuration options that feature the unique demands of the scientific applications, e.g.,~\citet{DBLP:conf/ics/TrumperBSCH23} encodes both static and dynamic features of parallel loop nests, and~\citet{wang2024slapp} design a subgraph-level performance prediction method that combines operator-level and graph-level features to model the application structure, enabling more effective configuration performance predictions. }

Another 19 studies focus on the modeling of \textbf{data management} systems that facilitate the storage, organization, retrieval, and manipulation of data, including software systems that provide structured data storage and query capabilities~\cite{DBLP:conf/sigmod/ZhangLZLXCXWCLR19, DBLP:conf/mascots/KarniavouraM17, DBLP:journals/pvldb/MarcusP19, DBLP:journals/pvldb/ZhouSLF20, DBLP:journals/concurrency/OuaredCD22}, I/O libraries that facilitate efficient input/output operations with data~\cite{DBLP:conf/cluster/IsailaBWKLRH15}, file systems that manage file storage and access~\cite{DBLP:journals/jsa/TangLLLZ22}, and compression tools that reduce data size for efficient storage and transmission~\cite{DBLP:conf/sigsoft/Gong023}. \textcolor{black}{The major challenges in modeling data management systems with deep learning models involve: (1) they often involve concurrent access to data and parallel processing, thereby, predicting the performance of these systems requires considering factors such as locking mechanisms, transaction management, and parallel query execution to accurately model system behavior under varying workloads, and (2) these systems often experience workload variability due to changing data volumes, query patterns, and access frequencies which requires adaptive modeling techniques that can adjust to varying workloads and resource demands. For instance,~\citet{DBLP:conf/sigmod/ZhangLZLXCXWCLR19} employ a reinforcement learning model (Q-learning network) to learn the interactions between the knobs and performance and predict the optimal knob settings through a try-and-error method.}

Furthermore, \textbf{Web service} software like web server systems, involving the utilization of internet-based technologies to provide a platform for hosting websites, is explored in 14 studies~\cite{DBLP:conf/sigsoft/Gong023, DBLP:conf/icse/HaZ19, DBLP:journals/access/NadeemAMFA19, DBLP:journals/concurrency/YinH00C23, DBLP:conf/kbse/BaoLWF19, DBLP:conf/icst/PorresARLT20}. \textcolor{black}{Notably, deep configuration performance prediction for Web service applications encounters challenges with fluctuating workloads based on user traffic, and uncertainty in web service environments, such as co-hosted services on the same server. To address these issues, multi-environment learning techniques like meta-learning have been utilized by~\citet{DBLP:conf/fse/GongC24}, which can discriminate the importance of each meta-environment and learn the environments in a tailored sequence, efficiently reusing the knowledge from the meta-environments.}

Additionally, \textbf{Programming tools} and \textbf{multimedia processing} are each used in eight studies. Programming tools include Python programs~\cite{DBLP:journals/infsof/WartschinskiNVK22} and compilers, which translate high-level programming languages into machine-readable instructions to allow the code to be executed by the computer~\cite{DBLP:journals/concurrency/RosarioSZNB23, DBLP:conf/sigsoft/Gong023, DBLP:conf/icse/HaZ19, DBLP:journals/access/NadeemAMFA19, DBLP:journals/tosem/ChengGZ23, DBLP:conf/esem/ShuS0X20}. {Multimedia processing} systems include image processing~\cite{DBLP:conf/kbse/ChenHL022, DBLP:journals/tosem/ChengGZ23, DBLP:conf/esem/ShuS0X20}, which focuses on techniques for enhancing, modifying, and analyzing images, as well as video encoding\cite{DBLP:conf/sigsoft/Gong023, DBLP:conf/icse/HaZ19, DBLP:journals/access/NadeemAMFA19, DBLP:journals/tosem/ChengGZ23}, which involves the compression and encoding of video data to optimize storage and transmission while maintaining high-quality playback. \textcolor{black}{Particularly, configuration performance learning of programming tools suffers from challenges such as domain-specific and language-specific features which require tailored feature construction for accurate predictions, and dependencies on external toolchains and libraries which can impact the performance of programming tools. A common practice among the primary studies is to leverage domain knowledge of the specific programming tool in feature selection, thereby modeling the interactions between the configuration options and software performance more efficiently ~\cite{DBLP:journals/concurrency/RosarioSZNB23}. Moreover, multimedia processing software often needs to deal with diverse workloads and types of input files, such as images, audio, or videos, which could significantly affect the performance behavior of the software. Thus, domain knowledge is required for designing workload-aware models, e.g.,~\citet{DBLP:conf/kbse/ChenHL022} propose \texttt{DeepPerform} to efficiently generate test samples to model the configuration performance and detect Input-Dependent Performance Bottlenecks for image processing applications.}

\textbf{Hardware computing systems}, explored in six studies, comprise diverse domains such as CPU computing~\cite{DBLP:journals/taco/WangLWB19, DBLP:journals/access/TousiL22}, GPU applications that leverage graphics processing units for general-purpose computing~\cite{DBLP:journals/mam/JooyaDB19}, chip multi-processors~\cite{DBLP:journals/cee/KundanSA21}, and multicore systems~\cite{DBLP:journals/jsa/ChengCWX17}, all of which contribute to enhancing computational capabilities, efficiency, and parallel processing within computer systems~\cite{DBLP:conf/wosp/CengizFAM23}. \textcolor{black}{Hardware computing systems face challenges in the complex architecture of GPUs, varying levels of parallelism in GPGPU applications, and the difficulty in tuning kernel and architectural parameters without a thorough examination of the design space. To mitigate these and enable more accurate performance predictions, one of the critical ways is to reduce the size of the configuration space by limiting parameter ranges, as performed by~\citet{DBLP:journals/mam/JooyaDB19}.}

Finally, six research works examine \textbf{embedded systems} and \textbf{network systems}, respectively. Especially, embedded computing includes mobile systems~\cite{DBLP:journals/access/ThraneZC20, DBLP:journals/access/WangXTW20}, which are designed for smartphones and tablets, Internet of Things (IoT) communication systems that facilitate connectivity and data exchange among IoT devices~\cite{said2021accurate}, large-scale commercial systems used in areas such as e-commerce and finance~\cite{DBLP:journals/ese/VituiC21a}, and operating systems~\cite{DBLP:conf/splc/Acher0LBJKBP22}; and network systems refer to the diverse domains associated with the establishment, operation, and administration of computer networks, enabling efficient communication and data exchange between devices. This includes traditional network infrastructure~\cite{DBLP:journals/tnse/CaoPC22, DBLP:journals/concurrency/CiciogluC22, DBLP:journals/tecs/TrajkovicKHZ22}, as well as emerging architectures like software-defined networking~\cite{sabbeh2016performance} and wireless sensor networks~\cite{DBLP:journals/comcom/AteeqAAK22}. \textcolor{black}{Embedded systems often have constraints such as a large number of connected devices, massive data exchange, high-speed network topology changes, and heterogeneous devices. Meanwhile, network systems like SDNs and WSNs usually operate in dynamic environments where network conditions can change rapidly, requiring adaptive performance models to account for fluctuations in traffic, topology, and interference. To bridge the gaps for both domains, several primary studies opt to leverage domain knowledge in incorporating environmental information and domain-specific features to increase the effectiveness in configuration performance modeling~\cite{jiang2024ml, said2021accurate, DBLP:journals/comcom/AteeqAAK22}.}

 \keybox{
\faIcon{search} \textit{\textbf{Finding \thefindingcount:} Most commonly, the primary studies on deep configuration performance learning evaluate one configurable software system only (43
out of 99 studies). Furthermore, among the various domains explored, distributed computing
emerges as the most popular domain for subject systems, utilized in 42 studies.}
\addtocounter{findingcount}{1}
}

\subsubsection{Good Practice} Another positive observation is that 56 studies have considered multiple subject systems in their experiments. This is significant because evaluating the proposed model on software from different domains, scales, and performance indicators demonstrates its generalizability and robustness~\cite{ DBLP:conf/icccnt/KumarMBCA20, DBLP:conf/noms/SanzEJ22, DBLP:conf/bigdataconf/GhoshalWPS19, DBLP:conf/icse/LiX0WT20, DBLP:conf/icse/HaZ19}. 

\textcolor{black}{Moreover, it is encouraging to see that the primary studies have examined a variety of domains with various challenges, and have leveraged domain-specific knowledge to address the challenges.} This diversity in domain coverage enhances the credibility and applicability of the findings across various contexts, providing broader insights for developers and practitioners.

\subsubsection{Bad Smell} It is worth noting that a significant proportion of studies (43\% of 99) evaluate their models on only one software, which can be considered a potential threat to the validity of the results and findings of the study. 
Indeed, software systems of different characteristics can exhibit unique performance behavior due to factors such as architecture, complexity, and usage patterns. Therefore, it is important for future studies to evaluate their models on multiple software systems, spanning various domains, scales, and performance indicators, to ensure the robustness and learning ability of the proposed performance models. 

\suggestbox{
\faIcon{thumbs-up} \textit{\textbf{Actionable Suggestion \thesuggestioncount:} It is crucial for future research to include multiple software systems in their evaluations, covering a diverse range of domains, scales, and performance indicators.}
\addtocounter{suggestioncount}{1}

\faIcon{thumbs-up} \textit{\textbf{Actionable Suggestion \thesuggestioncount:} Incorporate domain-specific knowledge into the model development process to tackle unique challenges of each domain, thereby enhancing the model’s adaptability and performance.}
\addtocounter{suggestioncount}{1}
}



%% file: Figures/RQ-evaluation.tex
\begin{tikzpicture}[scale=0.8]
\tikzstyle{every node}=[font=\large]
    \pie[
        /tikz/every pin/.style={align=center},
        text=pin,
        rotate=0,
        explode=0, 
        ]
        {
        36/\texttt{Bootstrap} (35),
        32/\texttt{Hold-out} (32),
        23/\texttt{Cross-validation} (23),
        9/\texttt{Benchmark} (9)
        }

    \end{tikzpicture}

%% file: Tables/metric.tex
\begin{table}[t!]
  \caption{Distribution of the accuracy metrics \textcolor{black}{(one study might use items from multiple categories)}.}
\centering
\footnotesize
\begin{adjustbox}{width=\linewidth,center}
\begin{tabular}{llp{4.5cm}lp{5.5cm}}
\toprule

\textbf{Category} & \textbf{Total \#} & \textbf{Example} & \textbf{\# Studies} & \textbf{Reference} \\ \hline
\multirow{8}{*}{Relative error metrics} & \multirow{8}{*}{44} & Mean absolute percentage error/Mean relative error & 34 & \cite{DBLP:journals/tnse/CaoPC22}, \cite{DBLP:journals/fgcs/LiLTWHQD19}, \cite{DBLP:conf/iccad/KimMMSR17}, \cite{DBLP:conf/sc/MaratheAJBTKYRG17}, \cite{DBLP:conf/sigsoft/Gong023}, \cite{DBLP:journals/taco/MoolchandaniKS22}, \cite{DBLP:conf/mascots/KarniavouraM17}, \cite{DBLP:conf/pldi/SinghHLS22}, \cite{DBLP:conf/splc/Acher0LBJKBP22}, \cite{DBLP:conf/sc/MalikFP09}, \cite{DBLP:journals/tompecs/MakraniSNDSMRH21}, \cite{DBLP:conf/ic2e/RahmanL19}, \cite{DBLP:conf/IEEEcloud/MarosMSALGHA19}, \cite{DBLP:conf/sc/MalakarBVMK18}, \cite{DBLP:conf/cf/LiuMCV20}, \cite{chai2023perfsage}, \cite{DBLP:journals/ese/VituiC21a}, \cite{DBLP:conf/msr/GongC22}, \cite{DBLP:journals/mam/JooyaDB19}, \cite{DBLP:conf/esem/ShuS0X20}, \cite{myung2021machine}, \cite{DBLP:journals/concurrency/FalchE17}, \cite{DBLP:journals/concurrency/OuaredCD22}, \cite{DBLP:journals/taco/LiL22}, \cite{DBLP:journals/tosem/ChengGZ23}, \cite{DBLP:conf/icse/HaZ19}, \cite{DBLP:conf/icse/GaoGZLY23}, \cite{betting2023oikonomos}, \cite{DBLP:conf/fse/GongC24}, \cite{jiang2024ml}, \cite{DBLP:conf/qrs/ZhangWZ23}, \cite{DBLP:journals/jss/LiBHWL24}, \cite{wang2024slapp}, \cite{DBLP:conf/sbac-pad/NemirovskyAMNUC17} \\ \cline{3-5} 
 &  & Relative error & 5 & \cite{du2013performance}, \cite{DBLP:journals/concurrency/JiZL22}, \cite{DBLP:conf/springsim/LuxWCBLXBBCH18}, \cite{DBLP:conf/cluster/AssogbaLRK23}, \cite{DBLP:journals/access/LiLSJ20} \\ \cline{3-5} 
 &  & Mean percentage error & 2 & \cite{DBLP:conf/dsrt/DuongZCLZ16}, \cite{betting2023oikonomos} \\ \cline{3-5} 
 &  & Symmetric MAPE & 1 & \cite{DBLP:journals/tse/ChenB17} \\ \cline{3-5} 
 &  & Relative absolute error & 1 & \cite{DBLP:journals/tcc/PhamDF20} \\ \cline{3-5} 
 &  & Root mean square relative error & 1 & \cite{DBLP:conf/nsdi/FuGMR21} \\ \cline{3-5} 
 &  & Relative percentage deviation & 1 & \cite{li2024pattern} \\ \cline{3-5} 
 &  & Mean relative error percentage & 1 & \cite{DBLP:journals/tnse/CaoPC22} \\ \hline
\multirow{2}{*}{Squared error metrics} & \multirow{2}{*}{27} & Root mean squared error & 16 & \cite{DBLP:journals/fgcs/LiLTWHQD19}, \cite{said2021accurate}, \cite{DBLP:conf/icpp/MadireddyBCLLRS19}, \cite{DBLP:journals/ese/VituiC21a}, \cite{DBLP:conf/icse/GaoGZLY23}, \cite{DBLP:journals/tecs/TrajkovicKHZ22}, \cite{DBLP:journals/tsc/KumaraACMHT23}, \cite{DBLP:conf/middleware/MahgoubWGMGHMGB17}, \cite{DBLP:journals/jsa/ZhangLWWZH18}, \cite{DBLP:journals/cee/KundanSA21}, \cite{DBLP:journals/access/ThraneZC20}, \cite{DBLP:conf/wosp/DidonaQRT15}, \cite{DBLP:conf/cluster/IsailaBWKLRH15}, \cite{DBLP:journals/concurrency/CiciogluC22}, \cite{jiang2024ml}, \cite{wyzykowski2024optimizing} \\ \cline{3-5} 
 &  & Mean squared error & 14 & \cite{DBLP:journals/tnse/CaoPC22}, \cite{DBLP:journals/ese/VituiC21a}, \cite{DBLP:conf/spects/KimK17}, \cite{DBLP:journals/jsa/TangLLLZ22}, \cite{sabbeh2016performance}, \cite{DBLP:journals/access/WangXTW20}, \cite{DBLP:conf/wosp/CengizFAM23}, \cite{chokri2022performance}, \cite{DBLP:journals/tecs/TrajkovicKHZ22}, \cite{DBLP:conf/icccnt/KumarMBCA20}, \cite{DBLP:journals/concurrency/CiciogluC22}, \cite{DBLP:journals/ipm/LeeSCP24}, \cite{DBLP:journals/pomacs/LiPFLWH22}, \cite{DBLP:journals/pvldb/ZhouSLF20} \\ \hline
\multirow{5}{*}{Absolute error metrics} & \multirow{5}{*}{25} & Mean absolute error & 20 & \cite{DBLP:journals/taco/WangLWB19}, \cite{DBLP:journals/access/NadeemAMFA19}, \cite{DBLP:conf/nsdi/LiangFXYLZYZ23}, \cite{DBLP:journals/corr/abs-2304-13032}, \cite{DBLP:journals/jsa/ChengCWX17}, \cite{DBLP:conf/ipps/FalchE15}, \cite{DBLP:journals/comcom/AteeqAAK22}, \cite{DBLP:journals/fgcs/KousiourisMKGV14}, \cite{DBLP:journals/pvldb/MarcusP19}, \cite{DBLP:journals/tnse/CaoPC22}, \cite{DBLP:journals/access/TousiL22}, \cite{DBLP:journals/fgcs/LiLTWHQD19}, \cite{said2021accurate}, \cite{DBLP:journals/ese/VituiC21a}, \cite{DBLP:conf/wosp/CengizFAM23}, \cite{DBLP:journals/tsc/KumaraACMHT23}, \cite{DBLP:journals/concurrency/CiciogluC22}, \cite{jiang2024ml}, \cite{wyzykowski2024optimizing}, \cite{DBLP:conf/ics/TrumperBSCH23} \\ \cline{3-5} 
 &  & Normalized MAE & 2 & \cite{DBLP:conf/im/JohnssonMS19}, \cite{DBLP:conf/noms/SanzEJ22} \\ \cline{3-5} 
 &  & Absolute error & 1 & \cite{DBLP:journals/access/WangXTW20} \\ \cline{3-5} 
 &  & Median absolute percentage error & 1 & \cite{DBLP:conf/icccnt/KumarMBCA20} \\ \cline{3-5} 
 &  & Median absolute error & 1 & \cite{DBLP:conf/icpp/MadireddyBCLLRS19} \\ \hline
\multirow{2}{*}{Correlation metrics} & \multirow{2}{*}{11} & Coefficient of determination (R$^2$) & 10 & \cite{DBLP:journals/access/TousiL22}, \cite{said2021accurate}, \cite{DBLP:journals/ese/VituiC21a}, \cite{DBLP:conf/wosp/CengizFAM23}, \cite{DBLP:conf/icccnt/KumarMBCA20}, \cite{DBLP:journals/tecs/TrajkovicKHZ22}, \cite{DBLP:journals/tsc/KumaraACMHT23}, \cite{DBLP:conf/cluster/IsailaBWKLRH15}, \cite{DBLP:journals/concurrency/CiciogluC22}, \cite{DBLP:conf/hpdc/YokelsonCL23} \\ \cline{3-5} 
 &  & Pearson correlation coefficient & 1 & \cite{DBLP:journals/ipm/LeeSCP24} \\ \hline
\multirow{3}{*}{Performance metrics} & \multirow{2}{*}{11} & Performance improvemet & 6 & \cite{DBLP:journals/concurrency/YinH00C23}, \cite{DBLP:conf/kbse/ChenHL022}, \cite{DBLP:conf/infocom/LiHZLT20}, \cite{DBLP:conf/kbse/BaoLWF19}, \cite{DBLP:conf/icpp/DouWZC22}, \cite{DBLP:conf/splc/GhamiziCPT19} \\ \cline{3-5} 
 &  & Optimal performance & 5 & \cite{wang2021morphling}, \cite{ding2021portable}, \cite{DBLP:journals/jiii/YuGLZIY22}, \cite{DBLP:conf/sigmod/ZhangLZLXCXWCLR19}, \cite{DBLP:conf/hipc/BettingZS23} \\ \hline
\multirow{4}{*}{Classification metrics} & \multirow{4}{*}{8} & F-measure & 5 & \cite{DBLP:journals/infsof/WartschinskiNVK22}, \cite{zain2022software}, \cite{DBLP:conf/middleware/GrohmannNIKL19}, \cite{DBLP:journals/pr/TranSWQ20}, \cite{DBLP:journals/jss/ZhuYZZ21} \\ \cline{3-5} 
 &  & Mean average precision & 1 & \cite{blott2018finn} \\ \cline{3-5} 
 &  & Positive predictive value & 1 & \cite{DBLP:conf/icst/PorresARLT20} \\ \cline{3-5} 
 &  & Percentage of right predictions & 1 & \cite{DBLP:journals/concurrency/RosarioSZNB23}
\\ \bottomrule
\end{tabular}
\end{adjustbox}
\label{tb:metric}
\end{table}

%% file: Tables/scenario_evaluation.tex
\begin{table}[t!]
  \caption{Summaries of the strengths, weaknesses, and best-suited usage scenarios for the model evaluation methods.}
\centering
\footnotesize
\begin{adjustbox}{width=\linewidth,center}
\begin{tabular}{p{1.5cm}p{4cm}p{4cm}p{4cm}}
\toprule

\textbf{Category} & \textbf{Strength} & \textbf{Weakness} & \textbf{Best-Suited Scenario} \\ \hline
Bootstrap & Accounts for sampling variability and provides robust performance estimates. & Prone to overfitting if the number of bootstrap samples is too high. & When the number of available samples is limited. \\ \hline
Hold-out & Simple and computationally efficient. & May not capture the full variability of model accuracy. & Quick evaluation of model performance. \\ \hline
Cross-validation & Provides a more stable estimate of model performance by averaging results. & Computationally expensive, especially with large datasets. & Evaluating the performance of models with high variance. \\ \hline
Benchmark & Allows for standardized evaluation. & Relies on the availability of well-defined benchmark datasets and metrics. & Assessing the performance of a new method against established baselines.
\\ \bottomrule
\end{tabular}
\end{adjustbox}
\label{tb:scenario_evaluation}
\end{table}

%% file: Figures/RQ-statistical.tex
\begin{tikzpicture}[scale=0.8]
\tikzstyle{every node}=[font=\large]
  
  
    \pie[
        /tikz/every pin/.style={align=center},
        text=pin,
        rotate=5,
        explode=0, 
        ]
        {
        92/\texttt{None} (91),
        4/\texttt{Scott-Knott} (4)
        }

    \pie[
        /tikz/every pin/.style={align=center},
        text=pin,
        before number=\phantom,
        after number=,
        rotate=350.6, 
        explode=0, 
        color={yellow!60, orange!60, red!60},
        ]
        {
        2/\texttt{2\% Wilcoxon signed-rank} (2),
        1/\texttt{1\% T-test} (1),
        1/\texttt{1\% Wilcoxon rank-sum} (1)
        }

    \end{tikzpicture}

%% file: Figures/effect_size.tex
\begin{tikzpicture}[scale=0.8]
\tikzstyle{every node}=[font=\large]
    \pie[
        /tikz/every pin/.style={align=center},
        text=pin,
        rotate=-200,
        explode=0, 
        ]
        {
        95/\texttt{None} (94),
        4/\texttt{Scott-Knott} (4)
        }
        
    \pie[
        /tikz/every pin/.style={align=center},
        text=pin,
        before number=\phantom,
        after number=,
        rotate=156.4, 
        explode=0, 
        color={yellow!60}, 
        ]
        {
        1/\texttt{1\% Cliff’s Delta} (1)
        }

    \end{tikzpicture}

%% file: Figures/RQ-subject_system.tex
\begin{tikzpicture}
\begin{axis}[
    xbar=0.05cm,
    width=10cm,
    height=6cm,
    xmax=64,
    bar width=0.2cm,
 enlarge y limits=0.09,
 enlarge x limits=0.01,
    legend style={at={(1.05,0.7)},anchor=north,legend columns=1,font=\tiny},
    legend cell align={left},
   xlabel={\# studies},
     x label style={at={(0.5,0.02)},font=\Large},
     nodes near coords,
          point meta=explicit symbolic,
          every node near coord/.append style={anchor=west,font=\large},
     symbolic y coords={
     L,
     K,
     J,
     I,
     H,
     G,
     F,
     E,
     D,
     C,
     B,
     A},
         y tick label style={font=\large},
          x tick label style={font=\large},
          yticklabels={,,},
         ytick=data
    ]


\addplot[fill=blue!50, draw=black] coordinates {(1,L)[1 (Twelve systems)] (1,K)[1 (Eleven systems)] (1,J)[1 (Ten systems)] (1,I)[1 (Nine systems)] (4,H)[4 (Eight systems)] (2,G)[2 (Seven systems)] (2,F)[2 (Six systems)] (2,E)[2 (Five systems)]
(6,D)[6 (Four systems)] (11,C)[11 (Three systems)] (23,B)[23 (Two systems)]  (43,A)[43 (One system)] 
};

\end{axis}
\end{tikzpicture}

%% file: Figures/domain_subject_system.tex
\begin{tikzpicture}
\begin{axis}[
    xbar=0.05cm,
    width=10cm,
    height=6cm,
    xmax=85,
    bar width=0.3cm,
 enlarge y limits=0.09,
 enlarge x limits=0.02,
    legend style={at={(1.05,0.7)},anchor=north,legend columns=1,font=\tiny},
    legend cell align={left},
   xlabel={\# studies},
     x label style={at={(0.5,0.02)},font=\Large},
     nodes near coords,
          point meta=explicit symbolic,
          every node near coord/.append style={anchor=west,font=\large},
     symbolic y coords={
     I,
     H,
     G,
     F,
     E,
     D,
     C,
     B,
     A},
         y tick label style={font=\large},
          x tick label style={font=\large},
          yticklabels={,,},
         ytick=data
    ]

\addplot[fill=blue!50, draw=black] coordinates {(6,I)[6 (Network systems)] (6,H)[6 (Embedded systems)] (6,G)[6 (Hardware computing systems)] (8,F)[8 (Multimedia processing)] (8,E)[8 (Programming tools)] (14,D)[14 (Web systems)] (19,C)[19 (Data management)] (19,B)[19 (Scientific computing)]  (42,A)[42 (Distributed computing)] 
};

\end{axis}
\end{tikzpicture}

%% file: RQ4.tex
\section{RQ4: How to Exploit the Deep Configuration Performance Model?}
\label{sec:rq4}

The successful application of deep learning models trained on configuration data relies on considering various factors, including application purposes, adaptability to dynamic software running environments, and the availability of materials for reproducibility. In this section, we study those factors involved in the exploitation and dissemination of the deep configuration performance models.

\subsection{Using the Model}
\label{sec:rq4.1}

\input{Tables/RQ-application}


A deep configuration performance model, once built, can be used in different stages of the software engineering practices. Table~\ref{tb:application} shows that out of the 99 papers surveyed, 65 studies emphasize the domain of \textbf{general prediction} where the key has been ensuring the accuracy and reliability of the deep configuration performance model, without focusing on any particular domain in which the model can be used. 


22 studies apply the deep learning models for \textbf{performance tuning}, or performance optimization, which aims at optimizing the performance by finding the optimal configuration. It involves tasks such as configuration auto-tuning~\cite{DBLP:journals/concurrency/FalchE17, DBLP:journals/concurrency/FalchE17, DBLP:journals/concurrency/RosarioSZNB23, DBLP:conf/icpp/DouWZC22, DBLP:conf/splc/GhamiziCPT19}, where software configurations are automatically adjusted to fit performance requirements, and self-adaptive systems which dynamically adapt to varying conditions to achieve optimal performance~\cite{DBLP:journals/tse/ChenB17}. In such a context, the deep configuration performance model directly serves as the surrogate for the tuning algorithm.

\textbf{Performance testing} tasks are considered by six studies. They involve assessing the system's behavior for specific configurations or under different environments to identify potential bottlenecks or performance issues. For example, configuration bug prediction seeks to identify potential bugs or issues that are caused by incorrect configuration, leading to performance degradation~\cite{DBLP:journals/jss/ZhuYZZ21, zain2022software}. Vulnerabilities detection focuses on identifying security vulnerabilities that could affect system performance or compromise its functionality~\cite{DBLP:journals/infsof/WartschinskiNVK22}. 

Additionally, tasks on \textbf{scheduling} have been addressed by five primary studies, which seek to allocate the available resources, such as CPU, memory, or network bandwidth, in an optimal manner to maximize system performance~\cite{DBLP:journals/concurrency/YinH00C23, DBLP:conf/sbac-pad/NemirovskyAMNUC17}. The unique property is that there are often low-level metrics, such as the CPU cycle or cache rate, that are considered together with the performance attributes. \textcolor{black}{Notably, while both resource scheduling and performance tuning involve optimization processes, they mainly differ in the following points: }

\begin{itemize}
    \item Performance tuning aims to directly optimize the software configuration for enhancing the performance, while resource scheduling focuses on selecting the most appropriate resources for a given sequence of workload/task~\cite{DBLP:conf/hipc/BettingZS23}. As a result, the latter involves an ordinal constraint while the former typically does not have such.

    \item Performance optimization can refer to the process of improving system performance at runtime (although offline scenarios are also possible)~\cite{DBLP:conf/saner/Chen22}, while resource scheduling mainly occurs before the execution of the software.
\end{itemize}

Meanwhile, one study applies performance models for \textbf{development}~\cite{DBLP:journals/pr/TranSWQ20}, involving designing and implementing the best set of configurations for developing the architecture of deep learning models or software.

 \keybox{
\faIcon{search} \textit{\textbf{Finding \thefindingcount:} The applications of deep configuration performance models are mainly for fulfilling five purposes, in which the most common one is the general prediction for performance analysis, as evidenced by 65 studies.}
\addtocounter{findingcount}{1}
}

\subsubsection{Good Practice} The best practice in the model applications is their widespread adoption across various domains and for different tasks. This diversity is critical as it showcases the potential of performance models in addressing diverse challenges, and provides practical insights for researchers in different fields~\cite{DBLP:journals/jss/PereiraAMJBV21}.

\subsubsection{Bad Smell} A potential concern is that 65 studies only consider pure prediction tasks, while these studies contribute valuable insights into performance prediction, it raises the question of whether these models can effectively be applied to other tasks beyond prediction, in other words, the applicability and generalizability of performance models across different domains and tasks remain unclear, especially when the usefulness of the model is merely measured by accuracy metrics.

\suggestbox{
\faIcon{thumbs-up} \textit{\textbf{Actionable Suggestion \thesuggestioncount:} Future studies should explore the applicability of performance models beyond prediction to ensure their relevance across a broader range of tasks.}
\addtocounter{suggestioncount}{1}
}





\subsection{Considering Multiple Runtime Environments}
\label{sec:rq4.2}



Configurable software systems would inevitably run under multiple, and potentially changing environments. For example, a database system may experience both read-heavy and write-heavy workload~\cite{DBLP:conf/kbse/JamshidiSVKPA17}. Similarly, the hardware between the testing and production infrastructure might be drastically different~\cite{DBLP:journals/toit/LeitnerC16,DBLP:journals/corr/BrunnertHWDHHHJ15}, especially during the modern DevOps era. It is therefore natural to understand how such a fact has been taken into account for deep configuration performance learning.

As in Figure~\ref{fig:environments}, our survey results indicate that, out of the 99 papers surveyed, 64 studies do not explicitly consider the challenges posed by multiple and dynamic software running environments in their application and design of performance models~\cite{DBLP:conf/icse/HaZ19, DBLP:conf/splc/GhamiziCPT19, DBLP:conf/sc/MaratheAJBTKYRG17, DBLP:conf/im/JohnssonMS19, DBLP:conf/sigsoft/Gong023, sabbeh2016performance}. In other words, they rely on using the data from a \textbf{single} environment to build the configuration performance models and predict therein.
This suggests that a considerable portion of the research in deep learning-based configuration performance learning has focused primarily on static or controlled settings, without accounting for the real-world complexities of dynamic, multiple environments. Nevertheless, among the 35 studies that acknowledge and consider dynamic environments in their application of performance models, 23 studies exclusively focus on exploring the impact of different \textbf{workloads} on configuration performance~\cite{DBLP:conf/mascots/KarniavouraM17, DBLP:journals/jsa/TangLLLZ22, DBLP:journals/concurrency/YinH00C23, chokri2022performance, DBLP:journals/pvldb/MarcusP19, DBLP:journals/jsa/ChengCWX17}. Additionally, eight studies concentrate solely on investigating the influence of different \textbf{hardware} settings~\cite{DBLP:conf/middleware/GrohmannNIKL19, DBLP:journals/concurrency/FalchE17, chai2023perfsage}, and one study specifically examines the impact of different \textbf{software versions}~\cite{DBLP:conf/noms/SanzEJ22}. Lastly, three studies research the implications of \textbf{mixed environment} types, i.e., one considers three workloads combined with two hardware settings~\cite{DBLP:conf/IEEEcloud/MarosMSALGHA19}, another considers 12 workloads combined with two hardware settings~\cite{DBLP:conf/icpp/DouWZC22}, and a recent work comprehensively addresses a total of 16 hardware, 26 workloads and 13 versions for nine software systems~\cite{DBLP:conf/fse/GongC24}.

 \keybox{
\faIcon{search} \textit{\textbf{Finding \thefindingcount:} More than half of the 99 primary studies (65\%) do not consider multiple and dynamic environments. Among others, workload stands out as the most commonly considered factor (23\%).}
\addtocounter{findingcount}{1}
}

\subsubsection{Good Practice} The most encouraging trend is that 35 out of 99 studies have actively considered multiple and dynamic environments, demonstrating the robustness of models. Acknowledging that software systems run under multiple and potentially varying environments is an important property as the behavior of software could be completely different across the environments~\cite{muhlbaueranalyzing,DBLP:conf/wcre/Chen22,DBLP:conf/wosp/PereiraA0J20}. By accounting for dynamic workloads, versions, and hardware in real-world scenarios, these studies contribute to a more comprehensive understanding of how performance models can be applied in practical settings.

\subsubsection{Bad Smell} Another bad smell is that 65\% of 99 primary studies have solely focused on datasets from a single environment without considering dynamic factors that can impact model performance, which means there is a risk of overestimating the model's capabilities or encountering unexpected issues when applied in real-world scenarios~\cite{muhlbaueranalyzing}. Therefore, it is essential for future research to address this limitation by actively incorporating dynamic environments into their evaluation processes.

\suggestbox{
\faIcon{thumbs-up} \textit{\textbf{Actionable Suggestion \thesuggestioncount:} Account for dynamic workloads, versions, and hardware to gain a comprehensive understanding of model performance in practical settings.}
\addtocounter{suggestioncount}{1}
}

\begin{figure}[!t]
  \centering
  \begin{minipage}{0.45\textwidth}
    \centering
\includestandalone[width=\columnwidth]{Figures/RQ-environments}
    \caption{Distribution of the environment types.}
 \label{fig:environments}
  \end{minipage}
~\hspace{0.2cm}
~\vspace{0.4cm}
  \begin{minipage}{0.5\textwidth}
    \centering
~\vspace{0.25cm}
\includestandalone[width=\columnwidth]{Figures/RQ-repo}
    \caption{Distribution of the availability of artifacts.}
 \label{fig:repo}
  \end{minipage}
\end{figure}

\subsection{Publishing Source Code and Data}
\label{sec:rq4.3}


The availability of public repositories plays an essential role in promoting replicability, transparency, and development in research. This is particularly crucial for practical research fields such as configuration performance learning. The results of our survey in Figure~\ref{fig:repo} reveal that a significant number of studies do not provide an open-access repository containing the necessary resources for replication. In particular, out of the 99 papers surveyed, a total of 78 studies \textbf{do not make their code, data, or experiment results publicly available}~\cite{DBLP:journals/ese/VituiC21a, DBLP:conf/icst/PorresARLT20, DBLP:journals/jss/ZhuYZZ21, zain2022software, DBLP:journals/infsof/WartschinskiNVK22}. This indicates a lack of emphasis on open science practices and hinders the ability of other researchers and practitioners to validate and build upon the findings. On the other hand, 21 studies offer an open-access repository, of which 17 provide \textbf{both source code and datasets}~\cite{DBLP:conf/splc/GhamiziCPT19, DBLP:conf/kbse/BaoLWF19, DBLP:conf/icst/PorresARLT20, DBLP:journals/concurrency/OuaredCD22, DBLP:conf/wosp/DidonaQRT15}, and four of them \textbf{only provide source codes}~\cite{DBLP:conf/sigmod/ZhangLZLXCXWCLR19, DBLP:conf/kbse/ChenHL022, wang2021morphling, DBLP:journals/infsof/WartschinskiNVK22}. 


 \keybox{
\faIcon{search} \textit{\textbf{Finding \thefindingcount:} Only 21\% (21 out of 99) of the primary studies provide public repositories. An even lower proportion of them, i.e., 17\%, make both source code and datasets publicly available.}
\addtocounter{findingcount}{1}
}

\subsubsection{Good Practice} 21 out of the total studies have embraced the principles of open science by providing a public repository that includes source codes and datasets. This is a good practice as it allows researchers to replicate or reproduce the experiment results, build further research based on the existing knowledge, and collaborate with each other~\cite{DBLP:journals/tosem/LiuGXLGY22}.

\subsubsection{Bad Smell} It is concerning to note that a significant number of studies (79\%) still missed the opportunity to promote open science by not providing public repositories. This is a limitation as it restricts the replicability, reproducibility, and development of the community, which could be addressed by future studies. Further, among the 21 studies that offer an open-access repository, only 17 of them provide both source codes and datasets.

\textcolor{black}{One possible explanation for this bad smell is that there is a lack of a universal standard, guidelines, or best practices for using and sharing source codes and data, which makes it expensive to construct and maintain a public repository. Additionally, a lack of awareness or understanding regarding the benefits associated with open science may also contribute to the limited availability of public repositories.}

\suggestbox{
\faIcon{thumbs-up} \textit{\textbf{Actionable Suggestion \thesuggestioncount:} Future studies should aim to establish and follow universal standards for sharing resources and raise awareness about the benefits of open science.}
\addtocounter{suggestioncount}{1}
}



%% file: Tables/RQ-application.tex
\begin{table}[t!]
  \caption{Distribution of the model application purpose.}
\centering
\footnotesize
\begin{adjustbox}{width=\linewidth,center}
\begin{tabular}{llp{10cm}}
\toprule
\textbf{Category} &
  \textbf{\# Studies} &
  \textbf{References} \\ \hline
Prediction &
  65 &
\cite{DBLP:journals/cee/KundanSA21}, \cite{DBLP:journals/tsc/KumaraACMHT23}, \cite{DBLP:journals/jiii/YuGLZIY22}, \cite{DBLP:journals/taco/MoolchandaniKS22}, \cite{DBLP:journals/tcc/PhamDF20}, \cite{DBLP:conf/dsrt/DuongZCLZ16}, \cite{DBLP:conf/ic2e/RahmanL19}, \cite{DBLP:journals/fgcs/LiLTWHQD19}, \cite{DBLP:journals/pvldb/ZhouSLF20}, \cite{DBLP:journals/pvldb/MarcusP19}, \cite{DBLP:conf/pldi/SinghHLS22}, \cite{chai2023perfsage}, \cite{DBLP:journals/mam/JooyaDB19}, \cite{chokri2022performance}, \cite{DBLP:conf/icpp/MadireddyBCLLRS19}, \cite{DBLP:conf/spects/KimK17}, \cite{DBLP:conf/sc/MaratheAJBTKYRG17}, \cite{DBLP:conf/springsim/LuxWCBLXBBCH18}, \cite{said2021accurate}, \cite{DBLP:journals/access/ThraneZC20}, \cite{DBLP:journals/access/WangXTW20}, \cite{DBLP:journals/concurrency/CiciogluC22}, \cite{DBLP:conf/splc/Acher0LBJKBP22}, \cite{DBLP:conf/cluster/IsailaBWKLRH15}, \cite{DBLP:conf/sc/MalikFP09}, \cite{DBLP:conf/im/JohnssonMS19}, \cite{DBLP:conf/noms/SanzEJ22}, \cite{DBLP:journals/taco/WangLWB19}, \cite{DBLP:journals/concurrency/OuaredCD22}, \cite{DBLP:journals/jsa/ChengCWX17}, \cite{DBLP:journals/tecs/TrajkovicKHZ22}, \cite{DBLP:conf/wosp/DidonaQRT15}, \cite{DBLP:conf/middleware/GrohmannNIKL19}, \cite{DBLP:journals/corr/abs-2304-13032}, \cite{DBLP:conf/IEEEcloud/MarosMSALGHA19}, \cite{DBLP:journals/fgcs/KousiourisMKGV14}, \cite{DBLP:journals/concurrency/JiZL22}, \cite{DBLP:journals/access/NadeemAMFA19}, \cite{DBLP:conf/iccad/KimMMSR17}, \cite{DBLP:conf/nsdi/LiangFXYLZYZ23}, \cite{DBLP:conf/sc/MalakarBVMK18}, \cite{DBLP:conf/msr/GongC22}, \cite{DBLP:journals/jsa/ZhangLWWZH18}, \cite{DBLP:conf/nsdi/FuGMR21}, \cite{DBLP:conf/esem/ShuS0X20}, \cite{DBLP:conf/icccnt/KumarMBCA20}, \cite{DBLP:conf/sigsoft/Gong023}, \cite{du2013performance}, \cite{DBLP:conf/cf/LiuMCV20}, \cite{DBLP:journals/access/TousiL22}, \cite{DBLP:journals/tosem/ChengGZ23}, \cite{DBLP:conf/icse/HaZ19}, \cite{DBLP:conf/wosp/CengizFAM23}, \cite{DBLP:conf/icse/GaoGZLY23}, \cite{DBLP:conf/mascots/KarniavouraM17}, \cite{sabbeh2016performance}, \cite{DBLP:journals/tnse/CaoPC22}, \cite{DBLP:conf/cluster/AssogbaLRK23}, \cite{DBLP:conf/fse/GongC24}, \cite{DBLP:conf/hpdc/YokelsonCL23}, \cite{jiang2024ml}, \cite{DBLP:conf/qrs/ZhangWZ23}, \cite{DBLP:journals/jss/LiBHWL24}, \cite{wang2024slapp}, \cite{DBLP:journals/pomacs/LiPFLWH22} \\ \hline
Tuning &
  22 &
\cite{DBLP:journals/concurrency/FalchE17}, \cite{DBLP:journals/concurrency/RosarioSZNB23}, \cite{DBLP:conf/icpp/DouWZC22}, \cite{DBLP:conf/splc/GhamiziCPT19}, \cite{DBLP:journals/access/LiLSJ20}, \cite{DBLP:journals/comcom/AteeqAAK22}, \cite{DBLP:conf/infocom/LiHZLT20}, \cite{DBLP:conf/middleware/MahgoubWGMGHMGB17}, \cite{DBLP:conf/kbse/BaoLWF19}, \cite{DBLP:journals/tompecs/MakraniSNDSMRH21}, \cite{wang2021morphling}, \cite{ding2021portable}, \cite{DBLP:conf/ipps/FalchE15}, \cite{DBLP:journals/tse/ChenB17}, \cite{DBLP:conf/sigmod/ZhangLZLXCXWCLR19}, \cite{myung2021machine}, \cite{DBLP:journals/jsa/TangLLLZ22}, \cite{DBLP:journals/taco/LiL22},  \cite{blott2018finn}, \cite{DBLP:journals/ipm/LeeSCP24}, \cite{wyzykowski2024optimizing}, \cite{DBLP:conf/ics/TrumperBSCH23} \\ \hline
Testing &
  6 &
  \cite{DBLP:journals/ese/VituiC21a}, \cite{DBLP:conf/icst/PorresARLT20}, \cite{DBLP:conf/kbse/ChenHL022}, \cite{DBLP:journals/jss/ZhuYZZ21}, \cite{zain2022software}, \cite{DBLP:journals/infsof/WartschinskiNVK22} \\ \hline
Scheduling &
  5 &
  \cite{DBLP:journals/concurrency/YinH00C23}, \cite{DBLP:conf/sbac-pad/NemirovskyAMNUC17}, \cite{betting2023oikonomos}, \cite{li2024pattern}, \cite{DBLP:conf/hipc/BettingZS23} \\ \hline
Development &
  1 &
  \cite{DBLP:journals/pr/TranSWQ20}
\\ \bottomrule
\end{tabular}

\end{adjustbox}
\label{tb:application}
\end{table}

%% file: Figures/RQ-environments.tex
\begin{tikzpicture}[scale=0.8]
\tikzstyle{every node}=[font=\large]
    \pie[
        /tikz/every pin/.style={align=center},
        text=pin,
        rotate=30,
        explode=0, 
        ]
        {
        65/\texttt{None} (64),
        23/\texttt{Workload} (23),
        8/\texttt{Hardware} (8),
        3/\texttt{Mixed} (3)
        }

    \pie[
        /tikz/every pin/.style={align=center},
        text=pin,
        before number=\phantom,
        after number=,
        rotate=386.4, 
        explode=0, 
        color={red!60}, 
        ]
        {
        1/\texttt{1\% Version} (1)
        }

    \end{tikzpicture}

%% file: Figures/RQ-repo.tex
\begin{tikzpicture}[scale=0.8]
\tikzstyle{every node}=[font=\large]
    \pie[
        /tikz/every pin/.style={align=center},
        text=pin,
        rotate=30,
        explode=0, 
        ]
        {
        79/\texttt{None} (78),
        17/\texttt{Codes \& Data} (17),
        4/\texttt{Only codes} (4)
        }

    \end{tikzpicture}

%% file: oppertunities.tex
\section{Future Research Opportunities}
\label{sec:gap}
Our survey has revealed several key knowledge gaps and promising directions that are worth further investigation in future studies, which we elaborate on below. 





\subsection{Model-Based Sampling for Deep Configuration Performance Learning} 
From RQ1 (Section~\ref{sec:rq1.3}), we see that while several sampling strategies have been employed in the studies reviewed, it is evident that a significant majority of them rely on random sampling, which highlights a significant opportunity for future research to explore and develop more effective sampling strategies. Random sampling, while straightforward, may not always yield the most informative or representative samples within the configuration space~\cite{DBLP:conf/icse/KalteneckerGSGA19, DBLP:conf/kbse/SarkarGSAC15, DBLP:conf/sigsoft/JamshidiVKS18}. 
By exploring alternative sampling techniques that select samples based on specific criteria or heuristics, researchers can potentially improve the efficiency and effectiveness of the models~\cite{DBLP:conf/wosp/PereiraA0J20}. This is especially important for deep learning configuration performance since it is known that the deep learning model is sensitive to the quality of training data~\cite{DBLP:conf/wosp/PereiraA0J20, DBLP:conf/sigsoft/JamshidiVKS18, DBLP:conf/kbse/SarkarGSAC15, DBLP:conf/icse/KalteneckerGSGA19}.

Among others, addressing this gap by investigating model-based sampling methods that can provide an explicit acquisition function to reliably and accurately guide the sampling process is a promising future direction. For instance, Sequential Model-Based Optimization (SMBO)~\cite{wang2021morphling}, such as Bayesian Optimization, effectively explores the sample space by quantifying the uncertainty of the samples to guide the search towards potentially optimal regions~\cite{DBLP:conf/icdcs/HsuNFM18, DBLP:conf/nips/SnoekLA12, DBLP:conf/eurosys/AlabedY21}. In particular, using a surrogate model, such as a Gaussian process, to sequentially select new samples based on the current model's predictions and uncertainty estimates, would enable efficient exploration of the sample space. However, to the best of our knowledge, there has been little work that investigates this thread of research.

\subsection{Explainable Deep Configuration Performance Learning.}

Through RQ2 (Section~\ref{sec:rq2.2}), we reveal that various types of deep learning models have been used to learn configuration performance, however, almost all of the proposed deep learning models lack explainability. Unlike analytical performance models, which enable researchers to analyze the importance of options and the interactions between them~\cite{DBLP:conf/icse/VelezJSAK21}, deep learning models learn the configuration performance in a black-box manner, which could be harmful to the reliability of the results. 

In contrast, certain statistical machine learning performance models are naturally interpretable. For instance, \texttt{SPLConqueror}~\cite{DBLP:conf/sigsoft/SiegmundGAK15} propose SPLConqueror leverages a linear model to learn configuration performance, enabling one to quantify the influence of each configuration;~\citet{DBLP:conf/msr/GongC22, DBLP:conf/sigsoft/Gong023, DBLP:journals/ese/GuoYSASVCWY18} have examined Random Forest (RF) and Classification and Regression Tree (CART) for performance learning, highlighting the importance of options and the decision boundaries to cluster the configurations. Several studies have explored Bayesian theory to analyze the uncertainty of predictions that aid the explainability~\cite{DBLP:conf/icdcs/HsuNFM18, DBLP:journals/pr/TranSWQ20, Agarwal19, zain2022software}.

Yet, addressing the above is challenging for deep learning model due to their nonlinear and high-dimensional representation nature. Luckily, explainable deep learning has been studied in various domains, such as pattern recognition tasks including medical diagnosis, face recognition, and self-driving cars~\cite{DBLP:journals/pr/0001WLLSSK21}, including those model-agnostic ones like Local Interpretable Model-agnostic Explanation (LIME)~\cite{DBLP:journals/corr/RibeiroSG16a} and SHapley Additive exPlanations (SHAP)~\cite{DBLP:conf/nips/LundbergL17}. We can directly leverage the advance of the current explainable extension of deep learning, and specialize them in the context of configuration performance. However, achieving explainable deep configuration performance models faces several challenges, e.g., 
\begin{itemize}
    \item How to quantify the internal elements of a deep learning model, e.g., activation function, with respect to their contributions according to the type and characteristics of the configuration options.

    \item How to equip a deep learning model with an intrinsic method that produces explanations on the configuration performance alongside its outputs.

    \item How to visualize the configuration options and their contributions toward the final prediction with a deep learning model.
\end{itemize}
Pushing the research toward addressing any of the above would be of immediate interest and benefit to the field.

\subsection{Configuration Performance Modeling with Few-Shot Deep Learning}

Deep neural networks for learning configuration performance often require a substantial amount of labeled data, which can be challenging and time-consuming to obtain, as the availability of labeled data is often limited due to factors such as the complexity of software systems, the cost of performance measurements, and the lack of expert knowledge.

A promising direction of most efficient deep learning approaches is to complement it with few-shot learning, which aims at training to quickly adapt and generalize to new tasks or classes with only a few labeled configuration samples or even a single configuration, e.g., siamese networks, memory-augmented networks, transfer learning, meta-learning, and augmentation strategies~\cite{DBLP:journals/csur/WangYKN20, DBLP:conf/isda/KadamV18}. For example, augmentation techniques like Generative Adversarial Networks have been used in three studies to generate additional training samples to augment the available configuration dataset~\cite{DBLP:conf/esem/ShuS0X20, DBLP:journals/access/LiLSJ20, DBLP:conf/kbse/BaoLWF19}, and siamese neural networks are able to differentiate between pairs of examples, making them effective for few-shot learning, which have been successfully used for image processing tasks~\cite{DBLP:conf/bigcom/GuoSGBJUK22}.


Despite the above, few-shot deep learning still imposes several challenges for learning configuration data due to the complexity and specialty of configurable software systems. For example, the selected most representative data samples are difficult to quantify, obscuring effective generalizability in the few-shot deep learning model. Therefore, further research is needed to unlock the full potential of few-shot deep learning for configuration performance modeling by enhancing robustness, generalizability, and transferability.

\subsection{Interactive Deep Configuration Performance Learning.}
While deep learning models have demonstrated effectiveness in configuration performance learning, they come with inherent challenges, notably the substantial training overhead when compared to other machine learning models. The resource-intensive nature of training deep learning models poses limitations on the adaptability and reliability, especially when quick adaptation to newly measured data or real-time updates is required. This issue underscores the urgent need for innovative solutions that can efficiently update and optimize deep learning models, paving the way for more updated and reliable systems.

To address these challenges, interactive deep learning techniques have emerged as a powerful solution, which enables continuous learning and adaptation optionally with the support of humans, overcoming the limitations associated with the static nature of traditional deep learning models and fully benefiting from the domain expertise. By incorporating real-time updates and leveraging user feedback, interactive deep learning ensures that models remain accurate and up-to-date using data filtered by human knowledge. This adaptability is particularly crucial in applications where the underlying data distribution may change over time as the new measurements become available~\cite{DBLP:journals/tompecs/MakraniSNDSMRH21, DBLP:journals/tse/ChenB17}. Despite being a promising direction related to the human-in-the-loop techniques for configuration performance modeling, interactive updates of deep configuration performance models are currently under-explored.

\textcolor{black}{Furthermore, an emerging technique in interactive deep learning is the use of pre-trained large language models (LLMs), which have demonstrated significant utility in numerous software engineering tasks, including code generation, bug detection, and requirement analysis. These models, trained on vast corpora of text, can provide powerful understanding and generation capabilities, making them potentially useful for software configuration performance learning. Potential usage scenarios for LLMs in this context include automatically generating optimal configuration settings based on system requirements and historical performance data, predicting the impact of configuration changes on system performance by reading the configuration setup files, and diagnosing performance bottlenecks by analyzing logs. }

However, the adoption of pre-trained LLMs in this domain is prevented by several  factors and challenges: 

\begin{itemize}

    \item Configuration performance learning is primarily based on relatively simple tubular data, while LLM is designed to handle complex modalities, such as texts or code. As such, how to better exploit the strengths of LLM is still unclear in the field. Of course, a potential way is to add extra modalities such as configuration code, documentation, and logs, but in that case, information fusion becomes a key challenge.

    \item There is a need to design sophisticated prompts to fine-tune these models to the specific problem of learning configuration performance which is not a straightforward task.

    \item Their generalizability to configurable software systems in diverse domains and environments such as database software vs. video encoding software and distributed computing vs. local computing, is still unknown. 

    \item The computational overhead associated with deploying and obtaining inference from LLMs could be the barrier to adopting them in real-time performance prediction scenarios.
\end{itemize}

Nevertheless, during our search, we did find two studies that leveraged LLMs for configuration tuning or performance modeling. Firstly, by leveraging the natural language processing capabilities of Generative Pre-trained Transformer (GPT),~\citet{DBLP:journals/corr/abs-2311-03157} propose GPTune to analyze, filter, summarize, and check the consistency of the domain knowledge extracted from various sources, ultimately enhancing the tuning process and optimizing the DBMS configuration effectively. 
\textcolor{black}{Yet, this study mainly concentrates on tuning the configuration and does not involve the process of modeling, and therefore lies outside the scope of this survey.} 
Secondly, in the work by~\citet{DBLP:journals/corr/abs-2306-17281}, an LLM is fine-tuned on a curated dataset containing HPC and scientific codes, and is employed to predict the relative performance impact of changes made to the source code. Specifically, given two versions of a code, the model can analyze and predict which version is likely to perform better in terms of execution time. \textcolor{black}{Nonetheless, they were filtered out in the stage when applying the exclusion criteria, since they had not been published in any peer-reviewed public venue by the end of our search procedures (till May 2024).}

As such, addressing these challenges is non-trivial and requires further research, which has not yet been achieved for deep configuration performance learning during the period covered by our survey. However, we anticipate that those challenges will be tackled in the community and we foresee a future where more research about LLM adoption will gradually appear. 



\subsection{Deep Configuration Performance Learning under Multiple Dynamic Environments}
From RQ4, it is evident that a majority, specifically 66\%, do not address the dynamic and varying nature of software running environments, which may significantly limit the effectiveness and generalizability of deep configuration performance models. In particular,~\citet{muhlbaueranalyzing} systematically analyzes the influence of workloads on configurable software systems and discloses that varying the workload does not only affect the performance value but also affects the relationships between the configurations and performance.

This observation highlights the need to address this gap by developing deep learning methodologies that can effectively handle configuration data across multiple environments. Little work has tackled the above by using, e.g., multi-task learning~\cite{DBLP:conf/iclr/YangH17, DBLP:journals/corr/abs-2106-02716, DBLP:conf/eurosys/AlabedY21, DBLP:conf/icml/StandleyZCGMS20}, which learns the common representations between multiple tasks simultaneously; transfer learning~\cite{DBLP:conf/icccnt/KumarMBCA20, DBLP:conf/noms/SanzEJ22, DBLP:conf/bigdataconf/GhoshalWPS19, DBLP:conf/icse/LiX0WT20, DBLP:conf/im/JohnssonMS19}, 
which seek to leverage the common information gained from related environments, enabling models to adapt and generalize to the target environments; or meta-learning~\cite{DBLP:conf/icml/YaoWHL19, DBLP:journals/air/VilaltaD02, DBLP:journals/pami/HospedalesAMS22, DBLP:conf/icse/ChaiZSG22, DBLP:conf/sigir/ZhangFW00LZ20}, 
which pretrain a set of model parameters via meta-training and can adapt quickly or unseen tasks or environments. Particularly, a recent work~\cite{DBLP:conf/fse/GongC24} proposes a sequential meta-learning framework, namely \texttt{SeMPL}, that trains a meta-model with multiple meta-environments in an optimal order such that the pre-trained model can generalize to new environments well. Yet, most of the existing work only assumes homogeneous configuration performance learning, i.e., for different environments, the configuration options need to remain the same. Further, those works do not cater to online learning~\cite{DBLP:conf/icse/Chen19b,DBLP:conf/wosp/0001BWY18}, where the data is continually used to update the model as it becomes available. As a result, deep learning models that can learn from multiple environments and are capable of self-updating at runtime are in high demand for configuration performance learning~\cite{DBLP:journals/tsc/ChenB17,DBLP:conf/saner/Chen22,DBLP:journals/pieee/ChenBY20}.

%% file: threats.tex
\section{Threats to validity}
\label{sec:threats}

When conducting a systematic literature review, it is important to consider potential threats to validity that could affect the reliability and generalizability of the findings. We now discuss the threats to the validity related to this work.


 

%

\subsection{Sampling Bias}
Sampling bias refers to the potential threats of the results due to the selection of studies. In this survey, sampling bias could arise from the selection of primary papers from indexing services or the inclusion and exclusion of certain papers. To mitigate this threat, we have codified specific rules, procedures and criteria in the search protocol, which is based on the guidance provided by~\citet{DBLP:journals/infsof/KitchenhamBBTBL09}. Specifically, the sampling of studies is done by a comprehensive automatic search via six popular indexes in software engineering to cover a wide range of papers, then the application of domain knowledge to remove the redundant and irrelevant studies, and filtering them with the carefully crafted inclusion and exclusion criteria to further remove potential bias, as summarized in Figure~\ref{fig:protocol}.

\subsection{Internal Threats}
Internal threats to validity relate to issues within the study design or data analysis that could affect the accuracy and reliability of the results. In this survey, potential internal threats could include inconsistencies in data extraction, subjectivity in data analysis, or biased interpretation of the findings. To address these threats, we have followed a systematic and rigorous approach to data extraction and analysis. By clearly defining research questions, it ensures that the data extraction process is focused and consistent. Additionally, we have limited this by employing multiple reviewers and by conducting three iterations of independent paper reviews among the authors. Error checks and investigations were also conducted to correct any issues found during the search. Any discrepancy in the results was discussed until an agreement can be reached.

\subsection{External Threats}
External threats to validity are usually related to the generalizability and applicability of the findings beyond the specific context of the survey. In this survey, external threats could arise from the limited time range from 2013 to 2023, or the limited scope of the papers using deep learning for performance modeling. To mitigate these threats, we have searched 948 studies from six indexing services, we clearly defined the inclusion and exclusion criteria for the selection of papers, based on which we extracted 85 prominent primary studies for detailed analysis. By including a diverse range of studies from different indexing services, we provide a broader perspective on deep configuration performance learning. 


%% file: conclusion.tex
\section{Conclusion}
\label{sec:conclusion}


\textcolor{black}{In this work, we present a comprehensive survey on the increasingly popular topic of deep configuration performance learning, covering 99 prominent works from 1,206 studies found on six indexing services.} We provide detailed taxonomy, together with discussions on the advantages, disadvantages, and best-suited scenarios of the various techniques involved, according to the key phases in building a deep configuration performance model, i.e., preparation, modeling, evaluation, and applications. Our results also reveal several good practices, bad smells, and actionable suggestions of the existing studies in the field.

More importantly, we highlight promising research opportunities for this particular research field, namely:

\begin{itemize}
    \item model-based sampling methods for deep configuration performance learning.
    \item explainable deep configuration performance model.
    \item interactively learned deep configuration performance model.
    \item deep few-shot learning for modeling configuration performance.
    \item deep configuration performance learning under multiple and dynamic environments.
\end{itemize}

By highlighting the identified trends and future directions of deep learning for configuration performance modeling---we hope to inspire sustainable growth on this particular topic.